\documentclass[useAMS,usenatbib,babel,superscriptaddress]{mnras}

\usepackage[english,english]{babel}
\usepackage[tbtags]{amsmath}
\usepackage{amssymb,amsfonts,textcomp}
\usepackage{array}
\usepackage{tabularx}
\usepackage{hhline}
\usepackage{hyperref}
\usepackage{soul}
\usepackage[usenames]{color}
\usepackage{balance}
\usepackage{times}
\usepackage{mathtools}
\usepackage{transparent}
\usepackage{bm}           
 \newcommand{\sfrac}[2]{#1/#2}
\usepackage{balance}

\usepackage{wasysym}
\newcommand{\SI}[2]{\ensuremath{#1~\mathrm{#2}}}
\newcommand{\num}[1]{\ensuremath{#1}}

\usepackage{ifthen}
\usepackage{pdflscape}
\usepackage{rotating}
\usepackage{geometry}
\usepackage{graphicx}

\def\gtrsim{\lower.5ex\hbox{$\; \buildrel > \over \sim \;$}}

\newcommand{\half}{{\sfrac{1}{2}} }
\newcommand{\form}{\mathrm{f}}
\newcommand{\nuf}{\nu_{\form}}

\newcommand{\D}{\mathrm{D}}
\newcommand{\G}{\mathrm{G}}

\newcommand{\dd}{\mathrm{d}}

\newcommand{\Ds}{\Delta\sigma}
\newcommand{\pd}{\partial}
\newcommand{\dc}{\delta_{\mathrm{c}}}
\newcommand{\nuc}{\nu_{\mathrm{c}}}
\newcommand{\nucS}{\nu_{\mathrm{c},\S}}
\newcommand{\nucv}{\nu_{\mathrm{c},v}}

\newcommand{\tnu}{\tilde\nu}
\newcommand{\Df}{D_{\mathrm{f}}}

\newcommand{\kk}{\mathbf{k}}
\newcommand{\rr}{\mathbf{r}}

\renewcommand{\S}{\mathcal{S}}
\newcommand{\tr}{\mathrm{tr}}
\newcommand{\erf}{\mathrm{erf}}
\newcommand{\sigmas}{\sigma_\S}
\newcommand{\Rs}{R_\S}
\newcommand{\nus}{\nu_{\!\S}}

\newcommand{\msun}{\ensuremath{M_\odot}}

\newcommand{\condmean}[2]{
  \langle #1
  |#2
  \rangle}
\newcommand{\mean}[1]{
  \langle #1
  \rangle}
\newcommand{\kron}{{\delta}}
\newcommand{\Var}[2]{
  \mathrm{Var}\!\left( #1 \middle |
    #2
  \right)}
\newcommand{\Cov}[2]{
  \mathrm{Cov}\!\left( #1 \middle |
    #2
  \right)}
\renewcommand{\d}{\mathrm{d}}             
\renewcommand{\P}{\mathbb{P}}             
\renewcommand{\vec}[1]{{\bm{#1}}}         

\newcommand{\shear}{q}
\newcommand{\fup}{f_\mathrm{up}}
\newcommand{\up}{\mathrm{up}}
\newcommand{\first}{\mathrm{first}}

\def \prd {{\it Phys. Rev. D}}

\def \apj {{\it ApJ}}
\def \apjl {{\it ApJ Let.}}
\def \apjs {{\it ApJ Sup.}}

\def \aap {{\it A\&A}}

\def \mnras {{\it MNRAS}}

\def \jcap {{\it JCAP} }

\definecolor{grey}{rgb}{0.75,0.75,0.75}
\definecolor{Orange}{rgb}{1.0,0.5,0.15}
\definecolor{brown}{rgb}{0.7,0.25,0.0}
\definecolor{pink}{rgb}{1.0,0.5,0.5}
\definecolor{darkerred}{rgb}{0.8,0,0}
\definecolor{darkerblue}{rgb}{0,0,0.8}
\definecolor{Blue}{rgb}{0,0.08,0.65}
\definecolor{Red}{rgb}{0.65,0.08,0.05}
\definecolor{Green}{rgb}{0.15,0.45,0.25}

\graphicspath{{schemes/}{plots/}}

\begin{document}
\author[M. Musso, C. Cadiou, C. Pichon, S. Codis,  K. Kraljic  and  Y. Dubois]{
\parbox[t]{\textwidth}{M. Musso$^{1,2}$\thanks{E-mail: mmusso@sas.upenn.edu}, C. Cadiou,$^{1}$\thanks{E-mail: cadiou@iap.fr}, C. Pichon$^{1,3}$,  S. Codis$^{4}$,
K. Kraljic$^5$  and  Y. Dubois$^1$}
\vspace*{6pt}\\
\noindent$^{1}$ Institut d'Astrophysique de Paris, CNRS \& UPMC, UMR 7095, 98 bis Boulevard Arago, 75014, Paris, France\\
\noindent$^{2}$ Institut de physique th\'eorique, Universit\'e Paris Saclay \& CEA, CNRS, 91191 Gif-sur-Yvette, France\\
\noindent$^{3}$ Korea Institute of Advanced Studies (KIAS) 85 Hoegiro, Dongdaemun-gu, Seoul, 02455, Republic of Korea\\
\noindent$^{4}$  Canadian Institute for Theoretical Astrophysics, University of Toronto, 60 St. George Street, Toronto, ON M5S 3H8, Canada\\
\noindent$^{5}$ {Aix Marseille Univ, CNRS, LAM, Laboratoire d'Astrophysique de Marseille, Marseille, France}
}

\title[How does the cosmic web impact assembly bias?]{How does the cosmic web impact assembly bias?}
\maketitle

\begin{abstract}
{
The mass, accretion rate and formation time of dark matter haloes near proto-filaments (identified as saddle points of the potential) are analytically predicted using a conditional version of the excursion set approach in its so-called ``upcrossing'' approximation.
The model predicts that at fixed mass, mass accretion rate and formation time vary with orientation and distance from the saddle,
demonstrating that assembly bias is indeed influenced by the tides imposed by the cosmic web.
Starved, early forming haloes of smaller mass lie preferentially along the main axis of filaments, while more massive and younger haloes are found closer to the nodes.
Distinct gradients for distinct tracers such as typical mass and accretion rate occur because the saddle condition is anisotropic, and because the statistics of these observables depend on both the conditional means and their covariances.
The theory is extended to other critical points of the potential field.
The response of the mass function to variations of the matter density field (the so-called large scale bias) is computed, and its trend with accretion rate is shown to invert along the filament.\\
The signature of this model should correspond at low redshift to an
excess of reddened galactic hosts at fixed mass along preferred directions, as recently reported in
spectroscopic and photometric surveys and in hydrodynamical simulations.
The anisotropy of the cosmic web emerges therefore as a significant ingredient to describe jointly
the dynamics and physics of galaxies, e.g. in the context of intrinsic alignments or morphological diversity.
}
\end{abstract}



\allowdisplaybreaks

\section{Introduction}

The standard paradigm of galaxy formation primarily assigns galactic properties to their host halo mass. While this assumption has proven to be very successful, more precise theoretical and observational considerations suggest other hidden variables must be taken into account.

The mass-density relation \citep{oemler74}, established observationally 40 years ago, was explained \citep[][]{Kaiser1984,efstathiouetal1988} via the impact of the long wavelength density modes of the dark matter field, allowing the proto-halo to pass earlier the critical threshold of collapse \citep{Bondetal1991}. This biases the mass function in the vicinity of the large-scale structure: the abundance of massive haloes is enhanced in overdense regions.

Numerical simulations have shown that denser environments display a population of smaller, older, highly concentrated `stalled' haloes,  which have stopped accreting and whose relationship with the environment is in many ways the opposite of that of large-mass actively accreting haloes that dominate their surroundings. This is the so-called  ``assembly bias'' \citep[e.g.][]{shethettormen2004,gaoetal2005,wechsleretal2006,dalaletal2008,ParanjPadma2017,Lazeyras2017}. More recently,
\cite{Alonso:2015eu,Tramonte2017,Braun-Bates2017}  have investigated the differential properties of haloes w.r.t.\ loci in the cosmic web. As they focused their attention to variations of the mass function, they also found them to vary mostly with the underlying density.
\cite{paranjapeetal2017} have shown that haloes
in nodes and in filaments behave as two distinct populations when a
suitable variable based on the shear strength on a scale of the order
of the halo's turnaround radius is considered.

In observations,
galactic conformity \citep{weinmannetal2006} relates quenching of centrals to the quenching of their satellite galaxies. It has been detected for low and high mass satellite galaxies up to high redshift   \citep[$z\sim2.5$,][]{kawinetal2016}  and fairly large separation  \citep[\SI{4}{Mpc},][]{2013MNRAS.430.1447K}.
 Recently, colour and type gradients driven specifically by the
 anisotropic geometry of the filamentary network have also been
 found in simulations and observations using SDSS
 (\citealt{Yan2013}; \mbox{\citealt{martinezetal2016}}; \citealt{poudeletal2016};
 \citealt{chenetal2017}), GAMA \citep[][Kraljic et al. submitted]{alpaslanetal2016} and, at higher redshift, VIPERS \citep[][]{Malavasi2016b} and COSMOS \citep{Laigle+17}.
This suggests that some galactic properties do not only depend on halo mass and density alone:
the co-evolution of conformal galaxies is likely to be connected to their evolution within the same large-scale anisotropic tidal field.

An improved model for galaxy evolution should explicitly integrate the diversity of the geometry of the environment on multiple scales
and the position of galaxies within this landscape
 to quantify the impact of its anisotropy on galactic mass assembly history.
From a theoretical perspective, at a given mass, if the halo is
sufficiently far from competing potential wells, it can grow by
accretion from its neighbourhood. It is therefore natural to expect,
at fixed mass, a strong correlation between the accretion rate of haloes and the density of their environment \citep[][]{Zentner2007,MussoShethmarkov2014}. Conversely, if this halo lies in the
vicinity of a more massive structure, it may stop growing earlier and
stall because its expected feeding will in fact recede towards the
source of anisotropic tide
\citep[e.g.][]{dalaletal2008,Hahnetal2009,Ludlow2011,wangetal2011}.

Most of the work carried out so far has focused on the role of the shear strength (a scalar quantity constructed out of the traceless shear tensor  which does not correlate with the local density) measured on the same scale of the halo: as tidal forces act against collapse, the strength of the tide will modify the relationship of the halo with its large-scale density environments, and induce distinct mass assembly histories by dynamically quenching mass inflow
\citep[][]{Hahnetal2009,Castorina:2016vg,2016arXiv161004231B}.
Such local shear strength should be added, possibly in the form of a modified collapse model that accounts for tidal deformations, so as to capture e.g. the effect of a central on its satellites' accretion rate.
This modified collapse model has been motivated in the literature on various grounds, e.g. as a phenomenological explanation of the scale-dependent scatter in the initial overdensity of proto-haloes measured in simulations \citep{Ludlow2011,ShethChanScocc2013} or as a theoretical consequence of the coupling between the shear and the inertia tensor which tends to slow down collapse \citep{BM1996,SMT2001,DelPop2001}.
Notwithstanding, the position within the large-scale anisotropic
cosmic web  also directly conditions the local statistics, even without a
modification of the collapse model, and affects different observables
(mass, accretion rate etc.) differently.

The purpose of this paper is to provide a mathematical understanding of how assembly bias is  indeed partially {\sl driven} by the anisotropy of large scale tides
imprinted in the so-called cosmic web.
To do so,  the formalism of excursion sets will be adapted to study the formation of structures in the vicinity of saddle points as a proxy for filaments of the cosmic web.
Specifically, various tracers of galactic assembly will be computed conditional to the presence of such anisotropic large-scale structure.
This will allow us to understand why haloes of a given mass and local density stall near saddles or nodes, an effect which is not captured by the density-mass relation,
as it is driven solely from the traceless part of the tide tensor.
This should have a clear signature in terms of the distinctions between contours of constant typical halo mass versus
those of constant accretion rate, which may in turn explain the distinct mass and colour gradients
recently detected in the above-mentioned surveys.

The structure of this paper is the following.
Section~\ref{sec:excursionsets} presents a motivation for
extended excursion set theory as a mean to compute tracers of assembly
bias.
Section~\ref{sec:unconditional} presents the unconstrained
expectations for the mass accretion rate and half-mass.
Section~\ref{sec:conditional} investigates the same statistics subject
to a saddle point of the potential and computes the induced map of
shifted mass, accretion rate, concentration and half mass time. It relies on the strong symmetry between 
the unconditional and conditional statistics. 
Section~\ref{sec:astro} provides a compact alternative to the previous
two sections for the less theoretically inclined reader and presents
directly the joint conditional and marginal probabilities of
upcrossings explicitly as a function of mass and accretion rate.
Section~\ref{sec:bias} reframes our results in the context of the theory of bias as
the response of the mass function to variations of the matter density field.
Section~\ref{sec:conclusion} wraps up and discusses perspectives.
Appendix~\ref{sec:definitions} sums up the definitions and conventions
used in the text.
Appendix~\ref{sec:validation}
tests these predictions on realizations of Gaussian random fields.
Appendix~\ref{sec:other-critical} investigates the conditional
statistics subject to the other critical points of the field.
Appendix~\ref{sec:pdf-saddles} presents the PDF of the eigenvalues at the saddle.
Appendix~\ref{sec:covariance}  presents the covariance matrix of the relevant
variables to the PDFs.
Appendix~\ref{sec:statistics} presents the relevant joint statistics
of the field and its derivatives (spatial and w.r.t. to filtering) and
the corresponding conditional statistics of interest.
Appendix~\ref{sec:moving-barr-gener} presents the generalization of
the results for a generic barrier.
Appendix~\ref{sec:speculations} speculates about  galactic colours.

\section{Basics of the excursion set approach}
\label{sec:excursionsets}

\allowdisplaybreaks

The excursion set approach, originally formulated by \cite{PressSchechter1974}, assumes that virialized haloes form from spherical regions whose initial mean density equals some critical value.
The distribution of late-time haloes can thus be inferred from the simpler Gaussian statistics of their Lagrangian progenitors. The approach implicitly assumes approximate spherical symmetry (but not homogeneity), and uses spherical collapse to establish a mapping between the initial mean density of a patch and the time at which it recollapses under its own gravity.

According to this model, a sphere of initial radius $R$ shrinks to
zero volume at redshift $z$ if its initial mean overdensity $\delta$
equals $\dc D(z_{\mathrm{in}})/D(z)$, where $D(z)$ is the growth rate
of linear matter perturbations, $z_\mathrm{in}$ the initial redshift,
and $\dc=1.686$ for an Einstein--de\ Sitter universe, or equivalently, if its mean
overdensity linearly evolved to $z=0$ equals $\dc/D(z)$, regardless of
the initial size. If so, thanks to mass conservation, this spherical
patch will form a halo of mass $M=(4\pi/3) R^3\bar\rho$
(where $\bar\rho$ is the comoving background density). The
redshift $z$ is assumed to be a proxy for its virialization time.

\cite{Bondetal1991} added to this framework the requirement that the
mean overdensity in all larger spheres must be lower than $\dc$, for outer shells to collapse at a later time. This condition
ensures that the infall of shells is hierarchical, and the selected
patch is not crushed in a bigger volume that collapses faster (the so-called \emph{cloud-in-cloud} problem).  The number density
of haloes of a given mass at a given redshift is thus related to the
volume contained in the largest spheres whose mean overdensity $\delta\equiv\delta(R)$
crosses $\dc$. The dependence of the critical value $\dc$ on departures from spherical collapse induced by initial tides was studied by \cite{BM1996}, and later by \cite{SMT2001}, who approximated it as a scale-dependent barrier. This will be further discussed in Section~\ref{sec:perspectives}.

As the variation of $\delta(R)$ with scale resembles random diffusion,
it is convenient to parametrize it with the variance
\begin{equation}
  \sigma^2(R)\equiv\mathrm{Var}(\delta(R)) =
  \int\d k \frac{k^2P(k)}{2\pi^2} W^2(kR) \label{eq:sigma2}
\end{equation}
of the stochastic process, smoothed with a real-space Top-Hat
filter $W$\footnote{The window function in Fourier space is
  $W(x)=3j_1(x)/x$, $j_1$ being the spherical Bessel function of order
  1.}, rather than with $R$ or $M$. In equation~\eqref{eq:sigma2}, $P(k)$ is the underlying power spectrum. The three quantities $\sigma$, $R$
and $M$ are in practice interchangeable. The mass fraction in haloes of
mass $M$ at $z$ is
\begin{equation}
  \frac{M}{\bar\rho} \frac{\d n}{\d M} =
  \left|\frac{\d \sigma}{\d M}\right| f(\sigma)\,,
\label{eq:massfrac}
\end{equation}
where $\d n/\d M$ is the number density of haloes per unit mass
(i.e. the mass function) and $f(\sigma)$ --  often called the halo
multiplicity -- is the probability distribution of the first-crossing
scale of the random walks, that is of the smallest $\sigma$ (largest
$R$) for which
\begin{equation}
 \delta(R,\rr) \equiv
 \int\frac{\d^3k}{(2\pi)^3}\delta_m(\kk) W(kR) e^{i \kk\cdot\rr}
 = \frac{\dc}{D(z)}\,,
\label{eq:cross}
\end{equation}
where $\delta_m$ is the (unsmoothed) matter density.
The first-crossing requirement avoids double counting and guarantees that $f(\sigma)$ is a well behaved probability distribution, and the resulting mass fraction is correctly normalized.
In equation~\eqref{eq:cross}, the linear growth factor, $D(z)$, is defined as a function of redshift via
\begin{equation}
D(z)=\frac{H(a)}{H_0}\!\int_0^a \!\!\frac{\dd a}{\sqrt{\Omega_m/a+\Omega_\Lambda a^2}^{3}}\,, \,\, {\rm with} \,\,  a=\frac{1}{1+z}\,.
\label{eq:defD}
\end{equation}
 At early time, $D(z)$ scales like $1/(1+z)$. Here $H(a) = H_0 \sqrt{\Omega_m/a+\Omega_\Lambda a^2} $ is the Hubble Constant.

\begin{figure}
  \begin{center}
    \includegraphics{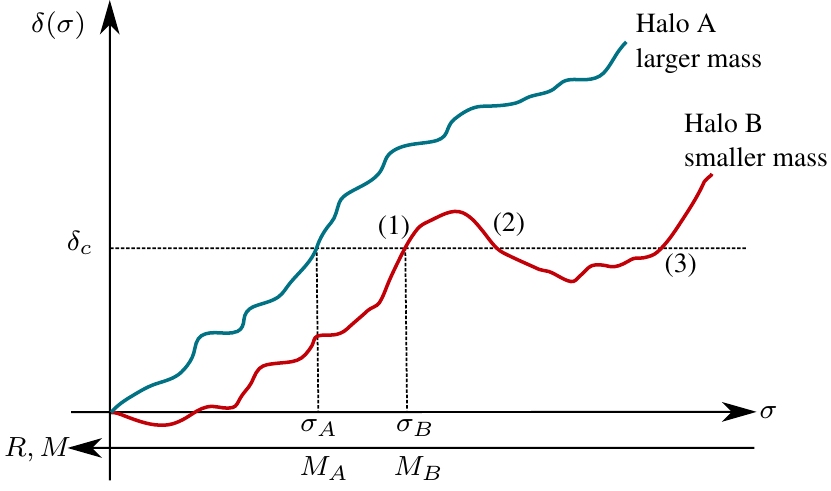}
      \caption{\label{fig:sketchup} Pictorial description of the
      first-crossing and upcrossing conditions to infer the halo mass
      from the excursion set trajectory. The first-crossing condition
      on $\sigma$ assigns at most one halo to each trajectory, with
      mass $M(\sigma)$. Upcrossing may instead assign several masses
      to the same trajectory (that is, to the same spatial location),
      thus over-counting haloes. Trajectory B in the figure has a first
      crossing (upwards) at scale $\sigma_B$ (1), a downcrossing (2)
      and second upcrossing (3), but the correct mass is only given by
      $\sigma_B$. However, the correlation of each step with the
      previous ones makes turns in small intervals of $\sigma$
      exponentially unlikely: at small $\sigma$ most trajectories will
      thus look like trajectory A. Thanks to the correlation between
      steps at different scales, for small $\sigma$ (large $M$) simply
      discarding downcrossings is a very good approximation.  }
  \end{center}
\end{figure}

The first-crossing probability, $f(\sigma)\Ds$, is the fraction of
walks that cross the threshold between $\sigma-\Ds$ and $\sigma$ for
the first time. Considering discretized trajectories with a large
number of steps $\sigma_1,\dots,\sigma_N$ of width
$\Ds\equiv\sigma_i-\sigma_{i-1}$ (corresponding to concentric spheres of radii $R_1,\cdots, R_N$),  the first-crossing probability is the joint probability that
$\delta_N>\dc$ and $\delta_i<\dc$ for $i<N$, with
$\delta_i\equiv\delta(\sigma_i)$ and $\sigma_N=\sigma=N\Ds$. Hence, the
distribution $f(\sigma)$ is formally defined as the limit
\begin{equation}
  f(\sigma) \equiv \lim_{\Ds\to0} \frac{1}{\Ds}
  \mean{\vartheta(\delta_N-\dc)\prod_{i}^{N-1}\vartheta(\dc-\delta_i)}\,,
\label{eq:ffirst}
\end{equation}
where $\vartheta(x)$ is Heaviside's step function, and the expectation value is evaluated with the multivariate distribution $p(\delta_i,\dots,\delta_N)$.
This definition discards crossings for which $\delta_i>\dc$ for any $i<N$, since $\vartheta(\dc-\delta_i)=0$, assigning at most one crossing (the first) to each trajectory. For instance, in Fig.~\ref{fig:sketchup}, trajectory B would not be assigned the crossing marked with (3), since the trajectory lies above threshold between (1) and (2). Since taking the mean implies integrating over all trajectories weighed by their probability, $f(\sigma)$ can be interpreted as a path integral over all allowed trajectories with fixed boundary conditions $\delta(0)=0$ and $\delta(\sigma)=\dc$ \citep{MaggioreRiotto2010}.

In practice, computing $f(\sigma)$ becomes difficult if the steps of the random walks are correlated, as is the case for real-space Top-Hat filtering with a $\Lambda$CDM power spectrum, and for most realistic filters and cosmologies.
For this reason, more easily tractable but less physically motivated sharp cutoffs in Fourier space have been often preferred, for which the correlation matrix of the steps becomes diagonal, treating the correlations as perturbations \citep{MaggioreRiotto2010,CorasAch2011}.
The upcrossing approximation described below can instead be considered as the opposite limit, in which the steps are assumed to be strongly correlated (as is the case for a realistic power spectrum and filter). This approximation is equivalent to constraining only the last two steps of equation~\eqref{eq:ffirst}, marginalizing over the first $N-2$.

\subsection{The upcrossing approximation to $f(\sigma)$.}
Indeed, \cite{MussoSheth2012} noticed that for small enough $\sigma$ (i.e. for large enough masses), the first-crossing constraint may be relaxed into the milder condition
\begin{equation}
  \delta'\equiv\frac{\dd\delta}{\dd\sigma}>0\,;
\label{eq:upcross}
\end{equation}
that is, trajectories simply need to reach the threshold with positive slope (or with slope larger than the threshold's if $\dc$ depends on scale). This upcrossing condition may assign several haloes of different masses to the same spatial location. For this reason, while first-crossing provides a well defined probability distribution for $\sigma$ (e.g. with unit normalization), upcrossing does not. However, since the first-crossing is necessarily upwards, and down-crossings are discarded, the error introduced in $f(\sigma)$ by this approximation comes from trajectories with two or more turns. \cite{MussoSheth2012} showed that these trajectories are exponentially unlikely if $\sigma$ is small enough when the steps are correlated.
The first-crossing and upcrossing conditions to infer the halo mass from excursion sets are sketched in Fig.~\ref{fig:sketchup}: while the trajectory A would be (correctly) assigned to a single halo, the second upcrossing of trajectory B in the figure would be counted as a valid event by the approximation, and the trajectory would  (wrongly) be assigned to two haloes. The probability of this event is non-negligible only if $\sigma$ is large.

Returning to equation~\eqref{eq:ffirst}, expanding $\delta_{N-1}$ around $\delta_N$    gives
\begin{equation}
  \vartheta(\dc-\delta_{N-1})\simeq \vartheta(\dc-\delta_{N})+\delta_\D(\dc-\delta) \delta'\Ds\,,
\end{equation}
where the crossing scale $\sigma$, giving the halo's final mass $M$, is
defined implicitly in equation~\eqref{eq:cross}, as the solution of
the equation $\delta(\sigma)=\dc/D$\footnote{A careful calculation
  shows that the step function should be asymmetric, so that
  $\vartheta(\delta-\dc)=1$ when $\delta=\dc$ instead of
  $1/2$. }.
The assumption that this upcrossing is first-crossing allows us to marginalize over the first $N-2$ variables in equation~\eqref{eq:ffirst} without restrictions. The first term has no common integration support with $\vartheta(\delta_{N}-\dc)$, and only the second one -- containing the Jacobian $(\delta'-\delta_c')$ -- contributes to the expectation value (throughout the text, a prime will denote the derivative $\d/\d\sigma$).
 Adopting for convenience the
normalized walk height $\nu\equiv\delta/\sigma$, for which
$\mean{\nu^2}=1$, the corresponding density of solutions in $\sigma$-space obeys
\begin{equation}
  \left|\nu'-\nuc'\right|\, \delta_\D(\nu-\nuc) =
  (|\delta'|/\sigma)\,
  \delta_\D(\nu-\nuc) \,,
\end{equation}
where $\nuc\equiv\dc/(\sigma D)$ is the rescaled threshold.
The probability of upcrossing at $\sigma$ in
equation~\eqref{eq:ffirst}  is therefore simply the expectation
value of this expression,
\begin{equation}
  f_\mathrm{up}(\sigma) \equiv
  p_\G(\nu=\nuc)
  \int_{0}^\infty \!\d\delta' \delta'
  p_\G(\delta'|\nuc)\,,
\label{eq:fup}
\end{equation}
where the integral runs over $\delta'>0$ because of the upcrossing
condition \eqref{eq:upcross}. Usually, one sets $D=1$ at $z=0$ for
simplicity so that $\nu_c=\delta_c/\sigma$.
For Gaussian initial conditions\footnote{No conceptual complications arise in dealing with a non-Gaussian distribution, which is nonetheless beyond the scope of this paper.}, the conditional distribution $p_\G(\delta'|\nuc)$ is a Gaussian with mean $\nu_c$ and variance $1/\Gamma^2$, where
\begin{equation}
  \Gamma^2= \frac{1}{\mean{\delta'^2}-1}
  =\frac{\gamma^2}{1-\gamma^2} = \frac{1}{\sigma^2\mean{\nu'^2}}\,,
\label{eq:Gamma}
\end{equation}
and $\gamma^2=\mean{\delta'\delta}^2\!/\mean{\delta'^2}\mean{\delta^2}$ is the cross-correlation coefficient between the density and its slope\footnote{recalling that $\mean{\delta'\delta}=\sigma$ so that $\gamma^2=1/\mean{\delta'^2}$.}.
Thanks to this factorization, integrating equation~\eqref{eq:fup}  over $\delta'$  yields the fully analytical expression
\begin{equation}
  \fup(\sigma)=p_{\rm G}(\nuc)
  \frac{\mu}{\sigma}F(X)\,,
  \label{eq:fup2}
\end{equation}
where $p_\G$ is a Gaussian with mean $\mean{\nu}=0$ and variance ${\rm Var}(\nu)=1$. For a constant barrier (see Appendix \ref{sec:moving-barr-gener} for the generalization to a non-constant case), the parameters  $\mu$ and $X$ are defined as
\begin{equation}
  \mu\equiv\condmean{\delta'}{\nuc}=\nuc\,, \quad {\rm and} \quad X\equiv\frac{\mu}{\sqrt{\Var{\delta'}{\nuc}}}=\Gamma \nuc\,,
\label{eq:X0}
\end{equation}
with
\begin{equation}
  F(x)
  \equiv \!
  \int_0^\infty\!\! \d y\, \frac{y}{x}\,
  \frac{e^{-(y-x)^2/2}}{\sqrt{2\pi}}
  \!=\! \frac{1\!+\!\erf(x/\sqrt{2})}{2}
   \!+\! \frac{e^{-x^2/2}}{x\sqrt{2\pi}}\,,
   \label{eq:defF}
\end{equation}
which is a function that tends to 1 very fast as $x\to\infty$, with
correction decaying like $\exp(-x^2/2)/x^3$.  It departs from one by $\sim 8\%$ for a typical  $\Gamma \nuc \sim1$.
Equation~\eqref{eq:fup2} can be written explicitly as
\begin{equation}
  f_\mathrm{up}(\sigma) = \frac{\nuc e^{-\nuc^2/2}}{\sigma\sqrt{2\pi}}
  F(\Gamma\nuc)\,,
\label{eq:fup3}
\end{equation}
where the first factor in the r.h.s. of equation~\eqref{eq:fup3} is the
result of \cite{PressSchechter1974}, ignoring the factor of 2 they
introduced by hand to fix the normalization. For correlated steps, their
non-normalized result
reproduces well the large-mass tail of $f(\sigma)$ (which is automatically
normalized to unit and requires to correcting factor), but it is too low
for intermediate and small masses. The upcrossing probability $f_{\rm up}(\sigma)$
also reduces to this result in the large mass limit, when
$\Gamma\nuc\gg1$ and $F(\Gamma\nuc)\simeq1$. However,
for correlated steps $f_{\rm up}(\sigma)$ is a very good approximation of $f(\sigma)$
on a larger mass range. For a $\Lambda$CDM power spectrum, the
agreement is good for halo masses as small as $\SI{10^{12}}{\msun/h}$, well below the peak
of the distribution.
The deviation from the strongly correlated regime is
parametrized by $\Gamma\nuc$, which involves a combination of mass and
correlation strength: the approximation is accurate for large masses
(small $\sigma$ and large $\nuc$) or strong correlations (large
$\Gamma$). Although $\Gamma$ mildly depends on $\sigma$, fixing
$\Gamma^2\sim1/3$ (or $\gamma\sim1/2$) can be theoretically motivated
\citep{MussoSheth2014} and mimics well its actual value for real-space
Top-Hat filtering in $\Lambda$CDM on galactic scales.
The limit of uncorrelated steps ($\Gamma=0$), whose exact solution is twice the result of \cite{PressSchechter1974}, is pathological in this framework, with $f_\mathrm{up}$ becoming infinite. More refined approximation methods can be implemented in order to interpolate smoothly between the two regimes \citep{MussoSheth2013}.

From equation~\eqref{eq:fup2}, a characteristic mass $M_\star$ can be
defined by requesting that the argument of the Gaussian be equal to
one, i.e. $\nuc=1$ or $\sigma(M_\star)=\dc/D$. This defines $M_\star$
implicitly via equation~\eqref{eq:sigma2} for an arbitrary cosmology. This quantity is
particularly useful because $\fup(\sigma)$ does not have well defined
moments (in fact, even its integral over $\sigma$ diverges). This is a
common feature of first passage problems \citep{redner2001}, not a
problem of the upcrossing approximation: even when the first-crossing
condition can be treated exactly, and $f(\sigma)$ is normalized -- it
is a distribution function --, its moments still diverge. Therefore,
given that the mean $\mean{M}$ of the resulting mass distribution
cannot be computed, $M_\star$ provides a useful estimate of a
characteristic halo mass.

\subsection{Joint and conditional upcrossing probability.}
\label{sec:multivar}
The purpose of this paper is to re-compute excursion set predictions such as equation~\eqref{eq:fup2} in the presence of additional conditions imposed on the excursions.
Adding conditions (like the presence of a saddle at some finite distance) will have an impact not only on the mass function of dark matter haloes, but also on other quantities such as their assembly time and accretion rate.

Let us present in full generality how the upcrossing probability is modified by such supplementary conditions.
When, besides $\delta(\sigma)=\dc$ and the upcrossing condition, a set of $N$ linear\footnote{indeed the saddle condition below imposes linear constraints on the contrast and the potential, since the saddle's height and curvature are fixed} functional constraints $\{\mathcal{F}_1[\delta],\dots,\mathcal{F}_N[\delta]\}=\{v_1,\dots,v_N\}$
on the density field is enforced, the additional constraints modify the joint distribution of $\nu$ and $\nu'$.
The conditional upcrossing probability
may
be obtained by replacing $p(\nu,\nu')$ with $p(\nu,\nu'|\{v\})$ in
equation~\eqref{eq:fup}. For a Gaussian process, when the functional constraints do not involve $\delta'$, this replacement
yields after integration over the slope
\begin{equation}
  \fup(\sigma,\{v\})=p_{\rm G}(\nuc,\{v\})
  \frac{\mu_v}{\sigma}F(X_v)\,,
  \label{eq:genfup}
\end{equation}
where $p_\G(\nuc|\{v\})$ is a Gaussian with mean $\condmean{\nu}{\{v\}}$ and variance $\Var{\nu}{\{v\}}$, while $\mu_v$ and $X_v$ are defined as
\begin{equation}
  \mu_v \equiv \condmean{\delta'}{\nuc,\{v\}}\,,
\quad
  X_v \equiv
  \frac{\mu_v}{\sqrt{\Var{\delta'}{\nuc,\{v\}}}}\,,
\label{eq:X}
\end{equation}
and $\condmean{\delta'}{\nuc,\{v\}}$ and $\Var{\delta'}{\nuc,\{v\}}$ are the mean and variance of the conditional distribution, $p_\G(\delta'|\nuc,\{v\})$ given by equations~\eqref{eq:genmeannuprime}-\eqref{eq:genvarnuprime}  and evaluated at $\delta=\dc$, while  $F$ is given by equation~\eqref{eq:defF}.
Equation~\eqref{eq:genfup} is formally the conditional counterpart to equation~\eqref{eq:fup2}, while incorporating extra constraints corresponding to e.g. the large-scale Fourier modes of the cosmic web.

The brute force calculation of the conditional means and variances entering equation~\eqref{eq:genfup} can rapidly become tedious. To speed up the process, and gain further insight, one can write the conditional statistics of $\delta'$ in terms of those of $\delta$ and their derivatives.
This is done explicitly in Appendix~\ref{sec:condstats}, which
 allows us to write explicitly the conditional probability of upcrossing at $\sigma$ given $\{v\}$, obtained by dividing equation~\eqref{eq:genfup} by $p(\{v\})$, as
\begin{equation}
  \fup(\sigma|\{v\}) = -\nucv'
  \frac{e^{-\nucv^2/2}}{\sqrt{2\pi}}
  F\left(-\frac{\nucv'}{\sqrt{\mathrm{Var}\!\left(\nu_v'\right)}}\right),
  \label{eq:fupjoint}
\end{equation}
given
\begin{equation}
  \nucv \equiv
  \frac{\dc-\condmean{\delta}{\{v\}}}{\sqrt{\Var{\delta}{\{v\}}}}
\,,  \quad\mathrm{and}\quad
  \nucv'\equiv\frac{\dd\nucv}{\d\sigma}\,,
\label{eq:genconstr}
\end{equation}
where these conditionals and variances can be expressed explicitly
in terms of the constraint via \eqref{eq:genmeannu}-\eqref{eq:genvarnuprime}.
Equation~\eqref{eq:fupjoint} is therefore also formally equivalent to equation~\eqref{eq:fup3}, 
upon replacing $\nuc\to\nucv$ and $\mean{\nu'^2}\to\mean{\nu_v'^2}$ to account for the constraint.
Remarkably, the conditional probability $\fup(\sigma|\{v\})$ is thus simply expressed as an unconditional upcrossing probability for the effective
  unit variance process obtained from the conditional density.

\begin{table}
\label{tab:method}
  \centering
  \begin{tabular}{c|cccc}
    & \multicolumn{2}{c}{without saddle} & \multicolumn{2}{c}{with saddle} \\
    \hline
             & height   & slope   & height  & slope \\
    \hline
  upcrossing ($\sigma$) & $\nuc$   & $\mu,X$ & $\nucS$ & $\mu_\S,X_\S$ \\
  accretion  ($\alpha$) &        & $Y_\alpha$ &         & $Y_{\alpha,\S}$ \\
  formation  ($\Df$) & $\nu_{\form,\mathrm{c}}$ & $\mu_\form,X_\form$ & $\nu_{\form,\mathrm{c},\S}$ & $\mu_{\form,\S},X_{\form,\S}$
\end{tabular}
  \caption{\label{tab:vars} List of variables for the three different probabilities studied in the text (upcrossing, accretion rate given upcrossing and formation time given upcrossing), conditioned or not to the presence of the saddle point, split by whether they relate to the height of the excursion set trajectory or its slope. Variables like $\mu$ and $X$ always appear as $\mu F(X)$ and describe the mean slope of the upcrossing trajectories given the different conditions (presence of the saddle and/or height $\nuf$ of the trajectory at formation). The unconditional case has $\mu=\nuc$ and $X=\Gamma\nuc$.
    The remaining variables appear as arguments of a Gaussian, and are used to define the typical values $\sigma_\star$, $\alpha_\star$ and $D_{\star}$ of the excursion set variables $\sigma$, $\alpha$ and $\Df$. The height-related variables describe the probability of reaching the collapse threshold $\nuc$ (unconditional or given the saddle), or the formation threshold $\nuf$ given $\nuc$ (with or without saddle).
 The slope-related ones describe the probability of having at upcrossing the slope corresponding to a given accretion rate.
 See also Table~\ref{tab:definitions}. }
\end{table}
The above-sketched formal procedure will be applied to practical constraints in the next section.
For convenience and consistency, Table~\ref{tab:vars} lists all the variables that are introduced in the following sections, for the combinations of the various constraints (on the slope at crossing, on the height of the trajectory at $\sigma(M/2)$, on the presence of a saddle) that will be imposed.

\section{Accretion rate and formation time }
\label{sec:unconditional}

Let us first present the tracers of galactic assembly
when there is no large-scale saddle.
Specifically, this section will consider the dark matter mass accretion rate and formation redshift.
It will compute the joint PDFs, the corresponding marginals, typical scales and expectations. Its main results are the derivation of the conditional probability of the accretion rate -- equation~\eqref{eq:condalpha} -- and formation time -- equation~\eqref{eq:condD12} -- for haloes of a given mass.
The emphasis will be on derivation in the language of excursion set. The reader only concerned with statistical predictions in terms of
quantities of direct astrophysical interest
may skip to Section~\ref{sec:astro}.

Following \cite{LaceyCole1993}, the entire mass accretion history of the halo is encoded in the portion of the excursion set trajectory after the first-crossing: solving the implicit equation~\eqref{eq:cross} at all $z$ allows to reconstruct $M(z)$. As the barrier $\dc/D(z)$ decreases with time (since $D(z)$ grows as $z$ decreases), the first-crossing scale moves towards smaller values (larger masses), thereby describing the accretion of mass onto the halo. Clearly, since $\delta(\sigma)$ is not monotonic, $M(z)$ is not a continuous function. Finite jumps of the first-crossing scale, corresponding to portions for which $\sigma$ is not a global maximum of the interval $[0,\sigma]$, can be interpreted as mergers (see trajectory B in Fig.~\ref{fig:sketchup}, or the portion marked with (1) in Fig.~\ref{fig:sketch-accretion}).
In the upcrossing approximation, the constraint $\delta'(\sigma)>0$ discards the downward part of each jump.

\subsection{Accretion rate.}

In the language of excursion sets, finding the mass accretion history
is equivalent to reconstructing the function $\sigma(D)$ (where $D$
was defined in equation~\eqref{eq:defD}): because the barrier grows as
$D$ decreases with $z$, the crossing scale $\sigma$ moves towards
larger values (smaller masses). Differentiating both sides of
equation~\eqref{eq:cross} w.r.t. $z$ gives
\begin{equation}
  \alpha \equiv
  -\frac{D}{\sigma} \frac{\dd\sigma}{\dd D}
  = \frac{\dc}{\sigma\delta'} =\frac{\nuc}{\sigma(\nu'-\nuc')}\,,
\label{eq:accrate}
\end{equation}
where $\alpha$ measures the fractional change of the first-crossing scale $\sigma(M)$ with $D(z)$, and is related to the instantaneous relative mass accretion rate by
\begin{equation}
  \frac{1}{M}\frac{\dd M}{\dd z}\equiv \frac{\dot M}{M}
  \label{eq:defmdot}
  =
  {\alpha}
  \frac{\dd\log D}{\dd z}\left(- \frac{\dd \log M}{\dd \log \sigma} \right)\,.
\end{equation}
The upcrossing condition implies that $\alpha>0$: excursion set haloes can only increase their mass since ${\dd\!\log\!M}/{\dd\!\log\sigma}<0$.
\begin{figure}
\begin{center}
  \includegraphics{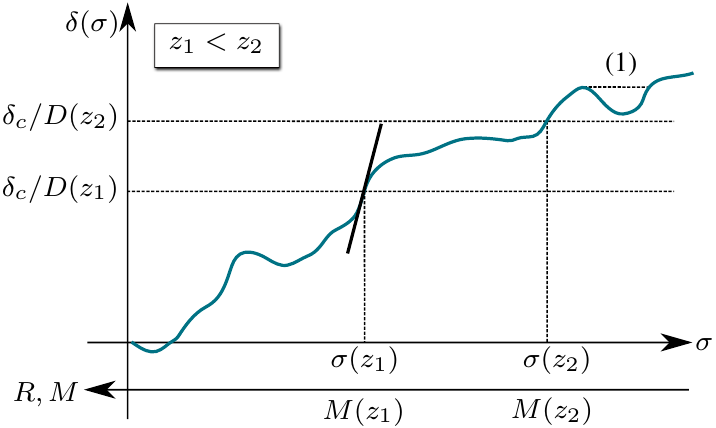}
  \caption{\label{fig:sketch-accretion} Pictorial representation of
    the procedure to infer accretion rates from excursion sets. As the
    redshift $z$ grows, the barrier $\dc/D(z)$ becomes higher and the
    first-crossing scale $\sigma(z)$ moves to the right, towards
    smaller masses. This procedure reconstructs the entire mass
    accretion history $M(z)$ from the first-crossing history
    $\sigma(D)$. As the two redshifts $z_1$ and $z_2$ in figure get
    close to each other, the difference between $\sigma(z_1)$ and
    $\sigma(z_2)$ is completely fixed by the slope of the
    trajectory. This deterministic relation connects the excursion set
    slope to the halo's instantaneous mass accretion rate. Finite jumps
    of the first-crossing $\sigma$ after a down-turn (where the
    inverse function $\sigma(\delta)$ becomes multi-valued, as in (1))
    cannot describe smooth accretion and are traditionally associated
    to large mergers.   }
\end{center}
\end{figure}

A pictorial representation of this procedure is given in Fig.~\ref{fig:sketch-accretion}.
Equation~\eqref{eq:accrate} gives a relation between the accretion rate of the final haloes and the Lagrangian slope of the excursion set trajectories, which is statistically meaningful in the framework of excursion sets with correlated steps (because the slope then has finite variance).
Note that $\alpha$ scales both like the inverse of the slope $\delta'$
and the logarithmic rate of change of $\sigma$ with $D$. It also
essentially scales like the relative accretion rate, ${\dot M}/M$
since in equation~\eqref{eq:defmdot} ${\dd\!\log\!D}/{\!\dd z} $ is
simply a time dependent scaling, while on galactic scales,
($n\!\sim\!2$),  ${\dd\!\log\! M}\!/\!{\dd\!\log\sigma} \sim -6$ (see
also Section~\ref{sec:astro} and Appendix~\ref{sec:covariance} for the
generic formula).

Fixing the accretion rate establishes a local bidimensional mapping
between $\{\nu,\nu'\}$, or $\{\delta,\delta'\}$, and $\{\sigma,\alpha\}$, defined as the solutions of the bidimensional constraint
\begin{equation}
  \mathcal{C}\equiv\{\nu(\sigma)-\nuc,\nu'(\sigma)-\nuc'-\nuc/\sigma\alpha\} =
  \mathbf{0}\,.
\end{equation}
The density of points in the $(\sigma,\alpha)$ space satisfying the
constraint is
\begin{equation}
  |\det\!\left(
    {\pd\mathcal{C}}/{\pd\{\sigma,\alpha\}}
  \right)|\,
  \delta_\D^{(2)}(\mathcal{C})\,. \label{eq:accrateconst}
\end{equation}
Since $\pd(\nu-\nuc)/\pd\alpha=0$, the determinant in equation~\eqref{eq:accrateconst} is simply $|(\nu'-\nuc')(\nuc/\sigma\alpha^2)| = \nuc^2/\sigma^2\alpha^3$, and is no longer a stochastic variable. Taking the expectation value of equation~\eqref{eq:accrateconst} gives
\begin{align}
f_\mathrm{up}(\sigma,\alpha) &=
  \frac{\nuc^2}{\sigma^2\alpha^3}
  p_\G(\nuc,\nuc'+\nuc/\sigma\alpha)\notag\,,  \\
  &= \frac{\Gamma\nuc^2}{\sigma\alpha^3}
  \frac{e^{-\nuc^2/2}}{\sqrt{2\pi}}
  \frac{e^{-Y_\alpha^2/2}}{\sqrt{2\pi}}\,,
  \label{eq:fupalpha}
\end{align}
with (using the conditional mean $\mu=\nuc$ from equation \eqref{eq:X0})
\begin{equation}
  Y_\alpha \equiv
  \frac{\nuc/\alpha-\mu}{\sqrt{\Var{\delta'}{\nuc}}}
  = \Gamma(\sigma\nuc'+\nuc/\alpha)\,,
\label{eq:Yalpha}
\end{equation}
which is the joint probability of upcrossing at $\sigma$ with accretion rate $\alpha$\footnote{As expected, marginalizing equation~\eqref{eq:fupalpha}  over $\alpha>0$ gives back equation~\eqref{eq:fup2}, upon setting $\Gamma\nuc/\alpha=x$. }.
This can be formally recovered setting $\condmean{\delta'}{\nuc,\alpha}=\nuc/\alpha$ and $\Var{\delta'}{\nuc,\alpha}\to0$ in equation~\eqref{eq:X} (because the constraint fixes $\delta'$ completely), which gives $F(X_\alpha)=1$ as needed.

\begin{figure}
  \includegraphics[width=\columnwidth]{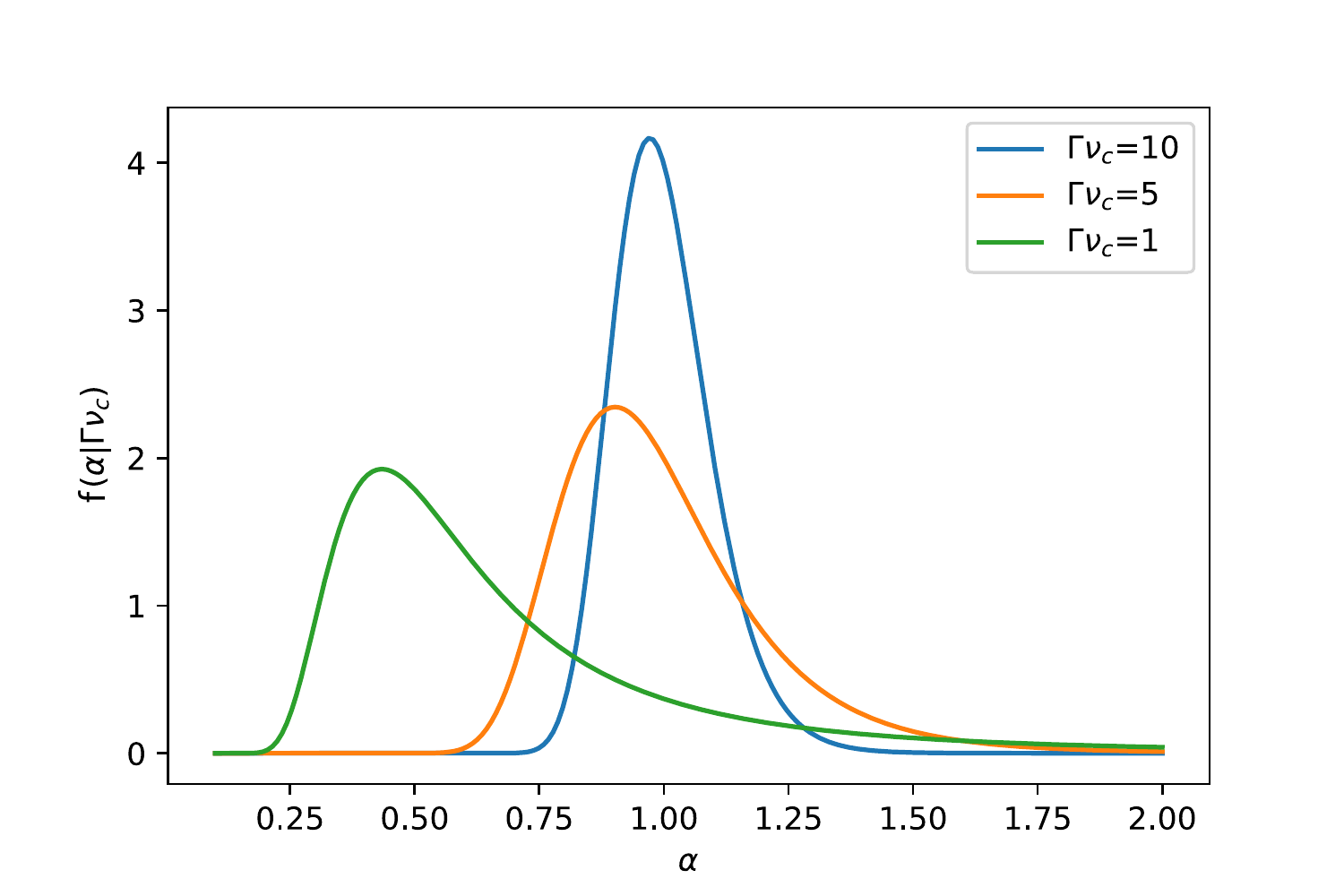}
  \caption{Plot of the conditional PDF
    $\fup(\alpha|\sigma)$ of the accretion rate for values of $\sigma$
    corresponding to $\Gamma\nuc=10, 5, 1$. As the mass gets smaller, so
    does $\Gamma\nuc$ and the conditional PDF moves towards smaller
    accretion rates $\alpha$. Therefore, haloes of smaller mass tend to
    accrete less.}
  \label{fig:pdfalpha}
\end{figure}

The conditional probability of having accretion rate $\alpha$ given upcrossing at $\sigma$ can be obtained taking the ratio of equations \eqref{eq:fupalpha} and \eqref{eq:fup3}, which gives
\begin{equation}
  f_\mathrm{up}(\alpha|\sigma)
  = \frac{\Gamma\nuc}{\alpha^3}
  \frac{e^{-Y_\alpha^2/2}}{\sqrt{2\pi}\,F(\Gamma\nuc)}\,,
\label{eq:condalpha}
\end{equation}
and represents the main result of this subsection. The exact form of $\fup(\alpha|\sigma)$ from equation~\eqref{eq:condalpha}, as $\sigma$ changes is shown in Fig.~\ref{fig:pdfalpha}.
 This conditional probability has a well defined mean value, which reads 
\begin{equation}
  \condmean{\alpha}{\sigma} =
  \int_0^\infty \d\alpha \,\alpha f_\mathrm{up}(\alpha|\sigma) =
  \frac{1+\erf(\Gamma\nuc/\sqrt{2})}{2F(\Gamma\nuc)}\,;
\label{eq:condmeanalpha}
\end{equation}
however, the second moment $\condmean{\alpha^2}{\sigma}$ and all higher order statistics are ill defined. The $n$-th moment is in fact proportional to the expectation value of $(1/\delta')^{n-1}$ (over positive slopes and given $\nuc$), which is divergent.
Equation~\eqref{eq:condalpha} shows that very small values of $\alpha$ (corresponding to very steep slopes) are exponentially unlikely, and very large ones (shallow slopes) are suppressed as a power law.
Unlike $\fup(\sigma)$, the conditional distribution $\fup(\alpha|\sigma)$ is a well defined normalized PDF. However, it is still an approximation to the exact PDF, as it assumes that the distribution of the slopes at first-crossing is a (conditional) Gaussian. This assumption is accurate for steep slopes, but overestimates the shallow-slope tail, for which the exact first-crossing condition would impose a boundary condition $p_{\rm G}(\delta'=0|\dc)=0$. The higher moments of the exact conditional distribution of accretion rates should be convergent. However, even if this were not the case, let us stress that these divergences would not represent a pathology of excursion sets, but are instead a rather common feature of first-passage statistics in a cosmological context.

Regardless of convergence issues, it remains true that the estimate \eqref{eq:condmeanalpha} of the mean $\condmean{\alpha}{\sigma}$ gets a significant contribution from the less accurate side of the distribution. One may therefore look for other more informative quantities.
In analogy with $M_\star$, defined as the value of $M$ for which $\nuc=1$, one can define the characteristic accretion rate $\alpha_\star$ as
the value for which $Y_\alpha$, the argument of the Gaussian in equation~\eqref{eq:condalpha}, equals one
\begin{equation}
  \alpha_\star(\sigma) = 
  \frac{\Gamma\nuc}{1+\Gamma\nuc}\,.
\label{eq:alphastar}
\end{equation}
For the above-mentioned typical value, it follows that $\alpha_\star(M_\star)= \left(\sqrt{3}-1\right)/2\approx 1/3$.
Another useful quantity is the most likely value of the accretion rate, corresponding to the maximum $\alpha_{\mathrm{max}}$ of $\fup(\alpha|\sigma)$. Requesting the derivative of the PDF to vanish, one gets
\begin{equation}
  \alpha_{\mathrm{max}}(\sigma) = \frac{(\Gamma\nuc)^2}{6}
  \left[\sqrt{1+\frac{12}{(\Gamma\nuc)^2}}-1\right].
\label{eq:amax}
\end{equation}
All three quantities $\condmean{\alpha}{\sigma}$, $\alpha_\star$ and $\alpha_{\mathrm{max}}$ tend to 1 in the large mass limit, and decrease for smaller masses. They thus contain some equivalent information on the position of the bulk of the conditional PDF of $\alpha$ at given mass.
Hence, haloes of smaller mass accrete less on average.

\subsection{Halo formation time}
\label{sec:form}

\begin{figure}
\begin{center}
  \includegraphics{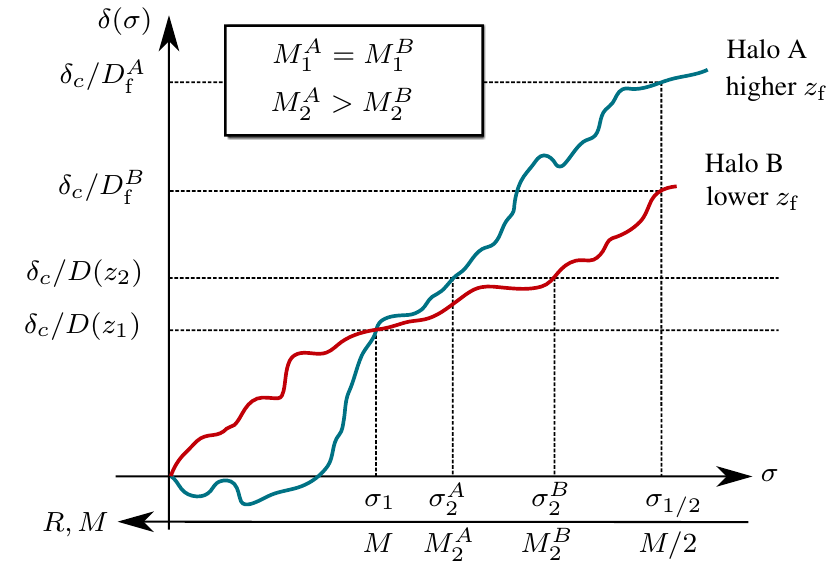}
    \caption{\label{fig:sketch-halfmass} Pictorial representation of the
    interplay between accretion rate and formation time as inferred
    from excursion sets. Two haloes $A$ and $B$ upcross the threshold
    $\dc/D(z_1)$ at the same scale $\sigma$. At redshift $z_1$, they
    have therefore the same mass. Halo $A$ has a steeper slope than
    halo $B$, and has thus a lower accretion rate.  At a slightly
    larger redshift $z_2$, halo $A$ crosses the higher threshold
    $\dc/D(z_2)$ at a lower $\sigma$, and its mass is thus larger than
    halo $B$'s: halo $A$ assembles its mass earlier, consistent with
    its lower accretion at $z_1$. At the half-mass scale
    $\sigma_\half=\sigma(M/2)$, the trajectory of halo $A$ is higher:
    its threshold $\dc/D_\form$ has a value of $D_\form$
    lower than halo $B$'s at the same $\sigma_\half$. Halo $A$ has
    thus assembled half of its mass at a redshift $z_\form$ higher than
    halo $B$.
    }
\end{center}
\end{figure}

The formation time is conventionally defined as the redshift $z_\form$ at which a halo has assembled half of its mass. It is thus related to the height of the excursion set trajectory at the scale $\sigma_{\half}\equiv\sigma(M/2)$ corresponding to the radius $R_{\half}=R/2^{1/3}$.
As the barrier $\dc/D(z)$ grows with $z$, and the first-crossing scale moves to the right towards higher values of $\sigma$, $z_{\form}$ is the redshift at which $\sigma_{\half}$ becomes the first-crossing scale for that trajectory, if it exists.
That is, neglecting for the time being the presence of finite jumps in the first-crossing scale (interpreted as mergers), one simply needs to solve for $z_\form$ the implicit relation $\delta(\sigma_{\half})=\dc/D(z_{\form})$, which makes $z_\form$ a stochastic variable.
As described in Fig.~\ref{fig:sketch-halfmass}, trajectories with the same upcrossing scale $\sigma$ but different heights at $\sigma_\half$ describe different formation times: a higher $\delta_\half$ corresponds to a smaller $D(z_\form)$ and thus to a halo with larger $z_\form$, which assembled half of its mass earlier.

In the language of excursion sets, it is convenient to work with $D_\form\equiv D(z_\form)$ rather than with $z_\form$. In terms of unit variance variables, haloes with formation time $D_{\form}$ correspond to trajectories satisfying
\begin{equation}
  \nu_{\half} \equiv \frac{\delta(\sigma_{\half})}{\sigma_{\half}}
  =  \frac{\dc}{\sigma_{\half}D_{\form}}  \equiv \nuf\,,
\label{eq:D12}
\end{equation}
where $\nu_{\half}$ is the Gaussian variable at $\sigma_{\half}$ and $\nuf$ is the threshold at $D_\form$. This constraint at $\sigma_{\half}$ imposes a second condition on the trajectory after $\nu=\nuc$, which selected the crossing scale $\sigma$.
One then needs to transform the bidimensional constraint
\begin{equation}
  \mathcal{\tilde C}\equiv\{\nu-\nuc,\nu_{\half}-\nuf\}
  = \mathbf{0}
\end{equation}
on $\{\nu,\nu_\half\}$ into one for $\{\sigma,D_\form\}$, which gives
\begin{equation}
  \left|\det\!\left(
    {\pd\mathcal{\tilde C}}/{\pd\{\sigma,D_\form\}}
  \right)\right|\,
  \delta_\D^{(2)}(\mathcal{\tilde C})
  = \left|\nu'-\nuc'\right|
  \frac{\nuf}{D_\form}
  \delta_\D^{(2)}(\mathcal{\tilde C})\,,
\label{eq:JacsigmaD}
\end{equation}
thanks to the fact that $\pd(\nu_c-\nu)/\pd\Df=0$.

The joint probability of upcrossing at $\sigma$ having formation time $D_\form$, denoted $\fup(\sigma,\!D_\form) $, is defined as the expectation value of equation~\eqref{eq:JacsigmaD} with the condition $\nu'>\nuc'$. That is,
\begin{align}
  \fup(\sigma,\!D_\form) &\equiv
  \frac{\nuf}{D_\form}\int_{\nuc'}^\infty \!\d\nu'
  (\nu'-\nuc') \, p_{\rm G}(\nuc,\nu',\nuf)\,, \notag\\
  &= \frac{\nuf}{D_\form}
  p_{\rm G}(\nuc,\nuf)
    \frac{\mu_\form}{\sigma} F(X_\form)\,,
\label{eq:fupD}
\end{align}
where the second equality follows from setting $\{v\}=\nuf$ in the general expression \eqref{eq:genfup},
while $\mu_{\form}$ and $X_{\form}$ are given by
\begin{equation}
  \mu_{\form}(D_{\form}) \equiv \condmean{\delta'}{\nuc,\nuf}\,,
\quad
  X_{\form}(D_{\form}) \equiv
  \frac{\mu_{\form}(D_{\form})}{\sqrt{\Var{\delta'}{\nuc,\nuf}}}\,,
\label{eq:Xf}
\end{equation}
as specified by equation~\eqref{eq:X}. The conditional mean $\condmean{\delta'}{\nuc,\nuf}$ and variance $\Var{\delta'}{\nuc,\nuf}$ are computed in equations \eqref{eq:truc} and \eqref{eq:varnucnuf}, which
give \begin{align}
  & \mu_{\form}(\Df) = \frac{\omega' \dc}{\sigma_{\half}\Df} +
  \frac{\sigma-\omega'\omega}{\sigma^2-\omega^2}
  \left(\dc-\frac{\omega\dc}{\sigma_{\half}\Df}\right)\,,
\label{eq:muF}  \\
  & X_\form(\Df)
  = \mu_{\form}(\Df)\bigg/\bigg[\langle\delta'^2\rangle - \omega'^2
    - \frac{(\sigma-\omega'\omega)^2}{\sigma^2-\omega^2}
    \bigg]^{1/2}
  \,,
\label{eq:cvarF}
\end{align}
where $\omega=\mean{\delta\nu_{\half}}$ and $\omega'=\mean{\delta'\nu_{\half}}$ are given by equations~\eqref{eq:omega} and~\eqref{eq:omegaprime} respectively.

The conditional probability of $D_\form$ given upcrossing at $\sigma$ -- the main result of this subsection -- is obtained dividing equation~\eqref{eq:fupD} by equation~\eqref{eq:fup2}
\begin{align}
  \fup(D_\form|\sigma) &= 
  \frac{\nuf}{D_\form} p_{\rm G}(\nuf|\nuc)
  \frac{\mu_{\form}F(X_{\form})}{\nuc F(X)} \notag\,, \\
  &= \frac{(\dc/\sigma_{\half}\Df^2) e^{-\nu_{\form,\mathrm{c}}^2/2}}{
         \sqrt{2\pi(1-\langle\nu\nu_\half\rangle^2)}}
  \frac{\mu_{\form}F(X_{\form})}{\nuc F(X)}\,,
\label{eq:condD12}
\end{align}
where $(\nuf/D_\form)p_{\rm G}(\nuf|\nuc)=p(\Df|\nuc)$, not surprisingly, is the conditional probability of the (non-Gaussian) variable $\Df$ given $\nuc$, and
\begin{equation}
  \nu_{\form,\mathrm{c}} \equiv 
  \frac{\nuf-\mean{\nu \nu_\half}\nuc}{\sqrt{1-\langle\nu \nu_\half\rangle^2}}
  = \frac{\dc}{\sigma_{\half}}
  \frac{1/\Df-\mean{\delta\delta_\half}\!/\sigma^2}{\sqrt{1-\langle\nu\nu_\half\rangle^2}}\,.
\end{equation}
 Recall also that $X=\Gamma\nuc$.
The conditional probability $\fup(D_\form|\sigma)$ depends on $\Df$ directly, through $\nu_{\form,\mathrm{c}}$ and through $\mu_\form$ (which appears also in $X_\form$). As both $\nu_{\form,\mathrm{c}}$ and $\mu_\form$ are proportional to $1/\Df$ in the small-$\Df$ limit, equation~\eqref{eq:condD12} scales like $e^{-\nu_{\form,\mathrm{c}}^2/2}/\Df^3$.
Hence, $\fup(D_\form|\sigma)$ is exponentially suppressed for small $D_\form$, that is for large formation redshift $z_\form$: it is exponentially unlikely for a halo to assemble half of its mass at very high redshift.

Like in the previous section, the Gaussian cutoff in equation~\eqref{eq:condD12} allows to define a characteristic value $D_{\star}(\sigma)$ of the formation time, below which $\fup(D_\form|\sigma)$ is exponentially suppressed, by requesting that $\nu_{\form,\mathrm{c}}=1$. This definition corresponds to
\begin{equation}
  D_\star(\sigma) = \frac{\dc/\sigma_\half}{
    \mean{\nu\nu_\half}\nuc + \sqrt{1-\langle\nu\nu_\half\rangle^2}}\,,
    \label{eq:defD*}
\end{equation}
which can then be solved for the typical formation redshift $z_\star$.
Similarly, one may define the most likely formation time $D_{\mathrm{max}}$ by finding the value of $D_\form$ that maximizes equation~\eqref{eq:condD12}. Because its expression is rather involved and not much more informative than $D_\star$, it is not reported here.

Expanding $D_\star$ in powers of $\Ds_{\half}\equiv\sigma_{\half}-\sigma$ (even though $\Ds_{\half}/\sigma\simeq-(1/2)\d\log\sigma/\d\log M$ may not be small, in which case this expansion may just give a qualitative indication), one gets
\begin{equation}
  D_{\star} \simeq 1-\frac{\Ds_{\half}}{\sigma}
  \bigg(1+\frac{\sqrt{\mean{\delta'^2}-1}}{\nuc}\bigg)
  \simeq 1-\frac{1}{\alpha_\star}\frac{\Ds_{\half}}{\sigma}\,,
\end{equation}
confirming the intuitive relation between accretion rate and formation time.
Haloes with smaller accretion rates today must have formed earlier, in order for their final mass to be the same. To derive this expression,  $\mean{\delta\delta_{\half}}$ was expanded up to second order in $\Ds$, using $\mean{\delta\delta'}=\sigma$ and $\mean{\delta\delta''}=1-\mean{\delta'^2}=\Gamma^{-2}$.
Let us stress that, strictly speaking, the conditional probability $\fup(D_\form|\sigma)$ is not a well defined probability distribution. For instance, just like $\fup(\sigma)$, equation~\eqref{eq:condD12} is not normalized to unity when integrated over $0<D_\form <D$. This is an artifact introduced by the upcrossing approximation to the first-crossing problem, because equation~\eqref{eq:D12} does not
require trajectories to reach $\dc/D_\form$ for the first time. As $D_\form$ gets close to $D$, most trajectories reaching $\dc/D_\form$ do so with negative slope, or after one or more crossings, which leads to overcounting.
For $D_\form=D$, trajectories that first crossed $\dc/D_\form$ at $\sigma$ cannot first cross again at $\sigma_\half$, since $\sigma_\half-\sigma$ remains finite: the true distribution should then have $f(D_\form|\sigma)=0$. This is clearly not the case for $\fup(D_\form|\sigma)$. In spite of these shortcomings, equation~\eqref{eq:condD12} approximates well the true conditional PDF for $D_\form\ll D_\star$, and the characteristic time $D_\star$ still provides a useful parametrization of the height of the tail.

A better approximation than equation~\eqref{eq:condD12} may be
obtained by imposing an upcrossing condition at $\sigma_{\half}$ as
well
\begin{equation}
  \frac{\dc}{D_\form^2}\int_0^\infty \!\d\delta' \,\delta'
  \int_0^\infty \!\d\delta_\half' \,
  p_{\rm G}(\dc,\delta',\dc/D_\form,\delta_\half')\,.
\end{equation}
Notice the absence in this expression of the Jacobian factor $\delta_\half'$: this is because the constraint at $\sigma_\half$ is not differentiated w.r.t. $\sigma_\half$, but only w.r.t. $D_\form$.
This reformulation, which unfortunately does not admit a simple analytical expression, would improve the approximation for values of $D_\form$ closer to $\D_\star$, but it would still not yield a formally well defined PDF.
Furthermore, the mean $\condmean{D_\form}{\sigma}$ and all higher moments would still be infinite: these divergences are in fact a common feature of first passage statistics, which typically involve the inverse of Gaussian variables. For all these reasons,  this calculation is not pursued further.

This section has formalized analytical predictions for accretion rates and formation times from the excursion set approach with correlated steps. It confirmed the tight correlation between the two quantities, according to which at fixed mass, early forming haloes must have small accretion rates today. 
Because the focus is here on accounting for the presence of a saddle of the potential at finite distance, for simplicity and in order to isolate this effect we have restricted our analysis to the case of a constant threshold $\dc$. More sofisticated models (e.g. scale dependent barriers involving other stochastic variables that account for deviations from spherical collapse) could however be accomodated without  extra conceptual effort (see Appendix \ref{sec:moving-barr-gener}).

\section{Halo statistics near saddles}
\label{sec:conditional}

\begin{figure}
  \centering
  \includegraphics{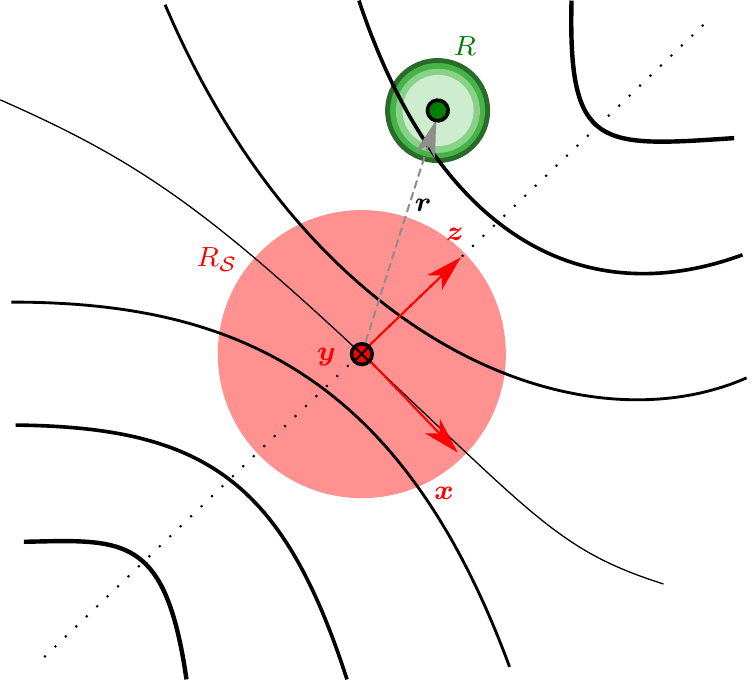}
  \caption{\label{fig:crit_point}%
    Illustration of the conditional excursion set smoothing on a
    few infinitesimally close scales around $R$ (in green) at 
    finite distance $\vec{r}$ from a
    saddle point of the gravitational potential smoothed on scale
    $\Rs \gg R$ (in red). The eigenvectors $\vec{e}_x$ and $\vec{e}_z$ of the
    tidal tensor at the saddle give the directions of steepest
    increase and decrease of the potential, corresponding to maximum
    inflow and outflow respectively. The region is compressed along
    $\vec{e}_x$ and $\vec{e}_y$ and stretched along $\vec{e}_z$, thus
    creating a filament. The solid lines are iso-contours of the mean
    density, the thickest the densest.
    The dotted line indicates a ridge of mean density (the filament), parallel
    to $\vec{e}_z$ near the saddle.   }
\end{figure}

Let us now quantify how the presence of a saddle of the large-scale gravitational potential affects the formation of haloes in its proximity.
To do so, let us study the tracers introduced in the previous section (the distributions of upcrossing scale, accretion rate and formation time) using conditional probabilities. The enforced condition is that the upcrossing point (the centre of the excursion set trajectories) lies at a finite distance $\rr$ from the saddle point. The focus is on (filament-type) saddles of the potential that describe local configurations of the peculiar acceleration with two spatial directions of inflow (increasing potential) and one of outflow (decreasing potential). See Appendix~\ref{sec:other-critical} for other critical points.
The vicinity of theses saddles will become filaments (at least in the Zel'dovich approximation), where particles accumulate out of the neighbouring voids from two directions, and the saddle points filament centres, where the gravitational attraction of the two nodes balances out. A schematic representation of this configuration is given in Fig.~\ref{fig:crit_point}.

The saddles are identified as points with null
gradient of the gravitational potential, smoothed on a sphere
of radius $\Rs$ (which is assumed to be larger than the halo's scale $R$).
This condition guarantees that the mean peculiar acceleration
of the sphere, which at first order is also the acceleration of
its centre of mass, vanishes. That is, the null condition (for
$i=1,\dots, 3$)
\begin{equation}
  g_{i} \equiv \frac{1}{R_\star}
  \int\frac{\dd^3k}{(2\pi)^3}\frac{ik_i}{k^2}\delta_m(\kk)
  \frac{W(k\Rs)}{\sigmas}  = 0,
  \label{eq:gi_def}
\end{equation}
where $\sigmas \equiv \sigma(\Rs)$,
is imposed on the mean gradient of the potential smoothed with a
Top-Hat filter on scale $\Rs$.  This mean acceleration is normalized
in such a way that $\mean{g_ig_j}=\delta_{ij}/3$ by introducing the
characteristic length scale\footnote{This scale is similar, but not
  equivalent, to the scale often defined in peak theory. Calling
  $\sigma_i^2$ the variance of the density field filtered with
  $k^{2i}W(kR)$, the $R_*$ defined here is $\sigma_{-1}/\sigma_0$,
  while the peak theory scale is $\sqrt{3}\sigma_1/\sigma_2$.}
\begin{equation}
  R^2_\star \equiv
  \int\d k \frac{P(k)}{2\pi^2} \frac{W^2(k\Rs)}{\sigmas^2}\,.
\label{eq:sadscale}
\end{equation}
Having null peculiar acceleration, the patch sits
at the equilibrium point of the attractions of what will become the
two nodes at the end of the filament\footnote{The mean gravitational acceleration $g_i$ includes an unobservable infinite wavelength mode, which should in principle be removed. A way to circumvent the problem would be to multiply $W(k\Rs)$ by a high-pass filter on some large-scale $R_0$ to remove modes with $k\lesssim 1/R_0$. Because $g_i$ is set to 0, it does not introduce any anisotropy, but simply affects the radial dependence of the conditional statistics through its covariance $\mean{g_ig_j}$, which however is not very sensitive to long wavelengths.
For this reason, this minor complication is ignored.}.

The configuration of the large-scale potential is locally described by the rank 2 tensor
\begin{equation}
  \label{eq:qij_def}
  q_{ij} \equiv \frac{1}{\sigmas}
  \int\frac{\dd^3k}{(2\pi)^3}\frac{k_{i}k_{j}}{k^2}\delta_m(\kk) W(k\Rs)\,,
\end{equation}
which represents the Hessian
of the potential smoothed on scale $\Rs$, normalized so that $\mean{\tr^2(q)}=1$. This tensor is the opposite of the so-called strain or deformation tensor.  The trace $\tr(q)=\nus$ of
$q_{ij}$ describes the average infall (or expansion, if negative) of
the three axes, while the anisotropic shear is given by the traceless
part $\bar q_{ij} \equiv q_{ij}-\delta_{ij}\nus/3$, which deforms the
region by slowing down or accelerating each axis.  By construction,
$\mean{\nus\bar q_{ij}}=0$.

For the potential to form a filament-type
saddle point, the eigenvalues $q_i$ of $q_{ij}$ must obey
$q_1<0<q_2<q_3$ (see also Fig.~\ref{fig:pdflambda}).
There is no clear consensus on what the initial density of a proto-filament should be for the structure to form at $z=0$ \citep[see however][]{Shenetal2006}. The value  $\nus=1.2$ was chosen here, corresponding to a mean density of 0.8 within a sphere of $\Rs=10$ Mpc$/h$, which is about one standard deviation higher than the mean value for saddle points of this type (see Appendix~\ref{sec:pdf-saddles} for details), and thus corresponds to a filament slightly more massive than the average (or to an average filament that has not completely collapsed yet). The qualitative results presented in this paper do not depend on the exact value of $\nus$ (even though they obviously do at the quantitative level).

\subsection{Expected impact of saddle tides}
\label{sec:tides}
The mean and covariance of $\delta$ and $\delta'$ at $\rr$ are
modified by the presence of the saddle at the origin. The zero mean density field
is replaced by $\delta -\langle  \delta| {\cal S }\rangle$,  where  (using Einstein's convention as usual)
\begin{equation}
 \langle  \delta| {\cal S }\rangle =
  \langle  \delta| {\cal S }\rangle \mean{\delta\nus}\nus + 3\mean{\delta g_{i}}g_{i}
  + \frac{15}{2}\mean{\delta \bar q_{ij}}\bar q_{ij}\,,
\label{eq:effdelta}
\end{equation}
where the correlation functions are evaluated at finite separation. Here ${\cal S}$ stands for a filament-type saddle condition of zero gradient
and two positive eigenvalues of the tidal tensor, see Fig.~\ref{fig:crit_point}.
The slope $\delta'$ is replaced by the derivative of this whole
expression w.r.t. to $\sigma$, which gives $\delta' -\langle  \delta'| {\cal S }\rangle$ since the correlation functions of $\delta'$ with the saddle quantities correspond to the derivatives of the  $\delta$ correlations.
These modified height and slope no longer correlate with any saddle
quantity. Thus, the abundance of the various tracers at $\rr$ can
be inferred from standard excursion sets of this effective
density field.
The building blocks of this effective excursion set problem -- the variance of the field and of its slope, height and slope of the effective barrier -- are derived in full in Appendix~\ref{sec:statistics}. The main text of this section discusses how the saddle condition affects the upcrossing statistics, and the excursion set proxies for accretion rate and formation time.

For geometrical reasons, since statistical isotropy is broken only by the separation vector, any angular dependence of the correlation functions may arise only as $r_i$ or $r_ir_j$. Let us thus write equation~\eqref{eq:effdelta} as
\begin{equation}
  \condmean{\delta}{\S}=\xi_{00}\nus + 3\xi_{11}\frac{r}{R_\star}\hat r_ig_{i}
  -5\xi_{20} \frac{3\hat r_i\bar q_{ij} \hat r_j}{2}\,,
\label{eq:effmean}
\end{equation}
where $\hat r_i\equiv r_i/r$  and the correlation functions $\xi_{\alpha\beta}(r,R,\Rs)$ -- whose exact form is given in equation~\eqref{eq:xi_xi} -- depend only on the radial separation $r=|\rr|$ and the two smoothing scales, and have positive sign. Notice the presence of a minus sign in the shear term.
In the frame of the saddle, oriented with the $\hat z$ axis in the direction of outflow,
\begin{equation}
{\cal Q}\equiv \hat r_i\bar q_{ij} \hat r_j =
  \bar q_3 \sin^2 \!\theta \cos^2 \!\phi +
  \bar q_2 \sin^2\!\theta\sin^2\!\phi +
  \bar q_1 \cos^2\!\theta\,,
  \label{eq:rqrhatinframe}
\end{equation}
where $\theta$ and $\phi$ are the usual cylindrical coordinates in the frame of the eigenvectors $({\bf e_{3}},{\bf e_{2}},{\bf e_{1}})$ of $\bar q_{ij}$ with eigenvalues $\bar q_3 > \bar q_2 > \bar q_1$.

\begin{figure}
\begin{center}
  \includegraphics{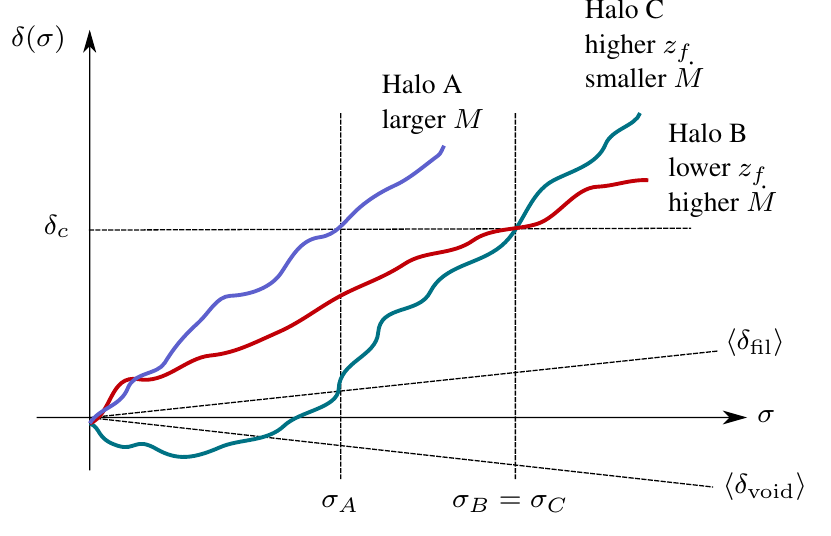}
  \caption{\label{fig:sketch-saddleES} Pictorial representation of the
    effect of the presence of saddle point on the excursion set
    trajectories at a finite distance from it. Halo $A$ and $B$ lie in
    the direction of the filament
    (${\cal Q}\equiv\hat r_i \bar{q}_{ij} \hat r_j <0$), where the mean density
    is higher than the average density. Halo $C$ lies in the direction
    orthogonal to it (${\cal Q}>0$), where
    the mean density is lower. haloes in the filament are likely to
    cross the collapsing threshold earlier, like halo $A$, than haloes
    in the voids. They thus tend to have larger mass.  At fixed
    crossing scale $\sigma_B=\sigma_C$, haloes in the filament are
    likely to cross with shallower slopes, like halo $B$, than halo in
    the voids. At their half-mass scale $\sigma_\half>\sigma_A$, their
    trajectories tend to be lower.  Hence, at fixed mass, haloes in
    the filaments tend to have larger accretion rates and to assemble
    half of their mass later. Conversely, haloes in the voids
    assemble their mass earlier, and then stop accreting.
  }
\end{center}
\end{figure}

When setting $g_i=0$, an angular dependence can only appear as a
functional dependence on ${\cal Q}(\hat\rr) =\hat r_i \bar{q}_{ij} \hat r_j$. That is, a
dependence on the direction $\hat\rr$ with respect to the eigenvectors
of the shear $\bar q_{ij}$.
As shown by equation~\eqref{eq:effmean}, a negative value of
${\cal Q}$ corresponds to a higher mean
density, which makes it easier for $\delta$ to reach $\dc$ and for
haloes to form.
At fixed distance from the saddle point, halo
formation is thus enhanced in the outflow direction with respect to
the inflow direction: haloes are naturally more clustered in the
filament than in the voids. Moreover, excursion set trajectories with
a lower mean will tend to cross the barrier with steeper
slopes than those crossing at the same scale but with a higher mean,
and will reach higher densities at smaller scales. Hence, haloes of the
same mass that form in the voids will form earlier and have a lower
accretion rates. These trends are shown in Fig.~\ref{fig:sketch-saddleES}.

To understand the radial dependence, one may expand equation~\eqref{eq:effmean} for small $r$ away from the saddle, obtaining
\begin{equation}
  \condmean{\delta}{\S} \simeq \mean{\delta\nus}_{r=0}\nus
  +\mean{\delta\nabla^2\nus}_{r=0}\frac{r^2}{2}
  \hat r_iq_{ij}\hat r_j\,;
\label{eq:smallr}
\end{equation}
whether the mean density increases or decreases with $r$ depends on the sign of the eigenvalues,  i.e the curvatures of the saddle, of the full $\mathbf{q}$ defined in equation~\eqref{eq:qij_def}.
Since $\mean{\delta\nabla^2\nus}<0$, the mean density grows quadratically with $r$ if $\hat r_i q_{ij}\hat r_j<0$, and decreases otherwise.
One thus expects the saddle point to be a maximum of halo number density, accretion rate and formation time in the two directions perpendicular to the filament, and a minimum in the direction parallel to it (corresponding to the negative eigenvalue $q_1$).

\subsection{Conditional halo counts}
\label{sec:condcounts}

The conditional distribution of the upcrossing scale $\sigma$ at finite distance $\rr$ from a saddle point of the potential can be evaluated following the generic procedure described in Section~\ref{sec:multivar}, fixing
\begin{equation}
  \{v_I\} =
  \{\nus,0,-\sqrt{5}(3{\cal Q}/2)\}
  \equiv \mathcal{S}(\rr)
\label{eq:csad}
\end{equation}
as the constraint. With this replacement,
equation~\eqref{eq:genfup} divided by $p_{\rm G}(\S)$ gives
\begin{equation}
  \fup(\sigma;\rr) =
  \frac{e^{-\nucS^2/2}}{\sqrt{2\pi\Var{\delta}{\S}}}\,
  \mu_\S F(X_\S)\,,
  \label{eq:condfup}
\end{equation}
which is the sought conditional distribution, with
\begin{equation}
    \mu_{\S}(\rr) \equiv \condmean{\delta'}{\nuc,\S}
\,,\quad
  X_{\S}(\rr) \equiv
  \frac{\mu_{\S}(\rr)}{\sqrt{\Var{\delta'}{\nuc,\S}}}\,,
\label{eq:XS}
\end{equation}
as in equation~\eqref{eq:X}.
The effective threshold $\nucS$ given the saddle condition is obtained replacing the generic constraint $v$ with $\S$ in equation~\eqref{eq:genconstr}.

The explicit calculation of the conditional quantities needed to compute $\nucS$, $\mu_{\cal S}$, $X_{\cal S}$
is carried out in Appendix~\ref{sec:statistics}. The results of Appendix~\ref{sec:dgivenS} (namely, equation~\eqref{eq:nu|S}) give
\begin{equation}
  \nucS(\rr) \equiv
  \frac{\dc-\condmean{\delta}{\S}}{\sqrt{\Var{\delta}{\S}}} =
  \frac{\dc-\xi_{00}\nus +\frac{15}{2}\xi_{20}{\cal Q}(\hat\rr)}{\sqrt{\sigma^2-\xi^2}}\,,
\label{eq:nucsad}
\end{equation}
consistently with equation~\eqref{eq:effmean},
where
\begin{equation} \label{eq:defxi2}
  \xi^2(r) \equiv
  \xi_{00}^2(r)+3\xi_{11}^2(r)r^2\!/R_\star^2+5\xi_{20}^2(r)
 \,.
\end{equation}
The effective slope parameters, obtained by replacing equations \eqref{eq:genmeannuprime} and \eqref{eq:genvarnuprime} into \eqref{eq:XS}, are
\begin{align}
  & \mu_\S(\rr) = \xi_I' \S_I +
  \frac{\sigma-\xi_I'\xi_I}{\sqrt{\sigma^2-\xi^2}}
  \nucS(\rr)\,,
\label{eq:muS}  \\
  & X_\S(\rr) = \mu_\S(\rr)\bigg/
  \bigg[\langle\delta'^2\rangle - \xi'^2
  - \frac{(\sigma-\xi_I'\xi_I)^2}{\sigma^2-\xi^2}\bigg]^{1/2} \,,
\label{eq:cvarS}
\end{align}
in terms of the vectors
\begin{align}
  \xi(r) &\equiv
  \{\xi_{00}(r),\sqrt{3}\xi_{11}(r)r/R_\star,\sqrt{5}\xi_{20}(r)\}
\label{eq:xisad}\,, \\
  \xi'\!(r) &\equiv
  \{\xi_{00}'(r),\sqrt{3}\xi_{11}'(r)r/R_\star,\sqrt{5}\xi_{20}'(r)\}\,.
\label{eq:xi'sad}
\end{align}
The correlation functions $\xi_{\alpha\beta}(r,R,\Rs)$ and their
derivatives $\xi_{\alpha\beta}'=\dd\xi_{\alpha\beta}/\dd\sigma$ are
given in equations~\eqref{eq:xi_xi} and ~\eqref{eq:xi_xiprime} respectively. Note that throughout the
text, $\xi_{\alpha\beta}$ or $\xi_{\alpha\beta}(r)$ will be used as a shorthand for
$\xi_{\alpha\beta}(r,R,\Rs)$.

Equation~\eqref{eq:condfup}, the main result of this subsection, is the conditional counterpart of equation~\eqref{eq:fup2}, and is formally identical to it upon replacing $\nuc$, $\nuc'$ and $X$ with $\nucS(\rr)$, $\nucS'(\rr)=-\mu_\S(\rr)/\sqrt{\sigma^2-\xi^2}$ and $X_\S(\rr)$.
The position dependent threshold $\nucS(\rr)$ and the slope parameter
$\mu_\S(\rr)$, given by equations \eqref{eq:nucsad} and \eqref{eq:muS}
respectively, contain anisotropic terms proportional to
${\cal Q}$ 
These terms account for all the
angular dependence of $\fup(\sigma;\rr)$. In the large-mass regime, as
$\{\xi_I'\}\simeq0$, $X_\S\simeq\nucS/(1-\xi^2)\gg1$ and
$F(X_\S)\simeq 1$. The most relevant anisotropic contribution is thus
the angular modulation of
$\nucS$, which raises or lowers the exponential tail of $\fup(\sigma;\rr)$ along or perpendicular to the filament. Upcrossing, and hence halo formation,
will be most likely in the direction that makes the threshold $\nucS$ smallest, as this makes it easier for the stochastic process to reach it.

In analogy to the unconditional case, when a characteristic mass scale could be
defined for which $\sigma=\dc$, equation~\eqref{eq:condfup} suggests
to define the characteristic mass scale $\sigma_\star=\sigma(M_\star)$ for haloes near the saddle as the one for which $\nucS=1$ in equation~\eqref{eq:nucsad}. In the language of
excursion sets, this request naturally sets the scale
\begin{equation}
  \sigma_\star^2(\rr)\equiv
  (\dc-\xi_{00}\nus +\frac{15}{2}\xi_{20} {\cal Q}
  )^2
  +\xi^2(r)
  \,.
\label{eq:Mstarcond}
\end{equation}
This is now an implicit equation for $\sigma_\star$, because the
right-hand side has a residual dependence on $\sigma_\star$ through
$\xi_{\alpha\beta}(r, R(\sigma_\star),\Rs)$, as shown in Appendix
\ref{sec:covariance}. This equation can be solved numerically for
$\sigma_\star$ and then for $M_\star$.

\begin{figure}
  \centering
  \includegraphics[width=\columnwidth]{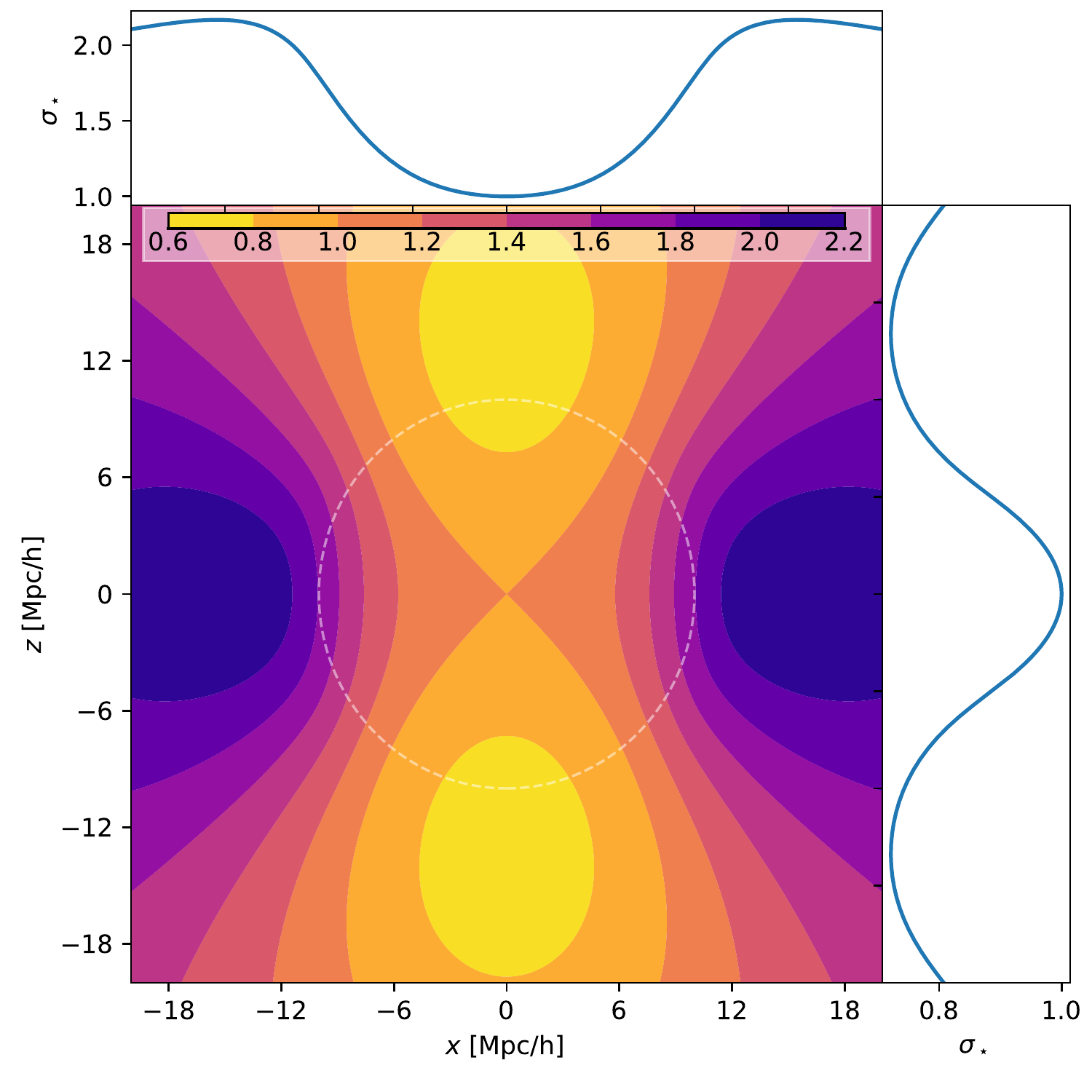}
   \caption{Isocontours in the $x-z$ plane of the typical upcrossing
    scale $\sigma_\star$ around a saddle point (at $(0, 0)$). The
    saddle point is defined using the values of
    table~\ref{tab:qijbar_nu_values}. The profiles in the direction of
    the filament ($z$ direction) and of the void ($x$ direction) are
    plotted on the sides. The smoothing scale is
    $R=\SI{1}{Mpc/h}$. They are obtained by solving
    equation~\eqref{eq:Mstarcond} for $\sigma_\star$ at each point,
    with a $\Lambda$CDM power spectrum, and normalized to the value at
    the saddle point. In the filament, haloes form at a smaller $\sigma$
    (higher mass) and conversely in the void.   }
  \label{fig:sigma_star}
\end{figure}

The angular dependence of $\sigma_\star(\rr)$ is entirely due to $\xi_{20}{\cal Q}$.
Since the prefactor of ${\cal Q}\equiv\hat r_i q_{ij}\hat r_j$ is positive, $\sigma_\star(\rr)$ will be smallest when $\rr$ aligns with the eigenvector with the smallest eigenvalue, and ${\cal Q}$ is most negative. This happens when $\theta=0$ in equation~\eqref{eq:rqrhatinframe}: that is, in the direction of positive outflow, along which a filament will form. Thus, in filaments haloes tend to be more massive than field haloes. The full radial and angular dependence of the characteristic mass scale $\sigma_\star$ is shown in Fig.~\ref{fig:sigma_star}.

\subsection{Conditional accretion rate}
\label{sec:condacrrate}

The abundance of haloes of given mass and accretion
rate at distance $\rr$ from a saddle is obtained by replacing
the probability distribution $p_{\rm G}(\nuc,\nuc'+\nuc/\sigma\alpha)$ in
equation~\eqref{eq:fupalpha} with its conditional counterpart
given the saddle constraint.
As shown by equation~\eqref{eq:gencondp}, this conditional distribution is equal to the distribution of the effective independent variables $\tnu$ and $\delta'-\condmean{\delta'}{\nuc,\S}$ introduced in Section~\ref{sec:multivar}, times a Jacobian factor of $\sigma/(1-\xi^2/\sigma^2)$. Furthermore, the relation \eqref{eq:accrate} giving the excursion set slope in terms of the accretion rate reads in these new variables
\begin{equation}
  \delta'- \condmean{\delta'}{\nuc,\S} =\frac{\nuc}{\alpha}-\mu_S\,.
\end{equation}
Putting these two ingredients together, equation~\eqref{eq:fupalpha} becomes
\begin{align}
  f_\mathrm{up}(\sigma,\alpha;\rr) & =
  \frac{\nuc^2}{\sigma^2\alpha^3} p_{\rm G}(\nuc,\nuc'+\nuc/\sigma\alpha|\S)
\notag\,, \\
  &=\frac{\nuc^2}{\alpha^3}
  \frac{e^{-\left(\nucS^2+ Y_{\alpha,\S}^2\right)/2}}{2\pi\sqrt{(\sigma^2-\xi^2)\Var{\delta'}{\nuc,\S}}}\,,
\label{eq:condjoint}
\end{align}
where ${\Var{\delta'}{\nuc,\S}}$  is given by equation~\eqref{eq:condvarx}  and
\begin{equation} \label{eq:defYalpha}
  Y_{\alpha,\S}(\rr) \equiv
  \frac{\nuc/\alpha-\mu_\S(\rr)}{\sqrt{\Var{\delta'}{\nuc,\mathcal{S}}}}\,,
\end{equation}
with $\mu_\S(\rr)$ given by equation~\eqref{eq:muS}.
Again, like equation~\eqref{eq:fupalpha}, this result could be obtained by taking $\condmean{\delta'}{\nuc,\alpha,\S}=\nuc/\alpha$ and the limit $\Var{\delta'}{\nuc,\alpha,\S}\to0$ in equation~\eqref{eq:X}, which would give $F(X_{\alpha,\S})=1$.

To investigate the anisotropy of the accretion rate for haloes of the same mass, one needs the conditional probability of $\alpha$ given upcrossing at $\sigma$, that is the ratio of equations \eqref{eq:condjoint} and \eqref{eq:condfup}. This conditional probability reads
\begin{equation}
  f_\mathrm{up}(\alpha|\sigma;\rr) = 
  \frac{\nuc e^{-Y_{\alpha,\S}^2\!/2}}{\alpha^3\sqrt{2\pi\Var{\delta'}{\nuc,\S}}}
  \frac{\nuc}{\mu_\S F(X_\S) } \,,
\label{eq:condacc}
\end{equation}
with $\mu_\S(\rr)$ and $X_\S(\rr)$ given by equation~\eqref{eq:muS} and~\eqref{eq:cvarS} respectively. The second fraction in this expression is thus a normalization factor that does not depend on $\alpha$, and which tends to 1 when $\nuc\gg1$ in the large-mass limit.
Equation~\eqref{eq:condacc}  is the main result of this subsection. It depends on the angular position $\hat\rr$ through the terms $\xi_{20}'{\cal Q}$ and $\xi_{20}{\cal Q}$ contained in $\mu_\S(\rr)$, and thus also in $Y_{\alpha\S}$ and $X_\S$. The angular dependence is now weighted by two different functions $\xi_{20}(r)$ and $\xi_{20}'(r)$, whose relative amplitude matters to determine the overall effect.

To understand the angular variation of the exponential tail of this distribution, let us focus on how $Y_\alpha(\rr)$ depends on $\hat\rr$. That is, on the anisotropic part of $-\mu_\S(\rr)$.
In the large mass limit, when $\sigma\xi'_{\alpha\beta}(r)\ll\xi_{\alpha\beta}(r)$, equation~\eqref{eq:muS} tells us that the anisotropic part of $Y_\alpha(\rr)$ is proportional to $-\xi_{20}{\cal Q}$, with a proportionality factor that is always positive and $\mathcal{O}(1)$.
Thus, the modulation has the opposite sign of the anisotropic part of $\nucS$, given in equation~\eqref{eq:nucsad}: for trajectories with the same upcrossing scale, the probability of having a given accretion rate is lowest in the direction of the eigenvector of $\bar q_{ij}$ with the lowest (most negative) eigenvalue, for which $Y_\alpha$ is largest. That is, for haloes with the same mass, the probability of having a given accretion rate is lowest along the ridge of the potential saddle, which will become the filament.

\begin{figure*}
  \begin{center}
    \includegraphics[width=\columnwidth]{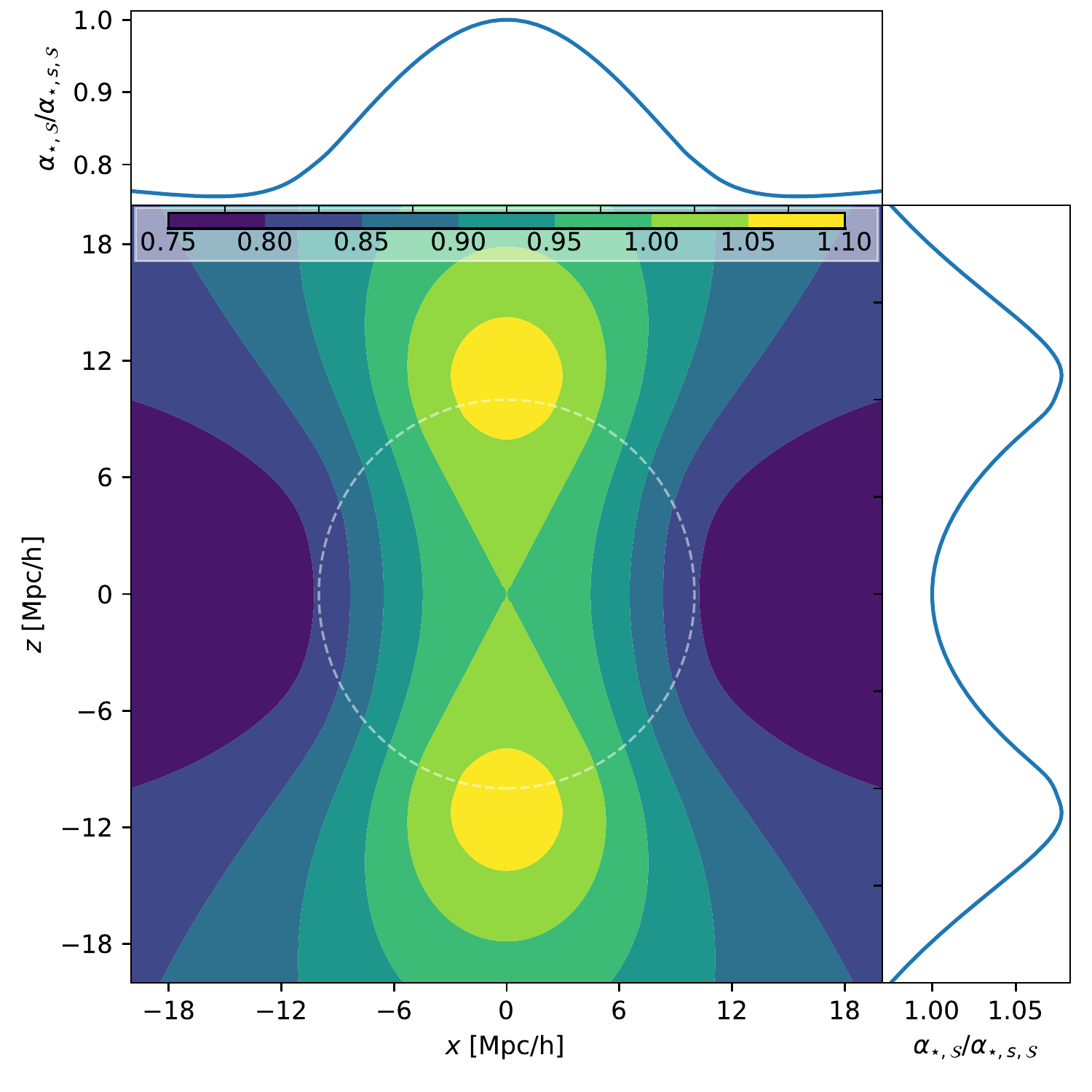}
    \includegraphics[width=\columnwidth]{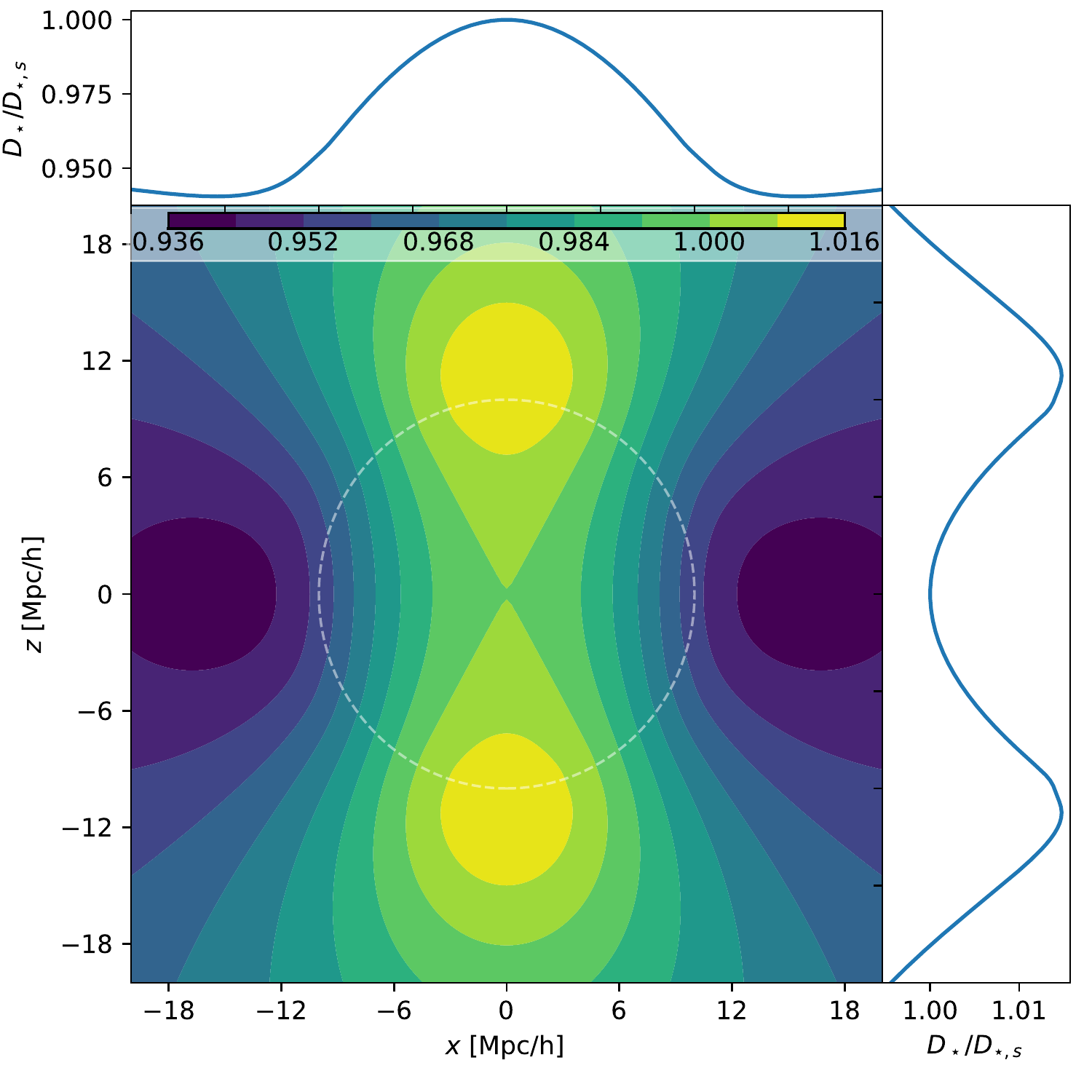}
  \includegraphics[width=\columnwidth]{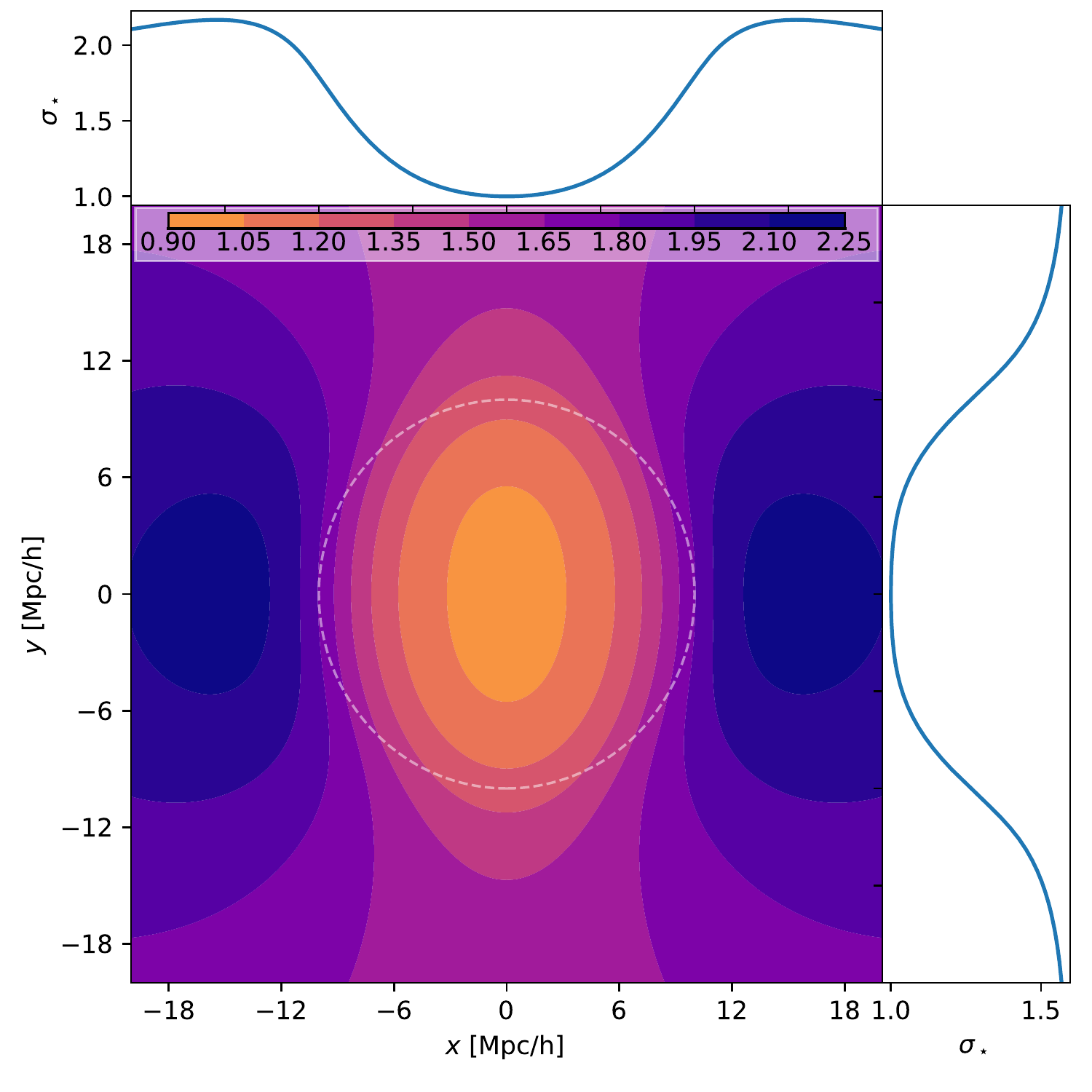}
  \includegraphics[width=\columnwidth]{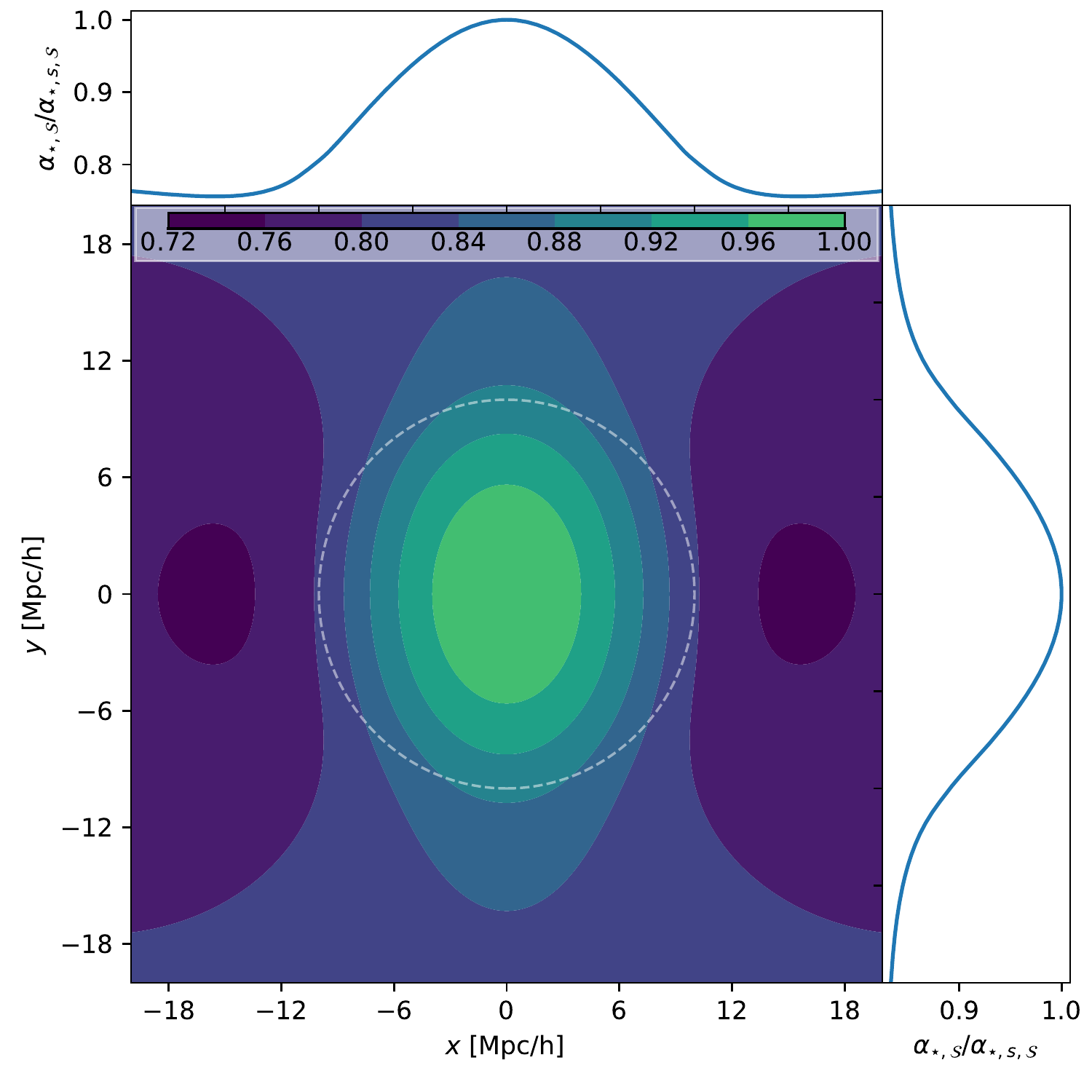}
  \caption{Isocontours in the $x-z$ plane of the typical accretion
    rate $\alpha_\star$ ({\em upper left}) and formation time
    $D_\star$ ({\em upper right}) around a saddle point (at $(0, 0)$)
    and in the $x-y$ plane of the characteristic upcrossing scale
    $\sigma_\star$ ({\em lower left}) and typical accretion rate ({\em
      lower right}). The saddle point is defined using the values of
    table~\ref{tab:qijbar_nu_values}. The profiles going through the
    saddle point in the $x-z$ ({\em upper panels}) and $x-y$ ({\em
      lower panels}) planes are plotted on the sides. The smoothing
    scale is $R=\SI{1}{Mpc/h}$. They were obtained with a $\Lambda$CDM
    power spectrum, and normalized to the value at the saddle point.
    Since the filament has higher mean density, excursion set
    trajectories upcrossing at a given $\sigma$ have shallower
    slopes. Hence, typical haloes are more massive in filaments and at
    fixed mass, haloes forming in the filament have larger accretion
    rates at $z=0$ and form later. The same hierarchy exists between
    the two perpendicular directions.   }
     \label{fig:astar}
    \label{fig:Dstar_xz}
  \end{center}
  \label{fig:sigma_xy}
  \label{fig:alphastar_xy_}
\end{figure*}

The typical accretion rate $\alpha_\star$ of the excursion set haloes described by the distribution \eqref{eq:condacc} corresponds to the condition $Y_{\alpha_\star,\S}=1$. This definition transforms equation~\eqref{eq:alphastar} into
\begin{equation}
  \alpha_\star(\sigma,\rr) \equiv
  \frac{\nuc}{\sqrt{\Var{\delta'}{\nuc,\S}} + \mu_\S(\rr)}  \,,
\label{eq:astarsad}
\end{equation}
where ${\Var{\delta'}{\nuc,\S}}$ and $ \mu_\S(\rr)$ are given by equations~\eqref{eq:condvarx} and \eqref{eq:muS}.
In the limit of small anisotropy, the angular variation of the typical accretion rate is
\begin{equation}
  \Delta\alpha_\star(\sigma,\rr) =
  \frac{\alpha_\star^2|_{\mathbf{\bar q}=0}}{\nuc} \frac{15}{2}
  \bigg[\xi_{20}'-\frac{\sigma-\xi'_I\xi_I}{\sigma^2-\xi^2}\xi_{20}\bigg]
  \hat{r}_i\bar q_{ij}\hat{r}_j\,,
\label{eq:Dalphastar}
\end{equation}
where $\alpha_\star|_{\mathbf{\bar q}=0}$ -- the value of
$\alpha_\star(\sigma,\rr)$ when $\bar q_{ij}=0$ -- is function of $r$
but not of the angles.  Therefore, at a fixed distance $r$ from the
saddle, haloes that form in the direction of the filament tend to have
higher accretion rates than haloes with the same mass that form in the
orthogonal direction.  The full dependence of the characteristic
accretion rate $\alpha_\star$ for haloes of the same mass on the
position with respect to the saddle point of the potential is shown in
Fig.~\ref{fig:astar}. The figure shows that the saddle point is a
local minimum of the accretion rate along the direction connecting two
regions with high density of final objects, that is two peaks of the
final halo density field.This is consistent with the result that the
accretion of haloes in filaments is suppressed by the effect of the
tidal forces \citep[as shown by,
e.g.,][]{Hahnetal2009,2016arXiv161004231B}. The threshold  $\delta\lesssim\delta_c$
is reached at smaller $\sigma$ in filaments than in void, hence the slope
is smaller at upcrossing. It is shown schematically in the top panel
of Fig.~\ref{fig:saddleES}. A verification with a constrained random
field is shown in the bottom panel of Fig.~\ref{fig:saddleES}. The
details of the method used are given in Appendix~\ref{sec:validation}.

One can also evaluate the mean of the conditional distribution \eqref{eq:condacc} following equation~\eqref{eq:condmeanalpha}, integrating $\alpha f_\mathrm{up}(\alpha|\sigma,\mathcal{S})$ over the range of positive $\alpha$.
This conditional mean value is
\begin{equation}
  \condmean{\alpha}{\sigma}\!(\rr) =
  \frac{\nuc}{\mu_\S(\rr)}
  \frac{1+\erf(X_\S(\rr)/\sqrt{2})}{2F(X_\S(\rr))}\,;
  \label{eq:alphaS*}
\end{equation}
in the large-mass regime, where $X_\S\gg1$ and the whole second fraction tends to 1, the position dependent conditional mean $\condmean{\alpha}{\sigma}(\rr)$ is essentially the same as $\alpha_\star(\rr)$ defined in equation~\eqref{eq:astarsad}. As for $\fup(\alpha|\sigma)$, all higher order moments are ill defined.
One can also find useful information in the most likely accretion rate
\begin{equation}
  \alpha_{\mathrm{max}}(\sigma,\rr) =
  \frac{\nuc^2}{6\Var{\delta'}{\nuc,\S}}
  \left[\sqrt{1+\frac{12}{X_\S^2(\rr)}}-1\right],
\end{equation}
which generalizes equation~\eqref{eq:amax} to the presence of a saddle point at distance $\rr$.
The same conclusion holds here namely the most likely accretion rate increases from voids to saddles and saddles to nodes.
The following only considers maps of $\alpha_\star(\sigma,\rr)$, since the information encoded in $\alpha_{\mathrm{max}}(\sigma,\rr)$ and $\condmean{\alpha}{\sigma}\!(\rr)$ is somewhat redundant.

\subsection{Conditional formation time}
\label{sec:condft}

The formation time in the vicinity of a saddle is obtained by fixing the saddle parameters $\S=\{\nus,\hat r_ig_i,\hat r_i\bar q_{ij}\hat r_j\}$, with $g_i=0$, besides $\nu=\nuc$ and $\nu_{\half}=\nuf$.
A 5-dimensional constraint on the Gaussian variables must now be dealt with, and mapped into $\{\sigma,\Df,\S\}$. Since the mapping of the saddle parameters is the identity, the Jacobian of the transformation still gives $|\nu'-\nuc'|\nuf/\Df$, like in Section~\ref{sec:form} (where there was no saddle constraint).
 The formalism outlined in Section~\ref{sec:multivar} still applies: the joint probability of upcrossing at $\sigma$ with formation time $\Df$ given the saddle is obtained replacing replacing $\{v\}$ with $\{\nuf,\S\}$ in \eqref{eq:X}, multiplying by the Jacobian $\nuf/\Df$ and dividing by the probability $p_{\rm G}(\S)$ of the saddle. The result is
\begin{equation}
  \fup(\sigma,D_\form;\rr)
  = \frac{\nuf}{D_\form} p_\G(\nuc,\nuf|\S) \,
  \frac{\mu_{\form,\S}}{\sigma} F(X_{\form,\S})\,
\label{eq:fupDS}
\end{equation}
which extends equation~\eqref{eq:fupD} by including the presence of a saddle point of the potential at distance $\rr$, with
\begin{equation}
  \mu_{\form,\S} \equiv \condmean{\delta'}{\nuf,\nuc,\S}
\,,\;\;
  X_{\form,\S} \equiv
  \frac{\mu_{\form,\S}}{\sqrt{\Var{\delta'}{\nuf,\nuc,\S}}}\,.
\label{eq:XfS}
\end{equation}
The conditional mean and variance of $\delta'$ given $\{\nuf,\nuc,\S\}$
are explicitly computed in Appendix~\ref{sec:condhalfmass}, equations \eqref{eq:meandp_nucnufS} and \eqref{eq:vardp_nucnufS}.

The conditional probability of the formation time $\Df$ given $\sigma$ at a distance $\rr$ from the saddle follows dividing equation~\eqref{eq:fupDS} by $\fup(\sigma|\rr)$, given by equation~\eqref{eq:condfup}. This ratio -- which is the main result of this section -- gives
\begin{align}
  \fup(\Df|\sigma;\rr)
  &= \frac{\nuf}{\Df} p_\G(\nuf|\nuc,\S) \,
  \frac{\mu_{\form,\S}}{\mu_\S} \frac{F(X_{\form,\S})}{F(X_{\S})}\,,
\notag \\
  &= \frac{(\dc/\Df^2)e^{-\nu_{\form,\mathrm{c},\S}^2/2}}{\sqrt{2\pi\Var{\delta_{\half}}{\nuc,\S}}} \,
  \frac{\mu_{\form,\S}}{\mu_\S} \frac{F(X_{\form,\S})}{F(X_{\S})}\,.
\label{eq:condDsad}
\end{align}
Equation~\eqref{eq:condDsad} provides the counterpart of equation~\eqref{eq:condD12} near a saddle point, in terms of the effective threshold
\begin{equation}
  \nu_{\form,\mathrm{c},\S}(\Df,\rr) \equiv 
  \frac{\dc/\Df -\condmean{\delta_{\half}}{\nuc,\S}}{\sqrt{\Var{\delta_{\half}}{\nuc,\S}}}\,,
\label{eq:tnufsad}
\end{equation}
with
\begin{align}
  &\condmean{\delta_{\half}}{\nuc,\S} =
  \xi_{\half}\!\cdot\!\mathcal{S}
  +\frac{\mean{\delta\delta_{\half}}-\xi\!\cdot\!\xi_{\half}}{\sigma^2-\xi^2}(\dc -\xi\!\cdot\!\S)
\label{eq:meannucS}\,,  \\
  &\Var{\delta_{\half}}{\nuc,\S} =
  \sigma_{\half}^2-\xi_{\half}^2-\frac{(\mean{\delta\delta_{\half}}-\xi\!\cdot\!\xi_{\half})^2}{\sigma^2-\xi^2}\,.
\label{eq:varnucS}
\end{align}
It also depends on the effective upcrossing parameters  $\mu_{\S}(\rr)$ and $X_{\S}(\rr)$, given in equations \eqref{eq:XS}-\eqref{eq:muS}.
The explicit forms of the functions $\mu_{\form,\S}(\Df,\rr)$, $X_{\form,\S}(\Df,\rr)$ are reported in Appendix~\ref{sec:condhalfmass}  for convenience
(equations~\eqref{eq:machin3} and ~\eqref{eq:bidule3}).

Note that in equation~\eqref{eq:condDsad}, $\fup(\Df|\sigma;\rr)$ depends on $\Df$ also through $\nu_{\form,\mathrm{c},\S}$ and $\mu_{\form,\S}$.
For early formation times ($\Df\ll1$), the conditional mean  $\condmean{\delta'}{\nuf,\nuc,\S}$ becomes large, since the trajectory must reach a very high value at $\sigma_{\half}$. Hence, $\mu_{\form,\S}(\Df,\rr)\propto 1/\Df$. In this limit, the last ratio in equation~\eqref{eq:condDsad} above tends to 1, and $\fup(\Df|\sigma;\rr)\propto (1/\Df^3)\exp(-\nu_{\form,\mathrm{c},\S}^2/2)$, with a proportionality constant that does not depend on the angle. Then, the probability decays exponentially for small $D_\form$ as $\nu_{\form,\mathrm{c},\S}$ grows. 
The typical formation time $D_{\star}=D(z_{\star})$ can be defined as that value for which $\nu_{\form,\mathrm{c},\S}=1$ and this exponential cutoff stops being effective, that is
\begin{equation}
  D_{\star}(\rr,\sigma) \equiv \frac{\dc}{
    \sqrt{\Var{\delta_\half}{\nuc,\mathcal{S}}} +
    \condmean{\delta_\half}{\nuc,\mathcal{S}} }\,,
\label{eq:defD*sad}
\end{equation}
which provides the anisotropic generalization of the expression given in equation~\eqref{eq:defD*}. The explicit expression for the conditional mean $\condmean{\delta_\half}{\nuc,\mathcal{S}}$ and variance $\Var{\delta_\half}{\nuc,\mathcal{S}}$  are given by equations \eqref{eq:meannucS} and \eqref{eq:varnucS} respectively. 

As the angular variation of $\condmean{\delta_\half}{\nuc,\mathcal{S}}$ is approximately
\begin{equation}
  \frac{15}{2}{\Ds_{\half}}
  \xi_{20}(r){\cal Q}(\hat\rr)\,,
\end{equation}
where ${\cal Q}(\hat r)\equiv\hat{r}_i\bar q_{ij}\hat{r}_j$, $\Ds_{\half}=\sigma_\half-\sigma>0$, the formation time $D_\star$ is larger when $\rr$ is aligned with the eigenvector with the most negative eigenvalue, corresponding to the direction of the filament. One has in fact
\begin{equation}
  \Delta D_{\star}(\rr,\sigma) =
  -\frac{D_{\star}^2|_{\mathbf{\bar q}=0}}{\dc}
  \frac{15}{2}{\Ds_{\half}}
  \xi_{20}(r){\cal Q}(\hat\rr)\,,
\label{eq:DDstar}
\end{equation}
where $D_{\star}$ depends only on the radial distance $r$,
which shows that at a fixed distance from the saddle point, haloes in the direction of the filament tend to form later (larger $D_{\star}$). The saddle point is thus a minimum of the half-mass time $D_\star$ along the direction of the filament, that is a maximum of $z_\star$: haloes that form at the saddle point assemble most of their mass the earliest.
Fig.~\ref{fig:Dstar_xz} displays a cross section of a map of $D_\star$ in the frame of the saddle.

\section{Astrophysical Reformulation}
\label{sec:astro}

The joint and conditional PDFs derived in Sections \ref{sec:excursionsets}, \ref{sec:unconditional} and \ref{sec:conditional} were expressed in terms of variables ($\sigma, \alpha$ and $\Df$) that are best suited for the excursion set theory.
Now, for the sake of connecting to observations and gathering a wider audience, let us write explicitly what the main results of those sections -- equations~\eqref{eq:fup3}, \eqref{eq:condalpha} and \eqref{eq:condD12}, and their constrained counterparts \eqref{eq:condfup}, \eqref{eq:condacc} and  \eqref{eq:condDsad} -- imply in terms of astrophysically relevant quantities like the distribution of mass, accretion rate and formation time of dark matter haloes.
\subsection{Unconditional halo statistics}
The upcrossing approximation provides an accurate analytical solution of the random walk problem formulated in the Extended Press--Schechter (EPS) model, for a Top-Hat filter in real space and a realistic power spectrum. In this framework, the mass fraction in haloes of mass $M$ is
\begin{equation}
  \frac{M}{\bar\rho} \frac{\d n}{\d M} =
  \left|\frac{\d \sigma}{\d M}\right| f_{\rm up}(\sigma(M))\,,
\label{eq:massfracup}
\end{equation}
with $f_{\rm up}(\sigma)$ given by equation~\eqref{eq:fup3} and
  is a function of mass via
equation~\eqref{eq:sigma2}. For instance, for a power-law power
spectrum $P(k)\propto k^{-n}$ with index $n=2$ one has
$M/M_\star =(\sigma/\sigma_\star)^{-6}$. The general power-law result $M\propto \sigma^{6/(n-3)}$ follows from equation \eqref{eq:sigmapowerlaw}.

The excursion set approach also establishes a natural relation between the accretion rate of the halo and the slope of the trajectory at barrier crossing. One can thus predict the joint statistics of $\sigma$ and of the excursion set proxy $\alpha\equiv\nuc/[\d(\delta-\dc)/\d\sigma]$ for the accretion rate. In order to get the joint mass fraction in haloes of mass $M$ and accretion rate $\dot M$, one needs to introduce the Jacobian of the mapping from $(\sigma,\alpha)$ to $(M,\dot M)$. Since $\sigma(M)$ does not depend on $\alpha$, this Jacobian has the simple factorized form $|\d\sigma/\d M||\d\alpha/\d\dot M|$. Since $\d\alpha/\d\dot M=\alpha/\dot M$  from equation~\eqref{eq:defmdot}, one can write the joint analog of equation~\eqref{eq:massfracup} as
\begin{equation}
 \frac{M\dot M}{\bar\rho} \frac{\d^2 n}{\d M \d {\dot M}} =
 \left|\frac{\d \log \sigma}{\d M}\right|\,
 \sigma\alpha f_{\rm up}(\sigma,\alpha)\,,
\label{eq:massfracalpha}
\end{equation}
where $ \fup(\sigma,\alpha)$ is now given by equation~\eqref{eq:fupalpha}, whereas $\sigma(M)$ and $\alpha(M,\dot M)$ are functions of $M$ and $\dot M$ via equations~\eqref{eq:sigma2} and \eqref{eq:defmdot} respectively.
From the ratio of equations~\eqref{eq:massfracalpha} and \eqref{eq:massfracup}, the expected mean density of haloes of given mass and accretion rate can be reformulated as
\begin{equation}
  \dot M \frac{\d^2 n}{\d M \d {\dot M}} =
 \alpha f_{\rm up}(\alpha|\sigma) \frac{\d n}{\d M }\,,
\label{eq:massfracalphacond}
\end{equation}
where $ f_{\rm up}(\alpha|\sigma)$ is given by equation~\eqref{eq:condalpha}. This expression relates analytically the number density of haloes binned by mass and accretion rate to the usual mass function.

Similarly, the joint mass fraction of haloes of mass $M$ and formation
time $z_\form$ (defined as the redshift
at which the halo has assembled half of its mass) can be inferred from
the joint statistics of $\sigma$ and
$D_\form\equiv\dc/\delta(\sigma_\half)$, where
$\sigma_\half\equiv\sigma(M/2)$ is the scale containing half of the
initial volume.  The redshift dependence of the growth function $D(z)$
is defined by \eqref{eq:defD}. Hence, the mass fraction in haloes of
given mass $M$ and formation time $z_\form$ is
\begin{equation}
  \frac{M}{\bar\rho}\frac{\d^2n}{\d M\d z_\form}
  = \frac{\d\sigma}{\d M} \frac{\d D_\form}{\d z_\form}
  \fup(\sigma,D_\form)\,,
\label{eq:massfracD12}
\end{equation}
and its conditional is
\begin{equation}
  \frac{\d^2n}{\d M\d z_\form} = \frac{\d D_\form}{\d z_\form}
  \fup(D_\form|\sigma) \frac{\d n}{\d M} \,,
\label{eq:massfracD12cond}
\end{equation}
where the joint and conditional distributions of $D_\form$ and $\sigma$ are given by equations~\eqref{eq:fupD} and \eqref{eq:condD12} respectively.

\begin{figure}
  \centering
  \includegraphics[width=1.1\columnwidth]{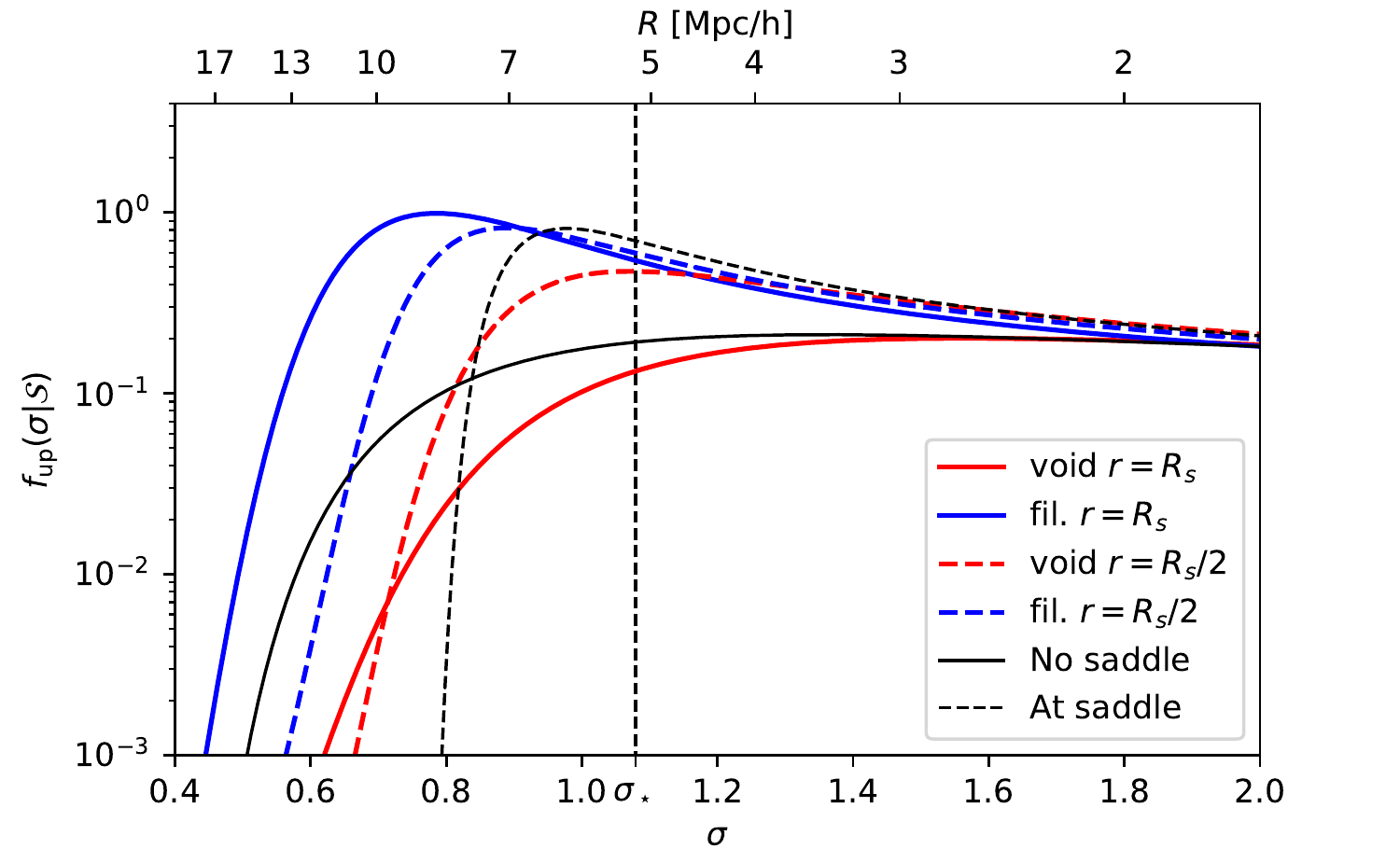}
  \caption{PDF of $\sigma$ at upcrossing given the saddle point in the
    $x$ (void, in red) and $z$ (filament, in blue) directions at
    distance $r = \SI{10}{Mpc/h}$ (solid lines) and
    $r = \SI{5}{Mpc/h}$ (dashed lines). The saddle point is defined
    using the values of table~\ref{tab:qijbar_nu_values}. The PDF
    without the saddle point is shown in black and at the saddle point
    in dashed black. The value of $\sigma_\star$ at the saddle point
    is shown by the vertical dashed line. In the filament, the PDF is
    boosted for small values of $\sigma$: there are more massive
    haloes in the filament. The opposite trend is seen in the void.
  }
  \label{fig:fup_sigma}
\end{figure}

Interestingly, while the excursion set mass function is subject to the limitation of upcrossing theory, the conditional statistics of accretion rate, or formation redshift, at given mass should be considerably more accurate.
This is because the main shortcoming of excursion sets is the lack of a prescription for where to centre in space each set of concentric spheres giving a trajectory. These spheres are placed at random locations, whereas they should insist on the centre of the proto-halo. However, choosing a better theoretical model (e.g. the theory of peaks) to set correctly the location of the excursion set trajectories would not dramatically modify the conditional statistics. Changing the model would modify the function $F(x)$, defined in equation~\eqref{eq:defF}, that modulates each PDF. In conditional statistics, only ratios of this function appear, which are rather model independent, whereas the probability of the constraint does not appear. The relevant part for our analysis -- the exponential cutoff of each conditional distribution given the constraint -- would not change.
Hence, even though equation~\eqref{eq:massfracup} does not provide a good mass function $\d n/\d M$, one may argue that the relations \eqref{eq:massfracalphacond} and \eqref{eq:massfracD12cond} are still accurate in providing the joint abundance statistics of mass and accretion rate, or mass and formation redshift, once a better model -- or even a numerical fit -- is used to infer $\d n/\d M$.

\subsection{Halo statistics in filamentary environments}

\begin{figure}
  \centering
  \includegraphics[width=\columnwidth]{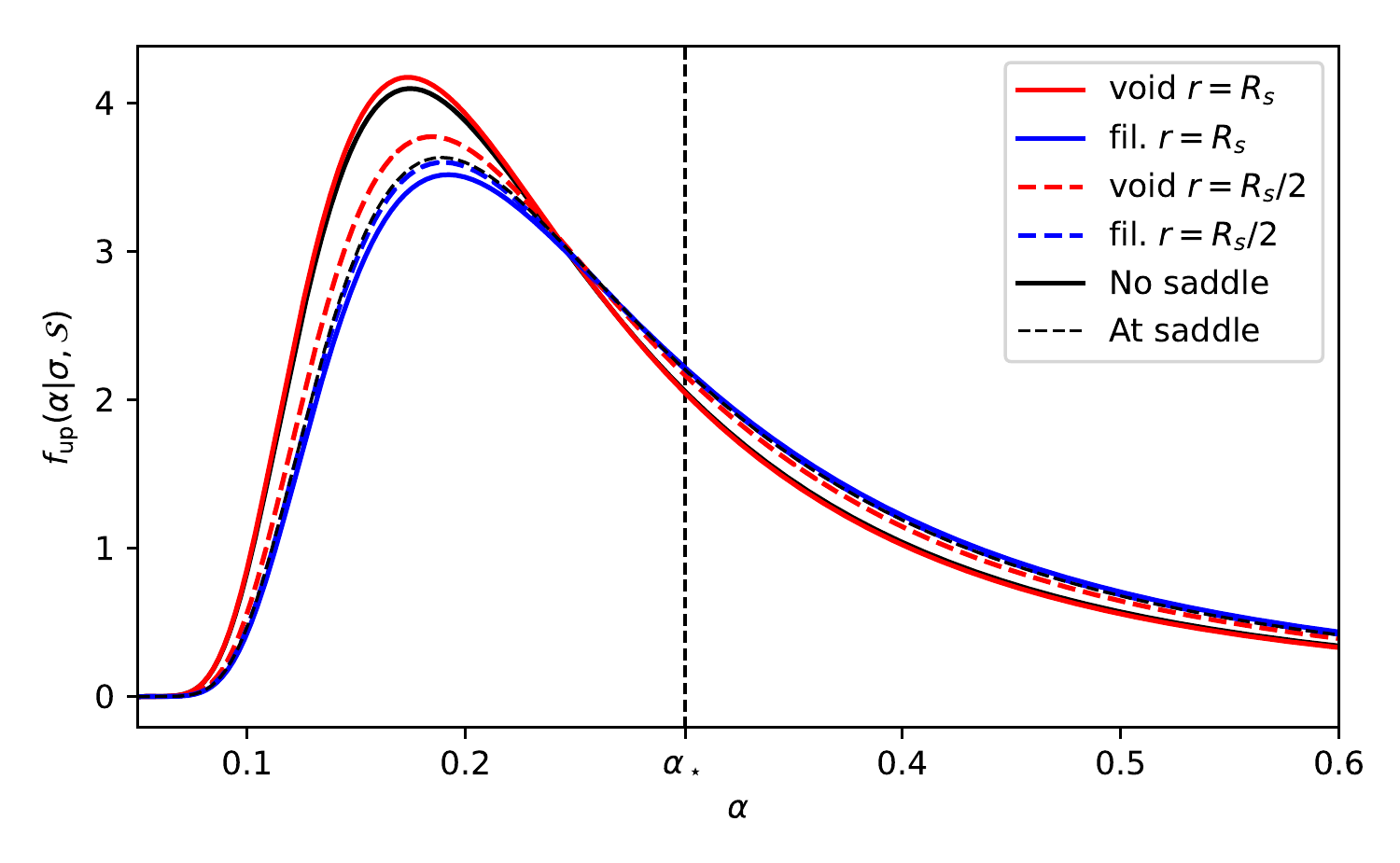}
    \includegraphics[width=\columnwidth]{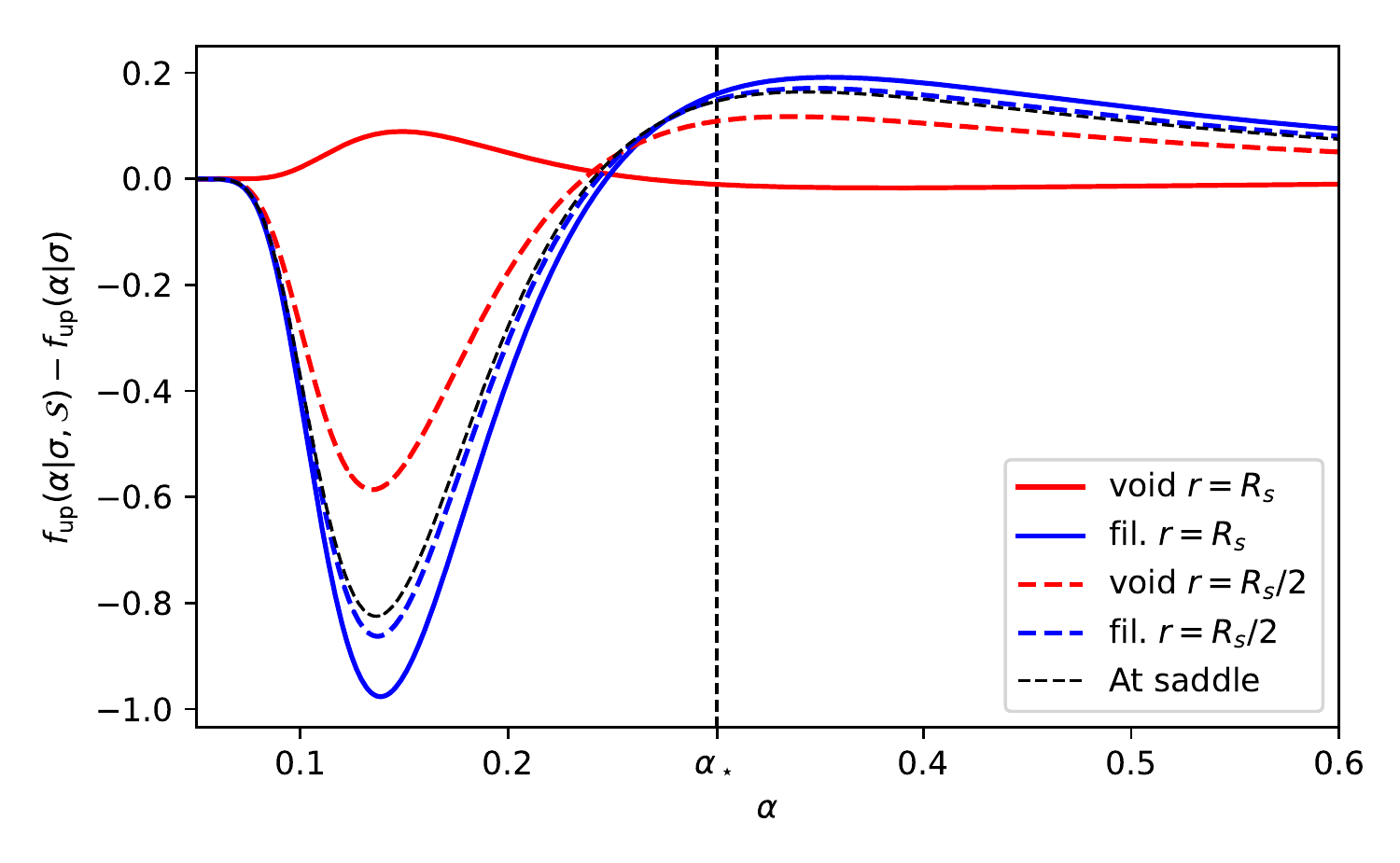}
    \caption{PDF of $\alpha$ at upcrossing given the smoothing scale
      and the saddle point in the $x$ (void, in red) and $z$
      (filament, in blue) directions at distance $r = \SI{10}{Mpc/h}$
      (solid lines) and $r = \SI{5}{Mpc/h}$ (dashed lines) ({\em upper
        panel}) compared to the PDF without the saddle point ({\em
        lower panel}). The saddle point is defined using the values of
      table~\ref{tab:qijbar_nu_values}. The PDF with no saddle point
      is shown in solid black and the PDF at the saddle point in
      dashed black. In the filament, the PDF is boosted at its high
      end: haloes accrete more. The opposite trend is seen in the
      void.   }
    \label{fig:fup_alpha}
\end{figure}
\begin{figure}
  \centering
    \includegraphics[width=\columnwidth]{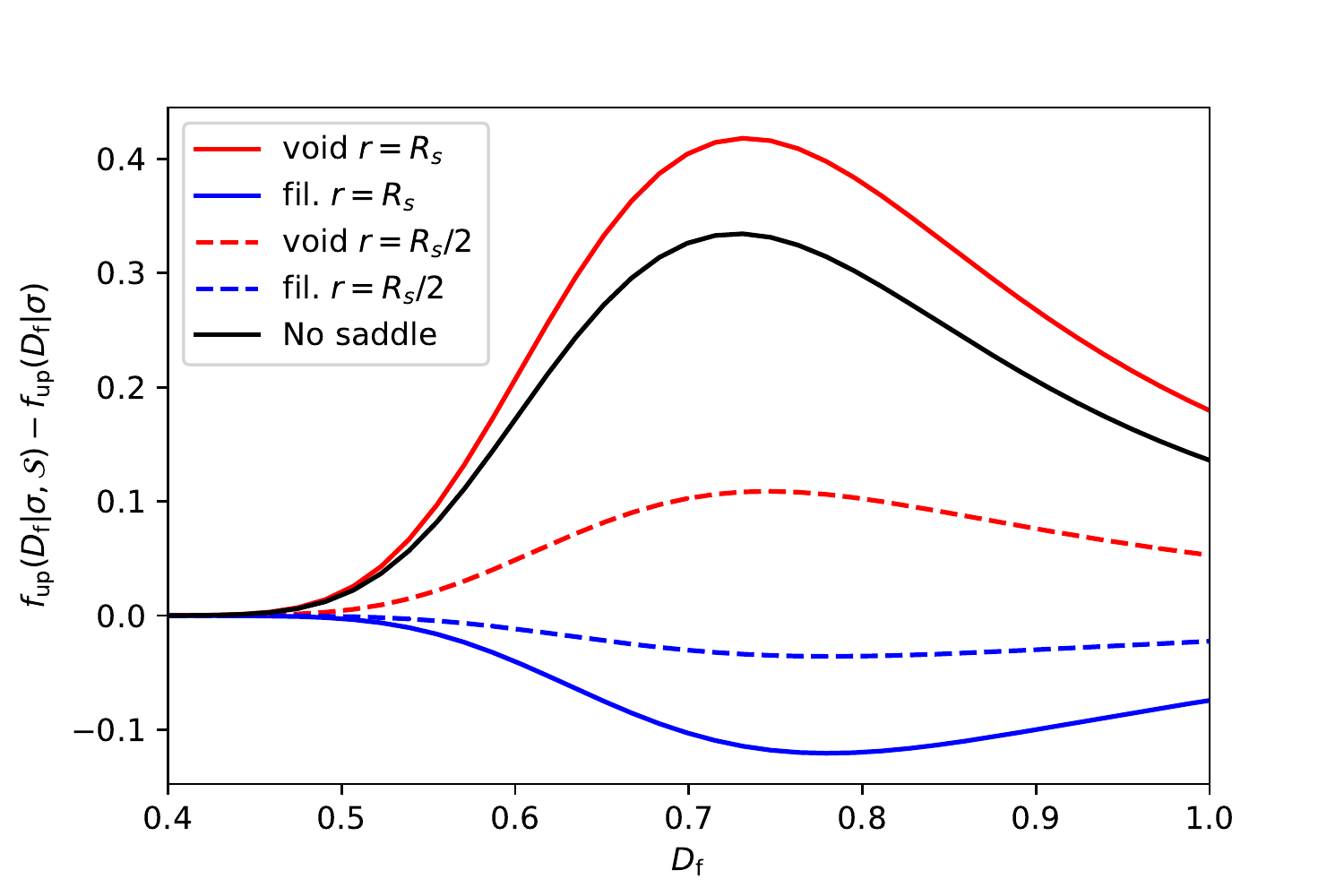}
  \caption{PDF of $D_\form$ at upcrossing given the smoothing scale
    and the saddle point in the $x$ (void, in red) and $z$ (filament,
    in blue) directions at distance $r = \SI{10}{Mpc/h}$ (solid
    lines) and $r = \SI{5}{Mpc/h}$ (dashed lines) and without saddle
    point (black) compared to the PDF at the saddle point.  The saddle
    point is defined using the values of
    table~\ref{tab:qijbar_nu_values}. In the filament, the PDF is
    boosted at the late formation end: haloes form later. The opposite
    trend is seen in the void.   }
  \label{fig:fup_Df}
\end{figure}

\allowdisplaybreaks

In the tide of a saddle of given height and curvature, equations~\eqref{eq:massfracup}, \eqref{eq:massfracalpha} and \eqref{eq:massfracD12}
remain formally unchanged, except for the replacement of $f_{\rm up}(\sigma)$, $ f_{\rm up}(\sigma,\alpha)$
and  $ f_{\rm up}(\sigma,D_\form)$ by their position dependent counterparts $f_{\rm up}(\sigma;\rr)$, $ f_{\rm up}(\sigma,\alpha;\rr)$
and  $ f_{\rm up}(\sigma,D_\form;\rr)$ conditioned to the presence of a saddle, given by \eqref{eq:condfup}, \eqref{eq:condjoint} and \eqref{eq:fupDS} respectively. Similarly, in equations
\eqref{eq:massfracalphacond} and \eqref{eq:massfracD12cond}
one should substitute the distribution $ f_{\rm up}(\alpha|\sigma)$ and  $ f_{\rm up}(D_\form|\sigma)$  by their conditional counterparts
$ f_{\rm up}(\alpha|\sigma;\rr)$ and  $ f_{\rm up}(D_\form|\sigma;\rr)$
of accretion rate and formation time at fixed halo mass, given by equations \eqref{eq:condacc} and \eqref{eq:condDsad}.

These functions depend on the mass $M$, accretion rate $\dot M$ and
formation time $z_\form$ of the halo through $\sigma(M)$,
$\alpha(M,\dot M)$ and $D_\form(z_\form)$, as before.  However,
conditioning on $\mathcal{S}$ introduces a further dependence on the
geometry of the environment (the height $\nus$ of the saddle and its
anisotropic shear $\bar\shear_{ij}$) and on the position $\rr$ of the
halo with respect to the saddle point. This dependence arises because
the saddle point condition modifies the mean and variance of the
stochastic process $(\delta,\delta')$ -- the height and slope of the
excursion set trajectories -- in a position-dependent way, making it
more or less likely to form haloes of given mass and assembly history
within the environment set by $\cal S$. The mean becomes anisotropic
through ${\cal Q}={\hat r_i\bar\shear_{ij}\hat r_j}$, and both mean and
variance acquire radial dependence through the correlation functions
$\xi_{\alpha\beta}$ and $\xi'_{\alpha\beta}$, defined in
equation~\eqref{eq:xi_xiprime}, which depend on $r,\Rs$ and $R$ (the variance remains isotropic because the variance of $\bar q_{ij}$ is still isotropic, see e.g. equation~(\ref{eq:varnucS}) and Appendix~\ref{sec:covariance}).

The relevant conditional distributions are displayed in
Figs.~\ref{fig:fup_sigma}, \ref{fig:fup_alpha} and
\ref{fig:fup_Df}. The plots show that haloes in the outflowing
direction (in which the filament will form) tend to be more massive,
with larger accretion rates and forming later than haloes at the same
distance from the saddle point, but located in the infalling direction
(which will become a void). This trend strengthens as the distance
from the centre increases. The saddle point is thus a minimum of the
expected mass and accretion rate of haloes, and a maximum of formation
redshift, as one moves along the filament. The opposite is true as one
moves perpendicularly to it. This behaviour is consistent with the
expectation that filamentary haloes have on average lower mass and
accretion rate, and tend to form earlier, than haloes in peaks.

\begin{figure*}
  \centering
  \includegraphics[width=\textwidth]{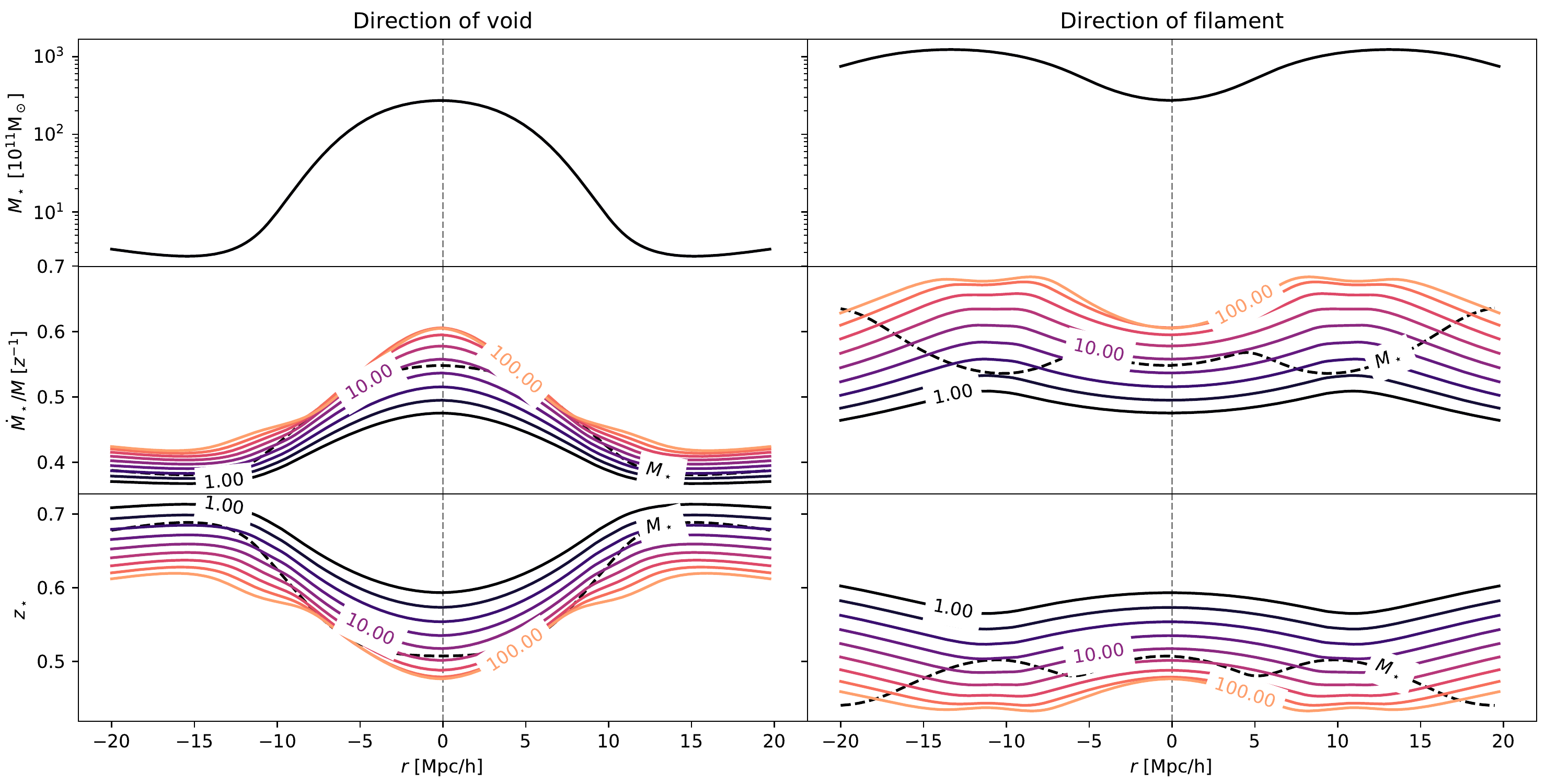}
  \caption{{\em Top:} Plot of the typical mass $M_\star$, {\em
      middle:} the typical specific accretion rates $\dot{M}/M$ and
    {\em bottom:} the formation redshifts $z_\star$ for different
    masses as a function of the distance to the saddle point {\em
      left:} in the direction of the void and {\em right:} in the
    direction of the filament. The colour of each line encodes the
    smoothing scale (hence the mass), from dark to light
    $M=\SI{10^{11}}{\msun/h}$ ($R=\SI{0.8}{Mpc/h}$) to
    $M=\SI{10^{13}}{\msun/h}$ ($R=\SI{3.7}{Mpc/h}$) logarithmically
    spaced; the dashed line is evaluated at $M=M_\star$. Labels are
    given in unit of \SI{10^{11}}{\msun/h}. The saddle point has been
    defined using the values given in
    table~\ref{tab:qijbar_nu_values}. More massive haloes accrete more
    and form later than less massive ones. At the typical mass, the
    space variation of the specific accretion rate and the formation
    redshift is smaller in the direction of the filament than in the
    direction of the void.}
  \label{fig:M_dotM_z}
\end{figure*}
To better quantify these trends let us define the tidally modified characteristic quantities
\begin{align}
  M_{\star}(\rr) &= M(\sigma_{\star}(\rr)) \,,
  \label{eq:M*r} \\
  {\dot M}_{\star}(\rr,M)
  &=  - \frac{\dd \log D}{\dd z} \frac{\dd M}{\dd \log \sigma}
  \alpha_{\star}(\rr,\sigma)\,, 
  \label{eq:mdot*r}\\
  z_{\star}(\rr,M) &= z(D_{\star}(\rr))
  \simeq 1/D_{\star}(\rr,\sigma)-1 \,, \label{eq:z*r}
\end{align}
giving the typical mass and the accretion rate and formation time at
given mass as a function of the position with respect to the centre of
the saddle.

The last approximation holds for haloes that assemble half of their mass before $z\sim2$, since at early times $D\simeq(1+z)^{-1}$.
These typical quantities are known functions of the position-dependent typical values of the excursion set parameters $\sigma_{\star}(\rr)$, $\alpha_{\star}(\rr,\sigma)$
and $D_{\star}(\rr,\sigma)$  given by equations~\eqref{eq:Mstarcond}, \eqref{eq:astarsad}
and \eqref{eq:defD*sad}  respectively.
They generalize the corresponding characteristic quantities obtained without conditioning on the the saddle, given by $\sigma_\star=\dc$, and by the functions $\alpha_{\star}(\sigma)$
and $D_{\star}(\sigma)$ defined in equations \eqref{eq:alphastar} and \eqref{eq:defD*}.

Taylor expanding equation~\eqref{eq:Mstarcond} in the anisotropy gives
the first-order angular variation of $M_\star$ at fixed distance
$r$ from the saddle
\begin{align} \label{eq:defDeltaM}
  \Delta M_\star( \mathbf{r}) &= -
  \frac{15}{2}\frac{\dc\,\xi_{20}(r)}{|(\d\sigma/\d M)_{M_\star}|}
{\cal Q}(\hat\rr)\,, 
\end{align}
where $\xi_{20}(r)$
is the radial part of the shear-height correlation function at finite separation.
Since $\xi_{20}$ is positive, this variation is largest when $\rr$ is parallel to the eigenvector with the
smallest eigenvalue. That is, in the direction of positive outflow
(with negative ${\cal Q}=\hat r_i\bar q_{ij}\hat r_j$),
along which a filament will form. Thus, in filaments haloes tend to be more massive, and haloes of large mass are more likely. The full dependence of the characteristic
mass $M_\star$ as a function of the position with respect to the saddle
point of the potential is shown in Fig.~\ref{fig:M_dotM_z}.

Similarly, like equations \eqref{eq:Dalphastar} and \eqref{eq:DDstar} for $\alpha_\star$ and $D_\star$, the first-order angular variations of $\dot M_\star$ and $z_\star$ are
\begin{align}
  \Delta \dot M_\star( \mathbf{r},M)
  &= - \frac{\dd \log D}{\dd z} \frac{\dd M}{\dd \log \sigma}
  \frac{\alpha_\star^2|_{\mathbf{\bar q}=0}}{\nuc} \notag \\
  &\times\frac{15}{2}
  \bigg[\xi_{20}'-\frac{\sigma-\xi'_I\xi_I}{\sigma^2-\xi^2}\xi_{20}\bigg]
 {\cal Q}(\hat\rr)\,, \label{eq:defDeltaDotM} \\
  \Delta z_\star( \mathbf{r},M)
  &= \bigg|\frac{\dd z}{\dd D}\bigg| \frac{D_{\star}^2|_{\mathbf{\bar q}=0}}{\dc}
  \frac{15}{2}\bigg|\frac{\dd\sigma}{\dd M}\bigg|\frac{M}{2}
  \xi_{20}(r){\cal Q}(\hat\rr)\,. \label{eq:defDeltaD}
\end{align}
These results confirm that in the direction of the filament, haloes have on average larger mass accretion rates and smaller formation redshifts than haloes of the same mass that form at the same distance from the saddle point, but in the direction perpendicular to it.
The space variation becomes larger with growing halo mass and fixed $\Rs$, as shown in Fig.~\ref{fig:M_dotM_z}, because the correlations become stronger as the difference between the two scales gets smaller. Conversely, for smaller masses haloes have on average smaller accretion rates (like in the unconditional case, see Fig.~\ref{fig:pdfalpha}) and later formation times, but also less prominent space variations.

Note that two estimators of delayed mass assembly, $\Delta \dot M_\star$ and $\Delta z_\star$ do not rely on the same property of the excursion set trajectory and do not
lead to the same physical interpretation.
In particular, when extending the implication of delayed mass assembly to galaxies and their
induced feedback, one should distinguish between the instantaneous accretion rate,
and the integrated half mass time as they trace different
components of the excursion hence different epochs.

\subsection{Expected differences between the iso-contours}
\label{sec:grad}

In order to investigate whether the assembly bias generated by the cosmic web and described in this work is purely an effect due to the local density (itself driven by the presence of the filament),  this section studies the difference between the isocontours of the density field and any other statistics (mass accretion rate for instance).  These contours will be shown to cross each others, which proves that the anisotropic effect of the nearby filament also plays a role.

The normals to the level surfaces of $ {\dot M_\star(\rr,M)}$, $ M_\star(\rr)$, $ z_{\star}(\rr,M) $  and $  \mean{\rho}(\rr)\equiv\bar\rho(1+\condmean{\delta}{\S})$ scale like the gradients of these functions.
First note that any mixed product (or determinant) such as  $\nabla {\dot M_\star}\cdot (\nabla M_\star \times \nabla \langle \rho\rangle)$ will be null by symmetry; i.e. all gradients are co-planar. This happens because the present theory focuses on scalar quantities (mediated, in our case, by the excursion set density and slope). In this context, all fields vary as a function of only two variables, $r$ and ${\cal Q}= \hat r_i\bar q_{ij} \hat r_j $, hence the gradients of the fields will all lie in the plane of the gradients of $r$ and $\cal Q$\footnote{In order to break this degeneracy one would need to look at the statistics of higher spin quantities. For instance the angular momentum of the halo would depend on the spin-one coupling $\varepsilon_{ijk} \hat r_j\bar q_{kl} \hat r_l  $,
with $\varepsilon_{ijk}$ the totally antisymmetric tensor
\citep[see, e.g.][]{codis2015}, or to consider a barrier that depends on the local shear at $\rr$ filtered on scale $R$ \citep[e.g.][]{Castorina:2016vg}, like e.g. $\dc + \beta \sigma\bar q_{ij}(\rr,R)\bar q_{ij}(\rr,R)$ with some constant $\beta$.}. Ultimately, if one focuses on a given spherically symmetric peak, then $\cal Q$ vanishes, so all gradients are proportional to each other and radial.
Let us now quantify the mis-alignments between two normals within that plane.
 In spherical coordinates, the Nabla operator reads
\begin{equation}
\nabla=\left(\frac{\partial}{\partial r},\frac{1}{r}\frac{\partial}{\partial \theta} ,
\frac{1}{r \sin \theta}\frac{\partial}{\partial \phi} \right)\equiv\left(\frac{\partial}{\partial r},\frac{1}{r}
\tilde \nabla\right)\,,
\end{equation}
so that for instance
 \begin{align}
\nabla \dot M_\star\!\propto\! \left(\frac{ \partial  \dot M_\star }{\partial r} ,\frac{1}{r} \frac{ \partial \dot M_\star }{ \partial{\cal Q}}  {\tilde \nabla} {\cal Q} \right)\nonumber\,,
\end{align}
where equation~\eqref{eq:rqrhatinframe} implies that
\begin{equation}
  {\tilde \nabla} {\cal Q}
  = \begin{pmatrix}
    &\!\!\!\!\! \sin 2\theta  \left({\bar q_3} \cos ^2 \phi +{\bar q_2} \sin^2 \phi -{\bar q_1}\right)\\
    &\!\!\!\!\!  \sin \theta({\bar q_2}-{\bar q_3}) \sin 2 \phi
  \end{pmatrix}\,.
\end{equation}
Hence, for instance the cross product  $\nabla M_\star \times \nabla  \dot M_\star$
reads
\begin{equation}
  \left(\frac{\partial  \dot M_\star}{\partial r} \frac{\partial
      M_\star }{\partial {\cal Q}}-\frac{\partial  \dot M_\star}{\partial {\cal Q}} \frac{\partial M_\star }{\partial r}\right) {\tilde \nabla}{\cal Q} \,. \label{eq:crossprod}
\end{equation}
It follows that the two normals are not aligned since the prefactor in
equation~\eqref{eq:crossprod} does not vanish: the fields are explicit
distinct and independent functions of both $r$ and $\cal Q$.  The
origin of the misalignment lies in the relative amplitude of the
radial and `polar' derivatives (w.r.t. $\cal Q$) of the field.
For instance, even at linear order in the anisotropy, since $\Delta \dot M_\star$ in
equation~\eqref{eq:defDeltaDotM} has a radial dependence in
$\xi'_{20}$ as a prefactor to $\cal Q$ whereas $M_\star$ has only
$\xi_{20}$ as a prefactor in equation~\eqref{eq:defDeltaM}, the
bracket in equation~\eqref{eq:crossprod} will involve the Wronskian
$\xi'_{20}\partial \xi_{20}/\partial r-\xi_{20}\partial
\xi'_{20}/\partial r$ which is non zero because $\xi_{20}$ and its
derivative w.r.t. filtering are linearly independent.  This
misalignment does not hold for $ M_\star$ and $\mean{\rho}$ at linear
order since $\Delta M_\star$ (equation~\ref{eq:defDeltaM}) and
$\mean{\rho}$ (equation~\ref{eq:effmean}) are proportional in this
limit.  Yet it does arises when accounting for the fact that the
contribution to the conditional variance in $M_\star$ also depends
additively on $\xi^2(r)$  in
equation~\eqref{eq:Mstarcond} (with  $\xi^2(r)$ given by equation~\eqref{eq:defxi2} as a
function of the finite separation correlation functions
$\xi_{\alpha\beta}$ computed in equation~\eqref{eq:xi_xiprime} for a
given underlying power spectrum).  Indeed, one should keep in mind that
the saddle condition not only shifts the mean of the observables but
also changes their variances.  Since the critical `star' observables
($M_\star$, $z_\star$ etc.) involve rarity, hence ratio of the shifted means to their
variances (e.g. entering equation~\ref{eq:defYalpha}), both impact the
corresponding normals. It is therefore a clear specific prediction of
conditional excursion set theory relying on upcrossing that the
level sets of density, mass density and accretion rates are distinct.

Physically, the distinct contours could correspond to an excess of
bluer or reddened galactic hosts at fixed mass along preferred directions
depending on how feedback translate inflow into colour as a function of redshift.
Indeed AGN feedback when triggered during merger events regulates cold gas inflow
which in turn impacts star formation: when it is active, at intermediate and low redshift, it may reverse the naive expectation (see Appendix~\ref{sec:speculations}).
This would be
in agreement with the recent excess transverse gradients (at fixed
mass and density) measured both in hydrodynamical simulation and those
observed in spectroscopic \citep[e.g. VIPERS or GAMA, ][Kraljic et
al. submitted]{Malavasi2016b} and photometric
\citep[e.g. COSMOS, ][]{Laigle+17} surveys: bluer central galaxies at high redshifts when
AGN feedback is not efficient and redder central galaxies at lower redshift.

\begin{figure}
  \centering
  \includegraphics[width=\columnwidth]{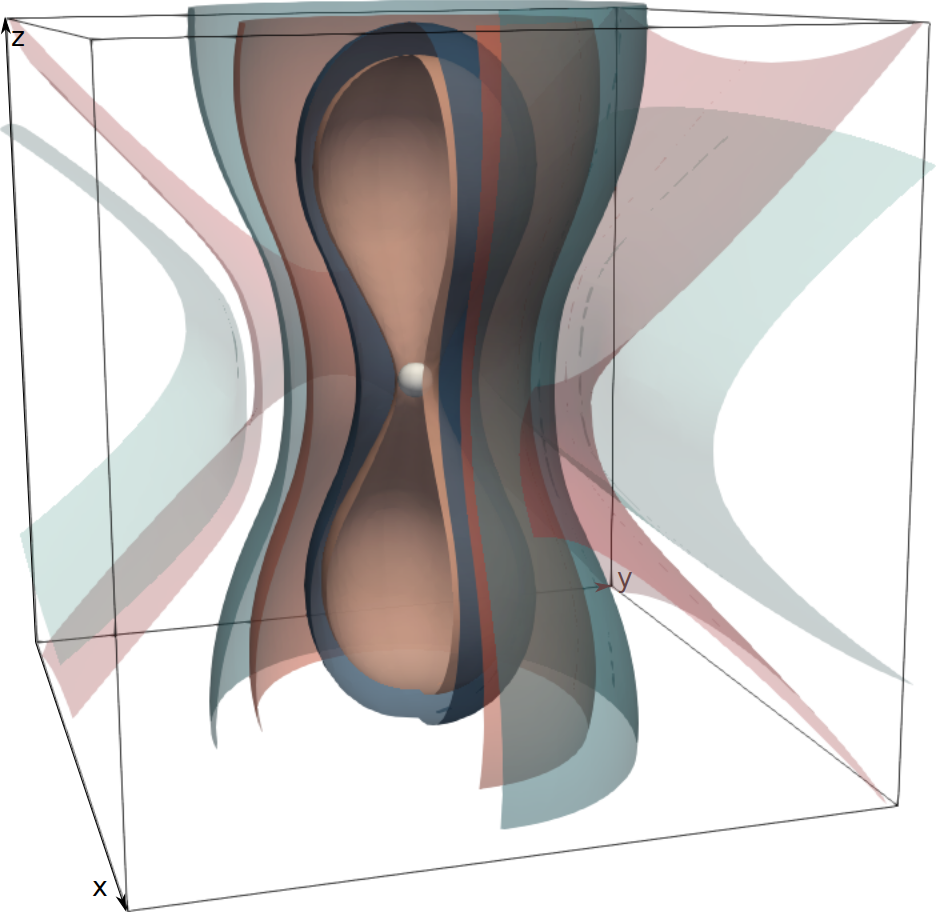}
  \includegraphics[width=\columnwidth]{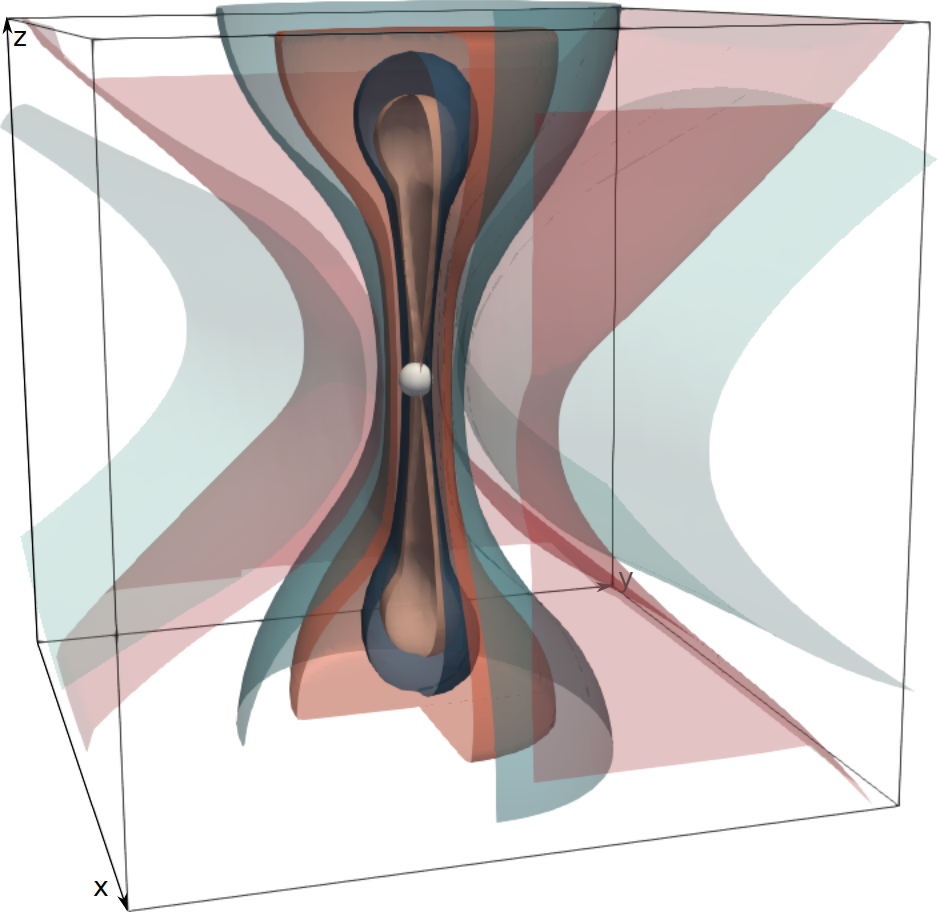}
    \caption{Level surfaces of the typical density $\rho_\star$ (light
    to dark blue) and of the accretion rate $\alpha_\star$ (light to
    dark red) with no Zel'dovich boost ({\em upper panel} and with a
    Zel'dovich boost ({\em lower panel}). The saddle is represented by
    a ball. Once boosted, the structure of the filament in the $z$
    direction is clearly seen and the isocontours align one with each
    other.   }
  \label{fig:3dview}
\end{figure}

These predictions hold in the initial conditions. However, one should
take into account a Zel'dovich boost to get the observable contours of
the quantities derived in the paper. Regions that will collapse into a
filament are expected to have a convergent Zel'dovich flow in the
plane perpendicular to the filament and a diverging flow in the filament's
direction. As such, the contours of the different quantities will be
advected along with the flow and will become more and more parallel
along the filament. This effect is clearly seen in
Fig.~\ref{fig:3dview} which shows the contours of
both the typical density and the accretion rate\footnote{Interactive versions can be found
  online \href{https://cphyc.github.io/research/assembly/with_boost.html}{with boost} and
  \href{https://cphyc.github.io/research/assembly/no_boost.html}{without boost}.}
(bottom panel) after the Zel'dovich boost  (having chosen the amplitude of the boost corresponding to the
formation of the filamentary structure).  The contours are
compressed towards the filament and become more and more parallel.
Hence the stronger the non-linearity the more parallel the contours.
This is consistent with the findings of Kraljic et
al. submitted.
%
\section{Assembly Bias}
\label{sec:bias}

The bias of dark matter haloes \citep[see][for a recent review]{biasreview2016} encodes the response of the mass function to variations of the matter density field. In particular, the Lagrangian bias function $b_1$ describes the linear response to variations of the initial matter density field.
For Gaussian initial conditions, the correlation of the halo overdensity with an infinite wavelength matter overdensity $\delta_0$ is then \citep{FryGazta1993},
\begin{equation}
  \mean{\delta_0\delta_h(\rr,M)} = \int\dd\rr_1
  \mean{\delta_0\delta_m(\rr_1)}b_1(\rr,\rr_1,M)\,,
\label{eq:biasdef}
\end{equation}
where formally $b_1(\rr,\rr_1,M)\equiv\mean{\partial[\delta_h(\rr,M)]/\partial[\delta_m(\rr_1)]}$ is the expectation value of the functional derivative of the local halo overdensity with respect to the (unsmoothed) matter density field $\delta_m(\rr)$ \citep{gammaexp2008}.
In the standard setup, because of translational invariance (which does not hold here), it is only a function of the separation $|\rr-\rr_1|$.

The dependence of the halo field on the matter density field can be parametrized with a potentially infinite number of variables constructed in terms of the matter density field, evaluated at the same point.
With a simple chain rule applied to the functional derivative, equation~\eqref{eq:biasdef} can be written as the sum of the cross-correlation of $\delta_0$ with each variable, times the expectation value of the ordinary partial derivative of the halo point process with respect to the same variable. The latter are the so-called bias coefficients, and are mathematically equivalent to ordinary partial derivatives of the mass function with respect to the expectation value of each variable.

The most important of these variables is usually assumed to be the density $\delta(\rr,R)$ filtered on the mass scale of the haloes, which mediates the response to the variation of an infinite wavelength mode of the density field, the so-called large-scale bias.
Because the smoothed density correlates with the $k=0$ mode of the density field, this returns the peak-background split bias.
Its bias coefficient is also equal to (minus) the derivative w.r.t. $\delta_c$.

Excursion sets make the ansatz that the next variable that matters is the slope $\delta'(\rr,R)$ \citep{MPS2012}.
In the simplest excursion set models with correlated steps and a constant density threshold, trajectories crossing $\dc$ with steeper slopes have a lower mean density on larger scales \citep{Zentner2007}. They are thus unavoidably associated to less strongly clustered haloes.
This prediction is in agreement with N-body simulations for large-mass haloes, but the trend is known to invert for smaller masses \citep{shethettormen2004,gaoetal2005,wechsleretal2006,dalaletal2008}. Although more sophisticated models are certainly needed in order to account for the dynamics of gravitational collapse, we will see that the presence of a saddle point contributes to explaining this inversion.

None of the concepts outlined above changes in the presence of a saddle point: the bias coefficients are derivatives of $\dd n/\dd M$, that is of the upcrossing probability through equation~\eqref{eq:massfracup}.
Because we are interested in the bias of the joint saddle-halo system, we must differentiate the joint probability $\fup(\sigma;\rr)p(\S)$, rather than just $\fup(\sigma;\rr)$, and divide by the same afterwards. Of course, the result picks up a dependence on the position within the frame of the saddle.
The relevant uncorrelated variables are
$\delta-\condmean{\delta}{\S}$, $\delta'-\condmean{\delta'}{\nu,\S}$,
$\nus$, $\hat r_ig_i=0$ and $\mathcal{Q}={\hat{r}_i\bar
  q_{ij}\hat{r}_j}$.
Differentiating equation~\eqref{eq:condfup}, the bias coefficients of the halo are
\begin{align}
  b_{10}(M;\rr)
  &\equiv
  \frac{\pd\log\left[ f_{\mathrm{up}}(\sigma;\rr)\right]}{\pd\condmean{\delta}{\S}}
  = \frac{\delta_c-\xi_I\S_I}{\sigma^2-\xi^2} \,,\\
  b_{01}(M;\rr)
  &\equiv
  \frac{\pd\log\left[ f_{\mathrm{up}}(\sigma;\rr)\right]}{\pd\condmean{\delta'}{\nuc,\S}}
  = \frac{1+\mathrm{erf}(X_\S(\rr)/\sqrt{2})}{2\mu_\S(\rr) F(X_\S(\rr))}\,,
\end{align}
which without saddle reduce to (a linear combination of) those defined by \cite{MPS2012}. The coefficients of the saddle are
\begin{align}
  b_{100}^{(\S)}
  &\equiv -\frac{\pd}{\pd\delta_s}\log p_\G(\S)
  = \frac{\nus}{\sigmas}\,, \\
  b_{010}^{(\S)}
  &\equiv -\frac{\pd}{\pd(\hat r_ig_i)}\log p_\G(\S)\bigg|_{g_i=0}
    = 0 \,,\\
  b_{001}^{(\S)}
  &\equiv -\frac{\pd}{\pd\mathcal{Q}}\log p_\G(\S)
  = \frac{15}{2} \frac{3\mathcal{Q}}{2}\,.
\end{align}

A constant $\delta_0$ does not correlate with $\bar q_{ij}$, since there is no zero mode of the anisotropy. One can see this explicitly by noting that $\xi_{20}(R_0,\Rs,r)\to 0$ as $R_0\to\infty$. The only coefficients that survive in the cross-correlation with $\delta_0$ are thus $b_{10}$, $b_{01}$ and $b_{100}^{(\S)}$, so that equation~\eqref{eq:biasdef} becomes
\begin{align}
  \mean{\delta_0\delta_h(\rr,M)} &=
  b_{100}^{(\S)}\mean{\delta_0\delta_s} +  b_{10} \Cov{\delta_0,\delta}{\S} \notag \\
  &\quad + b_{01} \Cov{\delta_0,\delta'}{\nuc,\S}.
\end{align}
Similarly, in this limit $\delta_0$ does not correlate with $g_i$ either, while $\mean{\delta_0\delta}$ becomes independent of $R$.  Thus $\mean{\delta_0\delta}\simeq\mean{\delta_0\delta_s}$ and $\mean{\delta_0\delta'}\simeq0$. Hence,
\begin{multline}
  \frac{\mean{\delta_0\delta_h}}{\mean{\delta_0\nus}} \simeq
  \nus + \frac{\delta_c-\xi_I\S_I}{\sigma^2-\xi^2}
  (\sigma_s-\xi_{00}) \\
  - b_{01} \bigg[\xi_{00}' +
  \frac{\sigma-\xi_I'\xi_I}{\sigma^2-\xi^2}(\sigma_s-\xi_{00})\bigg].
\label{eq:halomatter}
\end{multline}
Setting $\nus=\xi_{\alpha\beta}=\xi'_{\alpha\beta}=0$ recovers    \cite{MPS2012}'s results.

\begin{figure}
  \centering
  \includegraphics[width=\columnwidth]{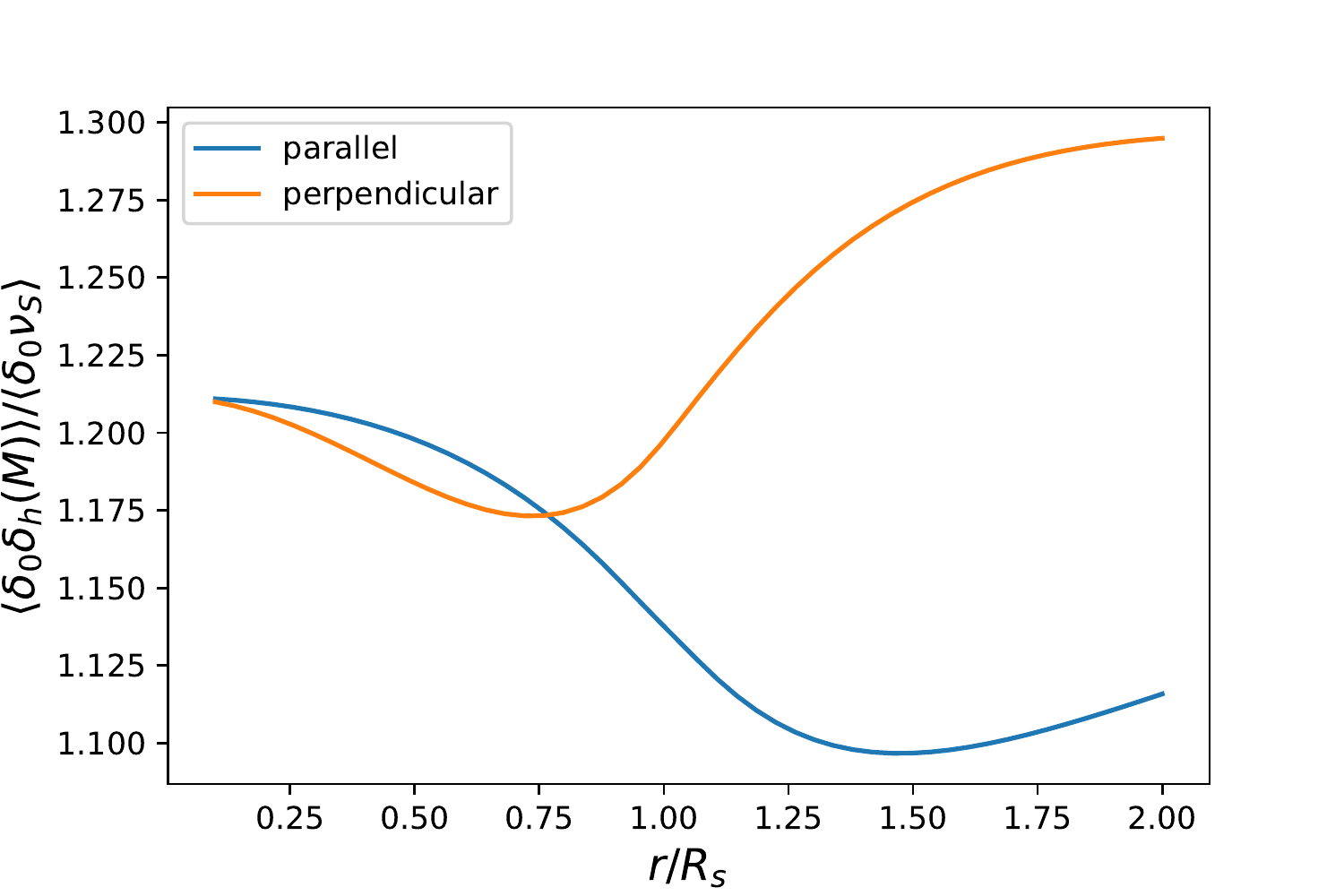}
  \includegraphics[width=\columnwidth]{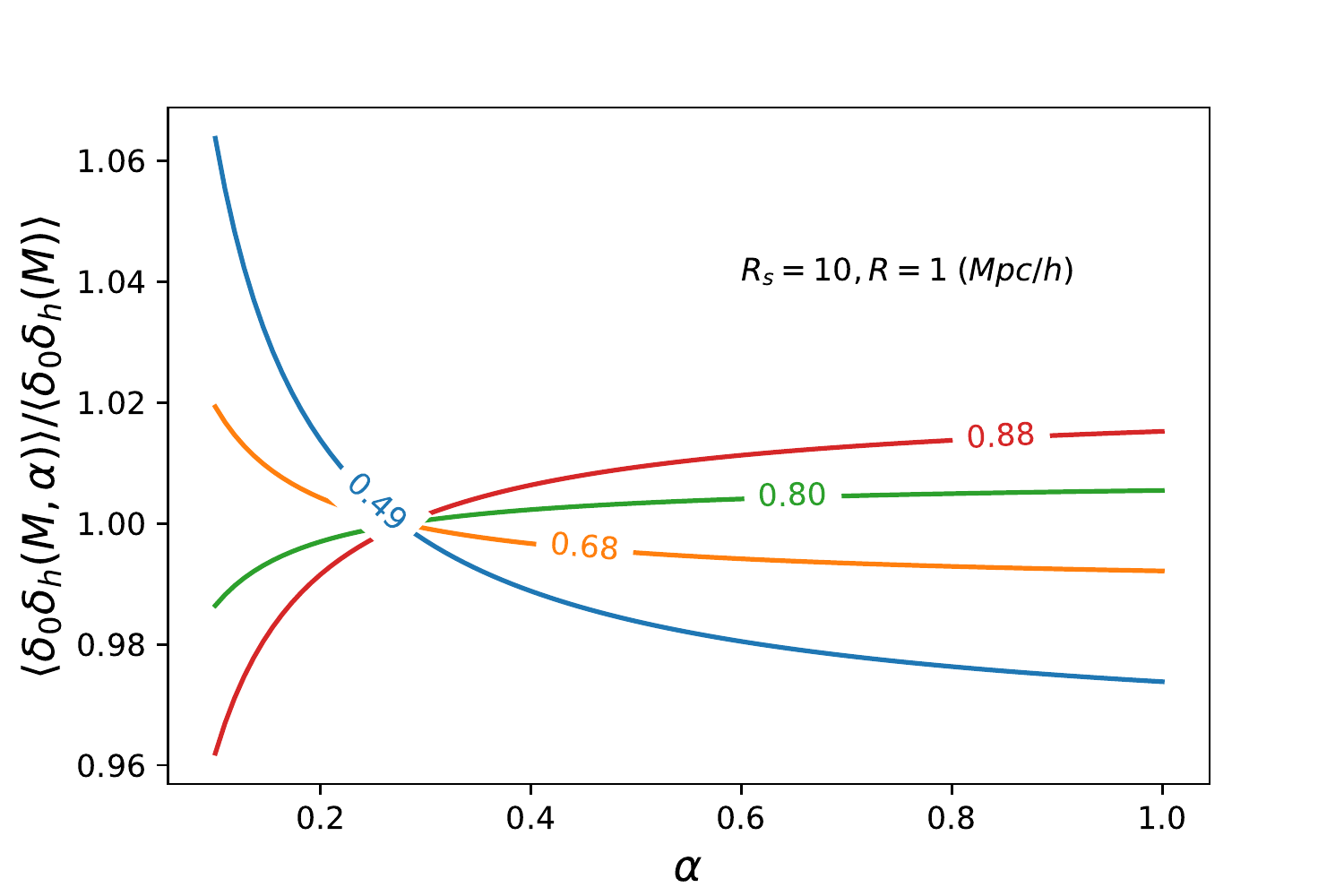}
  \caption{\emph{Upper panel}: Large-scale Lagrangian bias as a function of the distance from the saddle point, along the filament and perpendicularly to it, for haloes of mass $M = \SI{2.0 \times 10^{11}}{\msun/h}$ ($R=\SI{1}{Mpc/h}$). Haloes in the perpendicular direction are less biased at small separation, but the trend inverts at $r/\Rs\simeq 0.75$. \emph{Lower panel}: Bias as a function of accretion rate, for different values of the separation $r/\Rs$ in the direction of the filament. For haloes closer to the centre, bias decreases with accretion rate, but the trend inverts at $r/\Rs\simeq 0.75$. In the perpendicular direction the effect is 30\% smaller, but the relative amplitudes and the inversion point do not change appreciably. As discussed in the main text, both inversions depend on the fact that $\delta-\mean{\delta|\S}$ and $\delta_0$ correlate at large distance from the saddle, but they anti-correlate at small separation.}
   \label{fig:bias}
\end{figure}

The anisotropic effect of the saddle is easier to understand looking at the sign of the terms in the round and square brackets, corresponding to $\Cov{\delta_0,\delta}{\S}$ and $-\Cov{\delta_0,\delta'}{\nuc,\S}$ respectively.
One can check that for $R=\SI{1}{Mpc/h}$ and $\Rs=\SI{10}{Mpc/h}$ both terms are negative near $r=0$, but become positive at $r\simeq0.75 \Rs$.
This separation marks an inversion of the trend of the bias with $\nucS$, the parameter measuring how rare haloes are given the saddle environment. Far from the saddle, haloes with higher $\nucS$ are \emph{more biased}, which recovers the standard behaviour since $\nucS\to\nuc$ as $r\to\infty$. However, as $r/\Rs\lesssim 0.75$ the trend inverts and haloes with higher $\nucS$ become \emph{less biased}. Therefore, one expects that at fixed mass and distance from the saddle point haloes in the direction of the filament are less biased far from the saddle, but become more biased near the saddle point.
The upper panel of Fig. \ref{fig:bias}, displaying the exact result of equation~\eqref{eq:halomatter}, confirms these trends and their inversion at $r\simeq0.75 \Rs$. The height of the curves at $r=0$ depends on the chosen value for $\nus$, but the inversion at $r\simeq0.75 \Rs$ and the behaviour at large $r$ do not. Fig. \ref{fig:bias} also shows that a saddle point of the potential need not be a saddle point of the bias (in the present case, it is in fact a maximum).

The inversion can be interpreted in terms of excursion sets. Near the saddle, fixing $\nus$ at $\rr=0$ puts a constraint on the trajectories at $\rr$ that becomes more and more stringent as the separation gets small. At $r=0$, the value of the trajectory at $\Rs$ is completely fixed. Therefore, trajectories constrained to have the same height at both $\Rs$ and $R$, but lower $\condmean{\delta}{\S}$ at $R$, will tend to drift towards lower values between $\Rs$ and $R$, and thus towards higher values for $R_0\gg \Rs$. This effect vanishes far enough from the saddle point, since the constraint on the density at $\Rs$ becomes looser as the conditional variance grows. Hence, trajectories with lower $\condmean{\delta}{\S}$ at $R$ will remain lower all the way to $R_0$.
Note however that interpreting these trends in terms of clustering is not straightforward, because the variations happen on a scale $\Rs\ll R_0$ (they are thus an explicit source of scale dependent bias). The most appropriate way to understand the variations of clustering strength is looking at the position dependence of $\dd n/\dd M$, which is predicted explicitly through $\fup(\sigma;\rr)$ in equation~\eqref{eq:condfup}.

When one bins haloes also by mass and accretion rate, the bias is given by the response of the mass function at fixed accretion rate. That is, to get the bias coefficients one should now differentiate the joint probability $\fup(\sigma,\alpha;\rr)p_\G(\S)$ with respect to mean values of the different variables, with $\fup(\sigma,\alpha;\rr)$ given by equation~\eqref{eq:condjoint}. The only bias coefficient that changes is $b_{01}$, the derivative w.r.t. $\condmean{\delta'}{\nuc,\S}$, which becomes
\begin{equation}
  b_{01}(M,\dot M,\rr) \equiv
  \frac{\pd\log\left[ f_{\mathrm{up}}(\sigma,\alpha;\rr)\right]}{\pd\condmean{\delta'}{\nuc,\S}}
  = \frac{\nuc/\alpha-\mu_\S(\rr)}{\Var{\delta'}{\nuc,\S}}\,,
\end{equation}
with $\alpha$ defined by equation~\eqref{eq:defmdot}.
Inserting this expression in equation~\eqref{eq:halomatter}, returns the predicted large-scale bias at fixed accretion rate.
Notice that in this simple model the coefficient multiplying the $1/\alpha$ term is purely radial. The asymptotic behaviour of the bias at small accretion rates will then always be divergent and isotropic, with a sign depending on that of the square bracket in equation~\eqref{eq:halomatter}. If this term is positive, the bias decreases as $\alpha$ gets smaller, and vice versa. Clearly, the value of $\alpha$ for which the divergent behaviour becomes dominant depends on the size of all the other terms, and is therefore anisotropic.

\begin{figure}
  \centering
  \hskip -0.2cm \includegraphics[width=1.0\columnwidth]{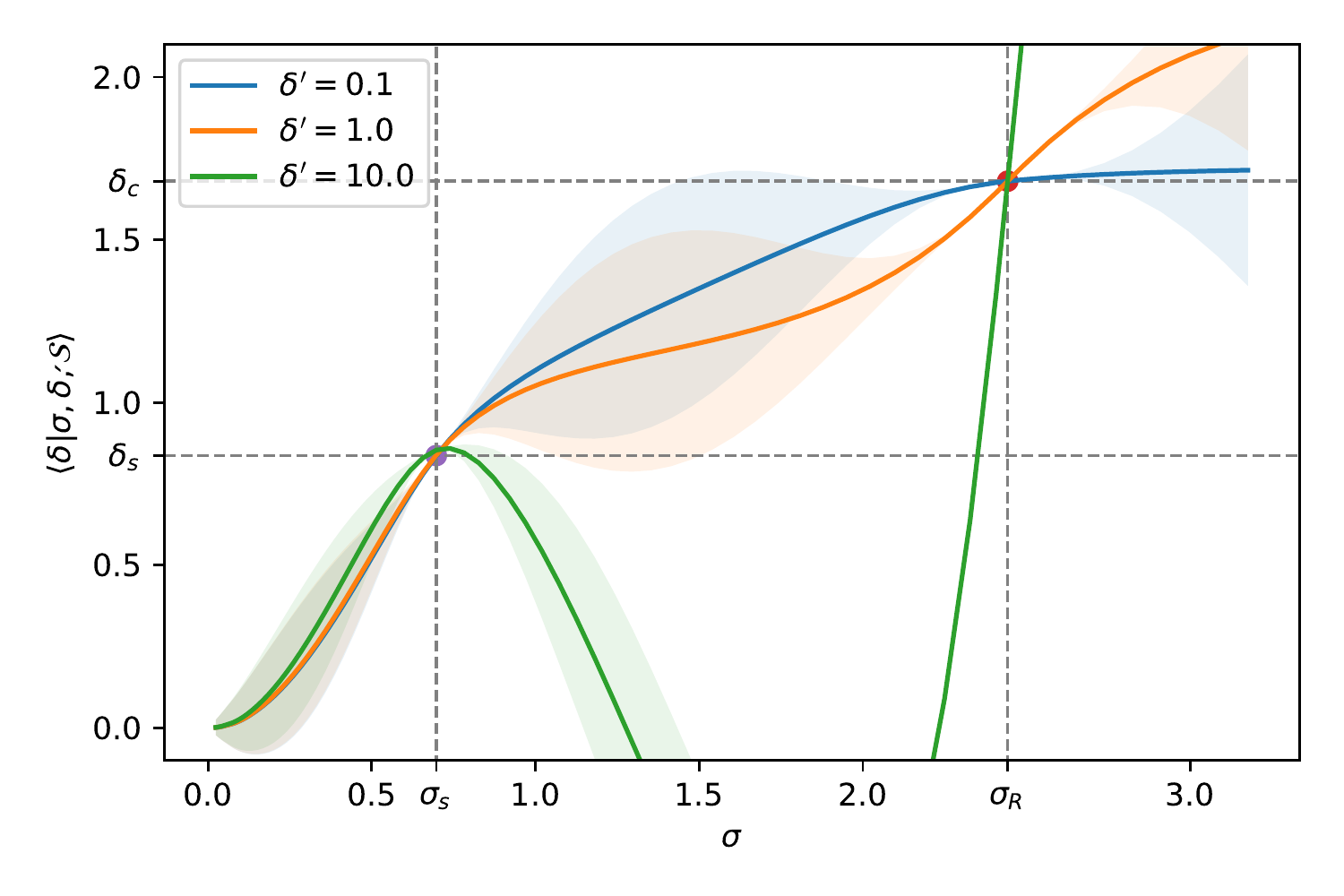}
  \caption{Plot of the mean of density given the saddle point, the
    upcrossing condition and the slope at $R$ for different
    slopes. The saddle point was defined using the values of
    table~\ref{tab:qijbar_nu_values}. The details of the calculation
    are provided in section~\ref{sec:validation}. For steep slopes
    (small accretion rate), the mean of the density overshoots at
    small $\sigma$, resulting in a larger bias. }
  \label{fig:constrained-slope}
\end{figure}

As one can see from Fig.~\ref{fig:bias}, the sign of the small-$\alpha$ divergence depends on the distance from the saddle point. It is negative for $r\gtrsim0.75 \Rs$, but it reverses closer to the centre.
This effect is again a consequence of the constraint on the excursion set trajectories at $\Rs$. Trajectories with steeper slopes at $R$ will sink to lower values between $\Rs$ and $R$, then turn upwards to pass through $\delta(\Rs)$, and reach higher values for $R_0\gg \Rs$. The haloes they are associated to are thus \emph{more biased}. This trend is represented in Fig.~\ref{fig:constrained-slope}.
This inversion effect is lost as the separation increases, and the constraint on the density at $\Rs$ becomes loose, and trajectories that reach $R$ with steeper slopes are likely to have low (or even negative) values at very large scales. These haloes are thus \emph{less biased}, or even anti-biased.

It follows that the bias of haloes far from structures grows with accretion rate (the usual behaviour expected from excursion sets), while the trend inverts for haloes near the centre of the filament. Because typical mass of haloes also depends on the position along the filament, with haloes towards the nodes being more massive, the different curves of Fig.~\ref{fig:bias} correlate with haloes of different mass.
This effect  explains why  low-mass haloes with small accretion rate (or early formation time, or high concentration) are more biased,  when measuring halo bias as a function of mass and accretion rate (or formation time or concentration, which strictly correlate with accretion rate), without knowledge of the position in the cosmic web. Conversely, the high-mass ones are less biased
\citep{shethettormen2004,gaoetal2005,wechsleretal2006,dalaletal2008,FaltenWhite2010,ParanjPadma2017}.
It is also intriguing to compare this result with the measurements by \cite{Lazeyras2017} (namely their fig. 7) which show the same trends (although their masses are not small enough to clearly see the inversion).

Note in closing that the conditional bias theory presented here does not capture changes in accretion rate and formation time presented in
Sections~\ref{sec:condacrrate} and~\ref{sec:condft}.

\section{Conclusion \& Discussion}
\label{sec:conclusion}

\subsection{Conclusion}

With the advent of modern surveys,   assembly bias has become the focus of renewed interest as a process which could explain some of the diversity of galactic morphology and clustering at fixed mass. It is also investigated as a mean to mitigate intrinsic alignments {in weak lensing survey such as Euclid or LSST}.
Both observations and simulations have hinted that the large-scale anisotropy of the cosmic web
could be responsible for stalling and quenching. This paper investigated this aspect in Lagrangian space within the framework of excursion set theory.
As a measure of infall, we
  computed quantities related to the slope of the contrast conditioned to the relative position of the collapsing
halo w.r.t. a critical point of the large-scale field.
We focused here on mass accretion rate and half mass redshift and found that
their expectation vary with the orientation and distance from saddle points, demonstrating
 that assembly bias is indeed {influenced} by the geometry of the tides imposed by the cosmic web.

More specifically, we derived the Press--Schechter typical mass,
typical accretion rate, and formation time of dark haloes in the vicinity of cosmic saddles by means of an extension of excursion set theory accounting for the effect of their large-scale tides.
Our principal findings are the following: we have computed the
(i)  \emph{Upcrossing PDF} for halo mass, accretion rate and formation time; they are given by equations~\eqref{eq:fup3}, \eqref{eq:fupalpha} and \eqref{eq:fupD}, and their constrained-by-saddles counterparts  equations~\eqref{eq:condfup}, \eqref{eq:condacc} and \eqref{eq:condDsad}. These PDFs allowed us to identify the
 (ii) \emph{typical halo mass}, and \emph{typical accretion rate and formation time at given mass} as functions of the position within the frame of the saddle via equations~\eqref{eq:defDeltaM}, \eqref{eq:defDeltaDotM} and~\eqref{eq:defDeltaD}. All quantities are expressed
 as a function of the geometry of the saddle for an arbitrary cosmology encoded in the underlying power spectrum via the correlations  $\xi_{\alpha\beta}$ and $\xi'_{\alpha\beta}$ given by equations~\eqref{eq:xi_xi} and~\eqref{eq:xi_xiprime}.
In turn this has allowed us to compute and explain the corresponding
(iii) \emph{distinct gradients} for the three typical quantities and for the local mean density (Section~\ref{sec:grad}). The misalignment of the gradients, defined as the normals to the their iso-surfaces, arises because
the saddle condition is anisotropic and because it does not only shift the local mean density and the mean density profile (the excursion set slope) but also their variances, affecting different observables in different way. Finally, we have presented
(iv) \emph{an extension of classical large-scale bias theory} to account for the saddle (Section~\ref{sec:bias}).

Our simple conditional excursion set model subject to filamentary tides makes intuitive predictions in agreement with the trends found in N-body simulations: haloes in filaments are less massive than haloes in nodes, and at equal mass they have earlier formation times and smaller accretion rates today. The same hierarchy exists for haloes in walls with respect to filaments. For the configuration we examined, the effect is stronger as one moves perpendicularly to the filament. The typical mass changes by a factor of 5 along the filament, and by two orders of magnitude perpendicularly.
The relative variation of accretion rates and formation times is of about 5-10\% along the filament, and of about 20-30\% in the perpendicular direction, for haloes of \SI{10^{11}}{\msun/h}.
Furthermore, our model predicts that at fixed halo mass the trend of the large-scale bias with accretion rate depends on the distance from the center of the filament. Far from the center the large-scale bias grows with accretion rate (which is the naive expectation from excursion sets) while near the center the trend inverts and haloes with smaller accretion rates become more biased. Since haloes near the center are also on average less massive, this effect should contribute to explaining why the trend of bias with accretion rate (or formation time) inverts at masses much smaller than the typical mass.

\begin{figure}
  \includegraphics[width=0.9\columnwidth]{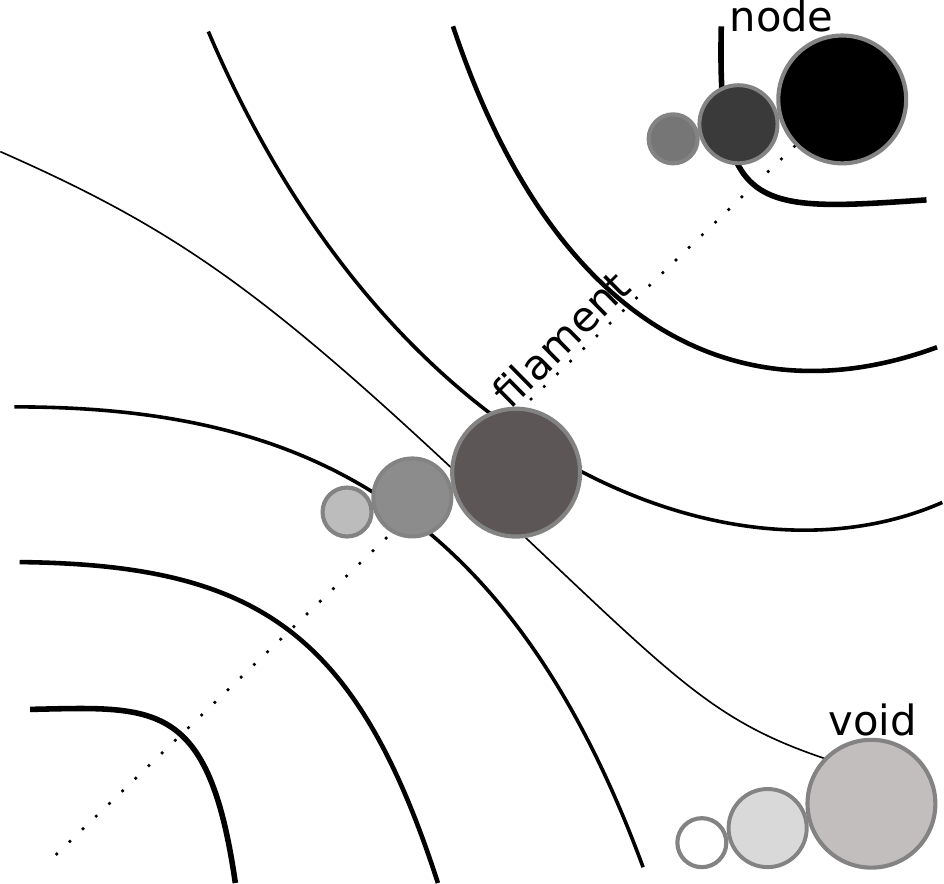}
  \caption{Scheme of the intensity of the accretion rate at different
    locations near a filament-type saddle for different final halo masses. The darkness of the colour
    encodes the intensity of the accretion rate (darker is more
    accretion). At fixed mass, the accretion rate increases from voids
    to saddle points and from saddle points to nodes (along dotted line which marks the filament's direction). At a given
    location, the accretion rate increases with mass.
    }
      \label{fig:conclusive-scheme}
\end{figure}

These findings conflict with the simplistic assumption that the properties of galaxies of a given mass are uniquely determined by the density of the environment.
The presence of distinct space gradients for the different typical quantities is also part and parcel of the conditional excursion set theory, simply because the statistics of the excursion set proxies for halo mass, accretion rate and formation time (the first-crossing scale and slope, and the height at the scale corresponding to $M/2$) are different functions of the position with respect to the saddle point. They have thus different level surfaces.
At the technical level, the contours depend on the presence of the conditional variance of $\delta(\rr)$, besides its conditional mean, and of the correlation functions of $\delta'(\rr)$. 
At finite separation, the traceless shear of the large-scale environment modifies in an anisotropic way the statistics of the local mean density $\delta(\rr)$ (and of its derivative $\delta'(\rr)$ w.r.t. scale).
The variations are modulated by $\mathcal{Q}={\hat{r}_i\bar q_{ij}\hat{r}_j}$, i.e. the relative orientation
of the separation vector in the frame set by the tidal tensor of the saddle. This angular modulation enters different quantities with different radial weights, which results in different angular variations of the local statistics of density, mass and accretion rate/formation time.
It provides a supplementary vector space, $ \tilde \nabla \cal Q$, beyond the radial direction over which to project the gradients, whose statistical weight depend on each specific observable.  These quantities have thus different iso-surfaces from each other and from the local mean density, a genuine signature of the traceless part of the tidal tensor. The qualitative differences in terms of mass accretion rate and galactic colour is sketched in Fig.~\ref{fig:conclusive-scheme}.

\subsection{Discussion and perspectives}
\label{sec:perspectives}

 In contrast to the findings of \cite{Alonso:2015eu,Tramonte2017,Braun-Bates2017} we focused our attention on variations of  {\sl mass accretion rates} w.r.t. the cosmic web rather than mass functions.
 We have found that, even in a very simple model like excursion sets, halo properties are indeed affected by the anisotropic tides of the environment (involving the traceless part of the tidal tensor), and not just by its density (involving the trace of the tidal tensor), and that this effect cannot be explained by a simple rescaling of the local mean density.

Although the excursion set approach is rather crude, and additional constraints (e.g. peaks) would be needed to pinpoint the exact location of halo formation in the initial conditions, we argued that the effect we are investigating does not strongly depend on the presence of these additional constraints. The underlying reason is that the extra constraints usually involve vector or tensor quantities evaluated {\sl at the same location $\rr$ as the excursion set sphere}, which do not directly correlate with the scalars considered here (they only do so through their correlation with the saddle point). They may add polynomial corrections to the conditional distributions, but will not strongly affect the exponential cutoffs on which we built our analysis. Our formalism may thus not predict exactly whether a halo will form (hence, the mass function), but it can soundly describe the secondary properties and the assembly bias of haloes that actually form. A more careful treatment would change our results only at the quantitative level. For this reason, we chose to prefer the simplicity of the simple excursion set approach. Furthermore, in order to describe the cosmic web we focused on saddle points of the initial gravitational potential, rather than of the density field, as these are more suitable to trace the {\sl dynamical} impact of filamentary structures in connection to the spherical collapse model.

The present Lagrangian formalism only aims at describing the behaviour of the central galaxy: it cannot claim to capture the strongly non-linear process of
dynamical friction of sub clumps within dark haloes, nor strong deviations from spherical collapse. We refer to  \cite{Hahnetal2009} which captures
the effect on satellite galaxies, and to \cite{Ludlow2011,Castorina:2016vg,2016arXiv161004231B} which study the effect of the local shear on halos forming in filamentary structures.
Incorporating these effects would require adopting a threshold for collapse that depends on the local shear, as discussed in the introduction.
Such a barrier would not pose a conceptual problem to our treatment\footnote{The details of the impact on the present derivation are given in Appendix~\ref{sec:moving-barr-gener}.}; technically, however it requires two extra integrations (over the amplitude of the local shear and its derivative w.r.t. scale), and cannot be done analytically.
The shear-dependent part of the critical density (and its derivative) would correlate with the shear of the saddle at $\rr=0$, and introduce an additional anisotropic effect on top of the change of mean values and variances of density and slope we accounted for. Evaluating this effect will be the topic of future investigation.

Our analysis demonstrated that the large-scale tidal field alone can induce specific accretion gradients,
distinct from mass and density ones. One would now like to translate those distinct DM gradients into colour and specific star formation rate gradients.
At high redshift, the stronger the accretion the bluer the central galaxy.
Conversely at low redshift, one can expect that the stronger the accretion, the stronger the AGN feedback, the stronger the quenching of the central.
Should this scaling hold true, the net effect in terms of gradients would be that colour gradients differ from mass and density ones.
The transition between these two regimes (and in general the inclusion of baryonic effects) is beyond the scope of this paper, 
but see Appendix~\ref{sec:speculations} for a brief discussion.

Beyond the dark matter driven processes described in this paper,
different explanations have been recently put forward to explain filamentary colour gradients. On the one hand, it has been argued  \citep{CWD2016}
that the large-scale turbulent flow within filaments may explain the environment dependence in observed physical properties.
Conversely the vorticity of gas inflow within filaments \citep{laigle2015} may be prevalent in feeding galactic discs coherently  \citep{pichonetal11,stewart11}.   Both processes will have distinct signatures in terms of the efficiency and stochasticity of
star formation.
 A mixture of both may in fact be taking place, given that
 the kinematic of the large-scale flow is neither strictly coherent nor fully turbulent. Yet,
 even if ram pressure stripping in filaments operate as efficiently as in clusters, it will remain that the anisotropy of the tides
will also impact the consistency of angular momentum advection, which is deemed important at least  for early type galaxies.
The amplitude of thermodynamical processes depends on the equation of state of the gas
and on the amplitude of feedback which are not fully calibrated today.
Recall that shock heating, AGN and stellar feedback are driven by cold gas infall, which in turn is set by gravity (as the dominant dynamical force). Since gravity has a direct effect through its tides, unless one can convincingly argue that its direct impact is negligible on galactic scales, it should be taken into account.

   \cite{codis2015}, following a formally related route,
    investigated the orientation of the spin of dark haloes in relation to their position w.r.t. the saddle points of
the (density) cosmic web.
Together with their predictions on spin orientation,
 the present work could be extended to model galaxy colours based on both spin and mass accretion.
It could also guide models aiming at mitigating the effect of intrinsic alignments \citep{Joachimi2011} impacting weak lensing studies
 while relying on colour gradients.
More generally, galactic evolution as captured by semi-analytical models will undoubtedly gain from a joint description of
involving both mass and spin acquisition as relevant dynamical ingredients.
Indeed, it has been recently shown in hydrodynamical simulation \citep[e.g.][]{zavala+16} that the assembly of the
inner dark matter halo and its history of specific angular momentum loss is correlated to the morphology of galaxies today.
One should attempt to explain the observed diversity at a given mass
 driven by anisotropic large-scale tides, which will impact gas inflow towards galaxies, hence their properties.
An improved model for galaxy properties should eventually explicitly integrate the geometry of the large environment \citep[following, e.g.~][]{Hanami2001}
and quantify the impact of its anisotropy on galactic mass assembly history.

Thanks to significant observational, numerical and theoretical advances,  the subtle connection between  
 the cosmic web and galactic evolution is on the verge of being understood.

\section*{Acknowledgements}
{\sl
Simulation where carried on the
 Horizon Cluster hosted by Institut d'Astrophysique de Paris. We thank S.~Rouberol for running it smoothly for us.
This research is part of Spin(e) (ANR-13-BS05-0005, \url{http://cosmicorigin.org}).
We are thankful to Stephane Arnouts, Francis Bernardeau, Oliver Hahn, Clotilde Laigle, Aseem Paranjape, Dmitri Pogosyan, Ravi Sheth, Marie Treyer, and Didier Vibert for helpful
discussions. MM is partially supported by the Programme Visiteur of the Institut d'Astrophysique de Paris.   CC is supported by the ILP LABEX (under reference ANR-10-LABX-63 and ANR-11-IDEX-0004-02).
}

\bibliography{author}

\begin{thebibliography}{}
\makeatletter
\relax
\def\mn@urlcharsother{\let\do\@makeother \do\$\do\&\do\#\do\^\do\_\do\%\do\~}
\def\mn@doi{\begingroup\mn@urlcharsother \@ifnextchar [ {\mn@doi@}
  {\mn@doi@[]}}
\def\mn@doi@[#1]#2{\def\@tempa{#1}\ifx\@tempa\@empty \href
  {http://dx.doi.org/#2} {doi:#2}\else \href {http://dx.doi.org/#2} {#1}\fi
  \endgroup}
\def\mn@eprint#1#2{\mn@eprint@#1:#2::\@nil}
\def\mn@eprint@arXiv#1{\href {http://arxiv.org/abs/#1} {{\tt arXiv:#1}}}
\def\mn@eprint@dblp#1{\href {http://dblp.uni-trier.de/rec/bibtex/#1.xml}
  {dblp:#1}}
\def\mn@eprint@#1:#2:#3:#4\@nil{\def\@tempa {#1}\def\@tempb {#2}\def\@tempc
  {#3}\ifx \@tempc \@empty \let \@tempc \@tempb \let \@tempb \@tempa \fi \ifx
  \@tempb \@empty \def\@tempb {arXiv}\fi \@ifundefined
  {mn@eprint@\@tempb}{\@tempb:\@tempc}{\expandafter \expandafter \csname
  mn@eprint@\@tempb\endcsname \expandafter{\@tempc}}}

\bibitem[\protect\citeauthoryear{Alonso, Eardley  \& Peacock}{Alonso
  et~al.}{2015}]{Alonso:2015eu}
Alonso D.,  Eardley E.,   Peacock J.~A.,  2015, Monthly Notices of the Royal
  Astronomical Society, 447, 2683

\bibitem[\protect\citeauthoryear{{Alpaslan} et~al.,}{{Alpaslan}
  et~al.}{2016}]{alpaslanetal2016}
{Alpaslan} M.,  et~al., 2016, \mn@doi [\mnras] {10.1093/mnras/stw134}, \href
  {http://adsabs.harvard.edu/abs/2016MNRAS.457.2287A} {457, 2287}

\bibitem[\protect\citeauthoryear{{Aragon-Calvo}, {Neyrinck}  \&
  {Silk}}{{Aragon-Calvo} et~al.}{2016}]{CWD2016}
{Aragon-Calvo} M.~A.,  {Neyrinck} M.~C.,   {Silk} J.,  2016, preprint, \href
  {http://cdsads.u-strasbg.fr/abs/2016arXiv160707881A} {} (\mn@eprint {arXiv}
  {1607.07881})

\bibitem[\protect\citeauthoryear{{Bernardeau}, {Crocce}  \&
  {Scoccimarro}}{{Bernardeau} et~al.}{2008}]{gammaexp2008}
{Bernardeau} F.,  {Crocce} M.,   {Scoccimarro} R.,  2008, \mn@doi [\prd]
  {10.1103/PhysRevD.78.103521}, \href
  {http://adsabs.harvard.edu/abs/2008PhRvD..78j3521B} {78, 103521}

\bibitem[\protect\citeauthoryear{{Bond} \& {Myers}}{{Bond} \&
  {Myers}}{1996}]{BM1996}
{Bond} J.~R.,  {Myers} S.~T.,  1996, \mn@doi [\apjs] {10.1086/192267}, \href
  {http://adsabs.harvard.edu/abs/1996ApJS..103....1B} {103, 1}

\bibitem[\protect\citeauthoryear{{Bond}, {Cole}, {Efstathiou}  \&
  {Kaiser}}{{Bond} et~al.}{1991}]{Bondetal1991}
{Bond} J.~R.,  {Cole} S.,  {Efstathiou} G.,   {Kaiser} N.,  1991, \mn@doi
  [\apj] {10.1086/170520}, \href
  {http://adsabs.harvard.edu/abs/1991ApJ...379..440B} {379, 440}

\bibitem[\protect\citeauthoryear{Borzyszkowski, Porciani, Romano-Diaz  \&
  Garaldi}{Borzyszkowski et~al.}{2016}]{2016arXiv161004231B}
Borzyszkowski M.,  Porciani C.,  Romano-Diaz E.,   Garaldi E.,  2016,
  arXiv.org, p. arXiv:1610.04231

\bibitem[\protect\citeauthoryear{Castorina, Paranjape, Hahn  \&
  Sheth}{Castorina et~al.}{2016}]{Castorina:2016vg}
Castorina E.,  Paranjape A.,  Hahn O.,   Sheth R.~K.,  2016, \apj

\bibitem[\protect\citeauthoryear{{Chen} et~al.,}{{Chen}
  et~al.}{2017}]{chenetal2017}
{Chen} Y.-C.,  et~al., 2017, \mn@doi [\mnras] {10.1093/mnras/stw3127}, \href
  {http://adsabs.harvard.edu/abs/2017MNRAS.466.1880C} {466, 1880}

\bibitem[\protect\citeauthoryear{{Codis}, {Pichon}  \& {Pogosyan}}{{Codis}
  et~al.}{2015}]{codis2015}
{Codis} S.,  {Pichon} C.,   {Pogosyan} D.,  2015, \mn@doi [\mnras]
  {10.1093/mnras/stv1570}, \href
  {http://adsabs.harvard.edu/abs/2015MNRAS.452.3369C} {452, 3369}

\bibitem[\protect\citeauthoryear{{Corasaniti} \& {Achitouv}}{{Corasaniti} \&
  {Achitouv}}{2011}]{CorasAch2011}
{Corasaniti} P.~S.,  {Achitouv} I.,  2011, \mn@doi [\prd]
  {10.1103/PhysRevD.84.023009}, \href
  {http://adsabs.harvard.edu/abs/2011PhRvD..84b3009C} {84, 023009}

\bibitem[\protect\citeauthoryear{{Dalal}, {White}, {Bond}  \&
  {Shirokov}}{{Dalal} et~al.}{2008}]{dalaletal2008}
{Dalal} N.,  {White} M.,  {Bond} J.~R.,   {Shirokov} A.,  2008, \mn@doi [\apj]
  {10.1086/591512}, \href {http://adsabs.harvard.edu/abs/2008ApJ...687...12D}
  {687, 12}

\bibitem[\protect\citeauthoryear{{Del Popolo}, {Ercan}  \& {Gambera}}{{Del
  Popolo} et~al.}{2001}]{DelPop2001}
{Del Popolo} A.,  {Ercan} E.~N.,   {Gambera} M.,  2001, Baltic Astronomy, \href
  {http://adsabs.harvard.edu/abs/2001BaltA..10..629D} {10, 629}

\bibitem[\protect\citeauthoryear{{Desjacques}, {Jeong}  \&
  {Schmidt}}{{Desjacques} et~al.}{2016}]{biasreview2016}
{Desjacques} V.,  {Jeong} D.,   {Schmidt} F.,  2016, preprint, \href
  {http://adsabs.harvard.edu/abs/2016arXiv161109787D} {} (\mn@eprint {arXiv}
  {1611.09787})

\bibitem[\protect\citeauthoryear{{Doroshkevich}}{{Doroshkevich}}{1970}]{doroshkevich70}
{Doroshkevich} A.~G.,  1970, \mn@doi [Astrophysics] {10.1007/BF01001625}, \href
  {http://adsabs.harvard.edu/abs/1970Ap......6..320D} {6, 320}

\bibitem[\protect\citeauthoryear{{Efstathiou}, {Frenk}, {White}  \&
  {Davis}}{{Efstathiou} et~al.}{1988}]{efstathiouetal1988}
{Efstathiou} G.,  {Frenk} C.~S.,  {White} S.~D.~M.,   {Davis} M.,  1988,
  \mn@doi [\mnras] {10.1093/mnras/235.3.715}, \href
  {http://adsabs.harvard.edu/abs/1988MNRAS.235..715E} {235, 715}

\bibitem[\protect\citeauthoryear{{Faltenbacher} \& {White}}{{Faltenbacher} \&
  {White}}{2010}]{FaltenWhite2010}
{Faltenbacher} A.,  {White} S.~D.~M.,  2010, \mn@doi [\apj]
  {10.1088/0004-637X/708/1/469}, \href
  {http://adsabs.harvard.edu/abs/2010ApJ...708..469F} {708, 469}

\bibitem[\protect\citeauthoryear{{Fry} \& {Gaztanaga}}{{Fry} \&
  {Gaztanaga}}{1993}]{FryGazta1993}
{Fry} J.~N.,  {Gaztanaga} E.,  1993, \mn@doi [\apj] {10.1086/173015}, \href
  {http://adsabs.harvard.edu/abs/1993ApJ...413..447F} {413, 447}

\bibitem[\protect\citeauthoryear{{Gao}, {Springel}  \& {White}}{{Gao}
  et~al.}{2005}]{gaoetal2005}
{Gao} L.,  {Springel} V.,   {White} S.~D.~M.,  2005, \mn@doi [\mnras]
  {10.1111/j.1745-3933.2005.00084.x}, \href
  {http://adsabs.harvard.edu/abs/2005MNRAS.363L..66G} {363, L66}

\bibitem[\protect\citeauthoryear{Gradshteyn \& Ryzhik}{Gradshteyn \&
  Ryzhik}{2007}]{gradshteyn2007}
Gradshteyn I.~S.,  Ryzhik I.~M.,  2007, Table of integrals, series, and
  products, seventh edn.
Elsevier/Academic Press, Amsterdam

\bibitem[\protect\citeauthoryear{{Hahn}, {Porciani}, {Dekel}  \&
  {Carollo}}{{Hahn} et~al.}{2009}]{Hahnetal2009}
{Hahn} O.,  {Porciani} C.,  {Dekel} A.,   {Carollo} C.~M.,  2009, \mn@doi
  [\mnras] {10.1111/j.1365-2966.2009.15271.x}, \href
  {http://adsabs.harvard.edu/abs/2009MNRAS.398.1742H} {398, 1742}

\bibitem[\protect\citeauthoryear{Hanami}{Hanami}{2001}]{Hanami2001}
Hanami H.,  2001, Monthly Notices of the Royal Astronomical Society, 327, 721

\bibitem[\protect\citeauthoryear{{Joachimi}, {Mandelbaum}, {Abdalla}  \&
  {Bridle}}{{Joachimi} et~al.}{2011}]{Joachimi2011}
{Joachimi} B.,  {Mandelbaum} R.,  {Abdalla} F.~B.,   {Bridle} S.~L.,  2011,
  \mn@doi [\aap] {10.1051/0004-6361/201015621}, \href
  {http://adsabs.harvard.edu/abs/2011A%26A...527A..26J} {527, A26}

\bibitem[\protect\citeauthoryear{{Kaiser}}{{Kaiser}}{1984}]{Kaiser1984}
{Kaiser} N.,  1984, \mn@doi [\apjl] {10.1086/184341}, \href
  {http://cdsads.u-strasbg.fr/abs/1984ApJ...284L...9K} {284, L9}

\bibitem[\protect\citeauthoryear{{Kauffmann}, {Li}, {Zhang}  \&
  {Weinmann}}{{Kauffmann} et~al.}{2013}]{2013MNRAS.430.1447K}
{Kauffmann} G.,  {Li} C.,  {Zhang} W.,   {Weinmann} S.,  2013, \mn@doi [\mnras]
  {10.1093/mnras/stt007}, \href
  {http://cdsads.u-strasbg.fr/abs/2013MNRAS.430.1447K} {430, 1447}

\bibitem[\protect\citeauthoryear{{Kawinwanichakij} et~al.,}{{Kawinwanichakij}
  et~al.}{2016}]{kawinetal2016}
{Kawinwanichakij} L.,  et~al., 2016, \mn@doi [\apj]
  {10.3847/0004-637X/817/1/9}, \href
  {http://adsabs.harvard.edu/abs/2016ApJ...817....9K} {817, 9}

\bibitem[\protect\citeauthoryear{Lacey \& Cole}{Lacey \&
  Cole}{1993}]{LaceyCole1993}
Lacey C.~G.,  Cole S.,  1993, Mon. Not. Roy. Astron. Soc., 262, 627

\bibitem[\protect\citeauthoryear{{Laigle} et~al.,}{{Laigle}
  et~al.}{2015}]{laigle2015}
{Laigle} C.,  et~al., 2015, \mn@doi [\mnras] {10.1093/mnras/stu2289}, \href
  {http://cdsads.u-strasbg.fr/abs/2015MNRAS.446.2744L} {446, 2744}

\bibitem[\protect\citeauthoryear{{Laigle} et~al.,}{{Laigle}
  et~al.}{2017}]{Laigle+17}
{Laigle} C.,  et~al., 2017, preprint, \href
  {http://adsabs.harvard.edu/abs/2017arXiv170208810L} {} (\mn@eprint {arXiv}
  {1702.08810})

\bibitem[\protect\citeauthoryear{{Lazeyras}, {Musso}  \& {Schmidt}}{{Lazeyras}
  et~al.}{2017}]{Lazeyras2017}
{Lazeyras} T.,  {Musso} M.,   {Schmidt} F.,  2017, \mn@doi [\jcap]
  {10.1088/1475-7516/2017/03/059}, \href
  {http://adsabs.harvard.edu/abs/2017JCAP...03..059L} {3, 059}

\bibitem[\protect\citeauthoryear{{Ludlow}, {Borzyszkowski}  \&
  {Porciani}}{{Ludlow} et~al.}{2011}]{Ludlow2011}
{Ludlow} A.~D.,  {Borzyszkowski} M.,   {Porciani} C.,  2011, preprint, \href
  {http://adsabs.harvard.edu/abs/2011arXiv1107.5808L} {} (\mn@eprint {arXiv}
  {1107.5808})

\bibitem[\protect\citeauthoryear{{Maggiore} \& {Riotto}}{{Maggiore} \&
  {Riotto}}{2010}]{MaggioreRiotto2010}
{Maggiore} M.,  {Riotto} A.,  2010, \mn@doi [\apj]
  {10.1088/0004-637X/711/2/907}, \href
  {http://adsabs.harvard.edu/abs/2010ApJ...711..907M} {711, 907}

\bibitem[\protect\citeauthoryear{{Malavasi} et~al.,}{{Malavasi}
  et~al.}{2016}]{Malavasi2016b}
{Malavasi} N.,  et~al., 2016, \mnras \ in press,\ arXiv1611.07045, \href
  {http://adsabs.harvard.edu/abs/2016arXiv161107045M} {}

\bibitem[\protect\citeauthoryear{{Mart{\'{\i}}nez}, {Muriel}  \&
  {Coenda}}{{Mart{\'{\i}}nez} et~al.}{2016}]{martinezetal2016}
{Mart{\'{\i}}nez} H.~J.,  {Muriel} H.,   {Coenda} V.,  2016, \mn@doi [\mnras]
  {10.1093/mnras/stv2295}, \href
  {http://adsabs.harvard.edu/abs/2016MNRAS.455..127M} {455, 127}

\bibitem[\protect\citeauthoryear{{Musso} \& {Sheth}}{{Musso} \&
  {Sheth}}{2012}]{MussoSheth2012}
{Musso} M.,  {Sheth} R.~K.,  2012, \mn@doi [\mnras]
  {10.1111/j.1745-3933.2012.01266.x}, \href
  {http://adsabs.harvard.edu/abs/2012MNRAS.423L.102M} {423, L102}

\bibitem[\protect\citeauthoryear{{Musso} \& {Sheth}}{{Musso} \&
  {Sheth}}{2014a}]{MussoSheth2013}
{Musso} M.,  {Sheth} R.~K.,  2014a, \mn@doi [\mnras] {10.1093/mnras/stt2387},
  \href {http://adsabs.harvard.edu/abs/2014MNRAS.438.2683M} {438, 2683}

\bibitem[\protect\citeauthoryear{{Musso} \& {Sheth}}{{Musso} \&
  {Sheth}}{2014b}]{MussoShethmarkov2014}
{Musso} M.,  {Sheth} R.~K.,  2014b, \mn@doi [\mnras] {10.1093/mnras/stu1222},
  \href {http://adsabs.harvard.edu/abs/2014MNRAS.443.1601M} {443, 1601}

\bibitem[\protect\citeauthoryear{{Musso} \& {Sheth}}{{Musso} \&
  {Sheth}}{2014c}]{MussoSheth2014}
{Musso} M.,  {Sheth} R.~K.,  2014c, \mn@doi [\mnras] {10.1093/mnras/stu1222},
  \href {http://adsabs.harvard.edu/abs/2014MNRAS.443.1601M} {443, 1601}

\bibitem[\protect\citeauthoryear{{Musso}, {Paranjape}  \& {Sheth}}{{Musso}
  et~al.}{2012}]{MPS2012}
{Musso} M.,  {Paranjape} A.,   {Sheth} R.~K.,  2012, \mn@doi [\mnras]
  {10.1111/j.1365-2966.2012.21903.x}, \href
  {http://cdsads.u-strasbg.fr/abs/2012MNRAS.427.3145M} {427, 3145}

\bibitem[\protect\citeauthoryear{{Oemler}}{{Oemler}}{1974}]{oemler74}
{Oemler} Jr. A.,  1974, \mn@doi [\apj] {10.1086/153216}, \href
  {http://adsabs.harvard.edu/abs/1974ApJ...194....1O} {194, 1}

\bibitem[\protect\citeauthoryear{{Paranjape} \& {Padmanabhan}}{{Paranjape} \&
  {Padmanabhan}}{2017}]{ParanjPadma2017}
{Paranjape} A.,  {Padmanabhan} N.,  2017, \mn@doi [\mnras]
  {10.1093/mnras/stx659}, \href
  {http://adsabs.harvard.edu/abs/2017MNRAS.468.2984P} {468, 2984}

\bibitem[\protect\citeauthoryear{{Paranjape}, {Hahn}  \& {Sheth}}{{Paranjape}
  et~al.}{2017}]{paranjapeetal2017}
{Paranjape} A.,  {Hahn} O.,   {Sheth} R.~K.,  2017, preprint, \href
  {http://adsabs.harvard.edu/abs/2017arXiv170609906P} {} (\mn@eprint {arXiv}
  {1706.09906})

\bibitem[\protect\citeauthoryear{{Pichon}, {Pogosyan}, {Kimm}, {Slyz},
  {Devriendt}  \& {Dubois}}{{Pichon} et~al.}{2011}]{pichonetal11}
{Pichon} C.,  {Pogosyan} D.,  {Kimm} T.,  {Slyz} A.,  {Devriendt} J.,
  {Dubois} Y.,  2011, \mn@doi [\mnras] {10.1111/j.1365-2966.2011.19640.x},
  \href {http://cdsads.u-strasbg.fr/abs/2011MNRAS.tmp.1739P} {pp 1739--+}

\bibitem[\protect\citeauthoryear{{Pogosyan}, {Bond}, {Kofman}  \&
  {Wadsley}}{{Pogosyan} et~al.}{1998}]{Pogosyanetal1998}
{Pogosyan} D.,  {Bond} J.~R.,  {Kofman} L.,   {Wadsley} J.,  1998, in
  {S.~Colombi, Y.~Mellier, \& B.~Raban} ed., Wide Field Surveys in Cosmology.
  p.~61 (\mn@eprint {} {arXiv:astro-ph/9810072})

\bibitem[\protect\citeauthoryear{{Poudel}, {Hein{\"a}m{\"a}ki}, {Tempel},
  {Einasto}, {Lietzen}  \& {Nurmi}}{{Poudel} et~al.}{2017}]{poudeletal2016}
{Poudel} A.,  {Hein{\"a}m{\"a}ki} P.,  {Tempel} E.,  {Einasto} M.,  {Lietzen}
  H.,   {Nurmi} P.,  2017, \mn@doi [\aap] {10.1051/0004-6361/201629639}, \href
  {http://adsabs.harvard.edu/abs/2017A%26A...597A..86P} {597, A86}

\bibitem[\protect\citeauthoryear{Press \& Schechter}{Press \&
  Schechter}{1974}]{PressSchechter1974}
Press W.~H.,  Schechter P.,  1974, \mn@doi [Astrophys. J.] {10.1086/152650},
  187, 425

\bibitem[\protect\citeauthoryear{Redner}{Redner}{2001}]{redner2001}
Redner S.,  2001, A Guide to First-Passage Processes.
Cambridge University Press, \mn@doi{10.1017/CBO9780511606014}

\bibitem[\protect\citeauthoryear{{Shen}, {Abel}, {Mo}  \& {Sheth}}{{Shen}
  et~al.}{2006}]{Shenetal2006}
{Shen} J.,  {Abel} T.,  {Mo} H.~J.,   {Sheth} R.~K.,  2006, \mn@doi [\apj]
  {10.1086/504513}, \href {http://adsabs.harvard.edu/abs/2006ApJ...645..783S}
  {645, 783}

\bibitem[\protect\citeauthoryear{{Sheth} \& {Tormen}}{{Sheth} \&
  {Tormen}}{2004}]{shethettormen2004}
{Sheth} R.~K.,  {Tormen} G.,  2004, \mn@doi [\mnras]
  {10.1111/j.1365-2966.2004.07733.x}, \href
  {http://adsabs.harvard.edu/abs/2004MNRAS.350.1385S} {350, 1385}

\bibitem[\protect\citeauthoryear{{Sheth}, {Mo}  \& {Tormen}}{{Sheth}
  et~al.}{2001}]{SMT2001}
{Sheth} R.~K.,  {Mo} H.~J.,   {Tormen} G.,  2001, \mn@doi [\mnras]
  {10.1046/j.1365-8711.2001.04006.x}, \href
  {http://adsabs.harvard.edu/abs/2001MNRAS.323....1S} {323, 1}

\bibitem[\protect\citeauthoryear{{Sheth}, {Chan}  \& {Scoccimarro}}{{Sheth}
  et~al.}{2013}]{ShethChanScocc2013}
{Sheth} R.~K.,  {Chan} K.~C.,   {Scoccimarro} R.,  2013, \mn@doi [\prd]
  {10.1103/PhysRevD.87.083002}, \href
  {http://adsabs.harvard.edu/abs/2013PhRvD..87h3002S} {87, 083002}

\bibitem[\protect\citeauthoryear{Sousbie, Pichon, Colombi  \& Pogosyan}{Sousbie
  et~al.}{2008}]{sousbie08}
Sousbie T.,  Pichon C.,  Colombi S.,   Pogosyan D.,  2008, \mnras, 383, 1655

\bibitem[\protect\citeauthoryear{{Stewart}, {Kaufmann}, {Bullock}, {Barton},
  {Maller}, {Diemand}  \& {Wadsley}}{{Stewart} et~al.}{2011}]{stewart11}
{Stewart} K.~R.,  {Kaufmann} T.,  {Bullock} J.~S.,  {Barton} E.~J.,  {Maller}
  A.~H.,  {Diemand} J.,   {Wadsley} J.,  2011, preprint, \href
  {http://adsabs.harvard.edu/abs/2011arXiv1103.4388S} {} (\mn@eprint {arXiv}
  {1103.4388})

\bibitem[\protect\citeauthoryear{Tramonte, Rubino-Martin, Betancort-Rijo  \&
  Dalla~Vecchia}{Tramonte et~al.}{2017}]{Tramonte2017}
Tramonte D.,  Rubino-Martin J.~A.,  Betancort-Rijo J.,   Dalla~Vecchia C.,
  2017, arXiv.org, p. arXiv:1702.01788

\bibitem[\protect\citeauthoryear{{Wang} et~al.,}{{Wang}
  et~al.}{2011}]{wangetal2011}
{Wang} J.,  et~al., 2011, \mn@doi [\mnras] {10.1111/j.1365-2966.2011.18220.x},
  \href {http://adsabs.harvard.edu/abs/2011MNRAS.413.1373W} {413, 1373}

\bibitem[\protect\citeauthoryear{{Wechsler}, {Zentner}, {Bullock}, {Kravtsov}
  \& {Allgood}}{{Wechsler} et~al.}{2006}]{wechsleretal2006}
{Wechsler} R.~H.,  {Zentner} A.~R.,  {Bullock} J.~S.,  {Kravtsov} A.~V.,
  {Allgood} B.,  2006, \mn@doi [\apj] {10.1086/507120}, \href
  {http://adsabs.harvard.edu/abs/2006ApJ...652...71W} {652, 71}

\bibitem[\protect\citeauthoryear{Weinmann, van~den Bosch, Yang  \& Mo}{Weinmann
  et~al.}{2006}]{weinmannetal2006}
Weinmann S.~M.,  van~den Bosch F.~C.,  Yang X.,   Mo H.~J.,  2006, \mnras, 366,
  2

\bibitem[\protect\citeauthoryear{Yan, Fan  \& White}{Yan
  et~al.}{2013}]{Yan2013}
Yan H.,  Fan Z.,   White S. D.~M.,  2013, Monthly Notices of the Royal
  Astronomical Society, 430, 3432

\bibitem[\protect\citeauthoryear{{Zavala} et~al.,}{{Zavala}
  et~al.}{2016}]{zavala+16}
{Zavala} J.,  et~al., 2016, \mn@doi [\mnras] {10.1093/mnras/stw1286}, \href
  {http://cdsads.u-strasbg.fr/abs/2016MNRAS.460.4466Z} {460, 4466}

\bibitem[\protect\citeauthoryear{{Zentner}}{{Zentner}}{2007}]{Zentner2007}
{Zentner} A.~R.,  2007, \mn@doi [International Journal of Modern Physics D]
  {10.1142/S0218271807010511}, \href
  {http://adsabs.harvard.edu/abs/2007IJMPD..16..763Z} {16, 763}

\bibitem[\protect\citeauthoryear{von Braun-Bates, Winther, Alonso  \&
  Devriendt}{von Braun-Bates et~al.}{2017}]{Braun-Bates2017}
von Braun-Bates F.,  Winther H.~A.,  Alonso D.,   Devriendt J.,  2017,
  arXiv.org, p. arXiv:1702.06817

\makeatother
\end{thebibliography}

\appendix

\section{Definitions and notations}
\label{sec:definitions}

\begin{table*}
  \renewcommand{\arraystretch}{1.2}
  \begin{tabularx}{\linewidth}{|p{1.2cm}|p{4.5cm}|X|}
    \hline
    Variable & Definition & Comment \\
    \hline\hline
    $\bar{\rho}_m$ & $\left(\SI{2.8\times10^{11}}{h^2 \msun/Mpc^3}\right)
                   \times \Omega_M$& Uniform matter background density \\
    $R, M, M_\star$ & $M=4/3\pi R^3\bar{\rho}_m$& Smoothing scale, mass, typical mass  \\
    $\delta_m$ & $\displaystyle(\rho_m-\bar{\rho}_m)/\bar{\rho}_m$ & Linear matter overdensity \\
    $W(x)$ & $3j_1(x)/x$ & Real-space Top-Hat filter (Fourier
                           representation) \\
    \hline
    $\delta$ & $\displaystyle\int\frac{\d^3k}{(2\pi)^3}\delta_m(\kk) W(kR) e^{i
               \kk\cdot\rr}$
                          & Linear matter overdensity smoothed at
                             scale $R$, position $\rr$ \\
    $\sigma^2$ & $\mathrm{Var}(\delta)$ & Variance of the overdensity
                                          at scale $R$ \\
    $\nu$ & $\delta/\sigma$ & Rescaled overdensity \\
    $\delta_c, \nu_c$ & \num{1.68}, $\delta_c/\sigma$ & Critical overdensity \\
    $\delta', \nu'$ & $\dd \delta/\dd \sigma, \dd\nu/\dd \sigma$  & Slope of the E.S. trajectories \\
    $\Gamma^{-2}$ & $\mathrm{Var}(\delta')-1 = \mean{(\sigma\nu')^2} = \Var{\delta'}{\nu}$ & Conditional variance of $\delta'$ at fixed $\nu$ \\
    \hline
    $\Rs,\sigmas$ & $\sigmas=\sigma(\Rs)$
                          & Smoothing scale used at the
                            saddle point \\
                              $R_\star^2$ & \eqref{eq:sadscale}
  $ \displaystyle  \int\d k \frac{P(k)}{2\pi^2} \frac{W^2(k\Rs)}{\sigmas^2}\,.
$ & Characteristic length scale of the saddle (squared)
  \\
    $g_i, q_{ij}, \nus$ & \eqref{eq:gi_def},\eqref{eq:qij_def}
                          & Mean acceleration, tidal tensor and
                            overdensity at saddle (see
                            Table~\ref{tab:qijbar_nu_values} for their
                            value) \\
    $\bar{q}_{ij}, \mathcal{Q}$ & $\bar{q}_{ij}=q_{ij}-\nus\delta_{ij}/3,
                                  \hat{r}_i\bar{q}_{ij}\hat{r}_j  $
                          & Traceless tidal tensor and anisotropy ellipsoidal-hyperbolic coordinate \\
    $\xi_{\alpha\beta},\xi'_{\alpha\beta}$& \eqref{eq:xi_xi}, \eqref{eq:xi_xiprime}; $\xi_{\alpha\beta}'=\d\xi_{\alpha\beta}/\d\sigma$
                          & Two point correlation functions at
                            separation $r$ and scales $R,\Rs$ \\
    \hline
    $\alpha, \alpha_\star$ & $\nu_c / [\sigma(\nu'-\nu'_c)]$; \eqref{eq:alphastar}, \eqref{eq:astarsad} & Accretion rate, typical accretion rate \\
    \hline
    $R_\half, \sigma_\half$ & $R/2^{1/3}, \sigma(R_\half)$
                          & Half-mass radius and variance \\
    $\delta_\half, \nu_\half$ & $\delta(\sigma_\half),\delta_\half/\sigma_\half$
                          & Overdensity at half-mass \\
    $D_\form,D_\star$ & $\dc/\delta_\half$; \eqref{eq:defD*}, \eqref{eq:defD*sad}& Formation time, typical formation time\\
    $\nu_\form$ & $\delta_c/(\sigma_{\half}D_\form)$
                          & Density threshold at formation time \\
    $\omega,\omega'$ & \eqref{eq:omega},~\eqref{eq:omegaprime}; $\omega'=\d\omega/\d\sigma$
                     & Zero-distance correlation functions between
                            scales $R$ and $R_\half$ \\
    $\Omega,\Omega'$ & \eqref{eq:Omega},~\eqref{eq:Omegap}; $\Omega'=\d\Omega/\d\sigma$
                     & Zero-distance conditional covariance
    between scales $R$ and $R_\half$ given the saddle point\\
    \hline
    $\delta_0$ & $\delta(R_0\gg R)$
                          & Large scale overdensity \\
    $\delta_h$ & & Local halo number density contrast \\
    \hline
  \end{tabularx}
  \caption{
    Summary of the variables used throughout the
    paper.
    }
  \label{tab:definitions}
\end{table*}
 Table~\ref{tab:definitions}  presents all the definitions introduced in
the paper.
 Table~\ref{tab:method} gives also the motivation behind the choice of variables.
 The following conventions is used throughout:
\begin{itemize}
\item unless stated otherwise, all the quantities evaluated at (halo) scale
  $R$ have their dependence on $R$ omitted (e.g.
  $\sigma = \sigma(R)$);
\item the quantities that have a radial dependence are evaluated at a
  distance $r$ when the radius is omitted. Sometimes, the full form is used to
  emphasize the dependence on this variable;
\item unless stated otherwise, the quantities are evaluated at
  $z=0, D(z)=1$ (e.g. $\delta_c=1.686$);
\item a prime denotes a derivative w.r.t $\sigma$ of the excursion set
  (e.g. $\delta'=\d\delta/\d\sigma$);
\item variables carrying a hat have unit norm (e.g. $|\hat\rr|=1$),
  matrices carrying an over-bar are traceless
  (e.g. $\mathrm{tr}(\bar{q}_{ij})=0$);
\item the Einstein's convention on repeated indexes is used throughout, except in Section~\ref{sec:dgivenS}.
\end{itemize}
\section{Validation with GRFs }
\label{sec:validation}
%
Let us first compare the prediction of section~\ref{sec:conditional}
to statistics derived from realization of Gaussian random fields (GRF) while
imposing a saddle point condition. The values used at the saddle point
are reported in table~\ref{tab:qijbar_nu_values}. We further imposed
the saddle point's eigenframe to coincide with the $x, y, z$ frame,
which in practice has been done by imposing $\bar{q}_{ij}$ to be
diagonal.
We have used two different methods to validate our results, by
generating random density cubes (section~\ref{sec:valid-sigm}) and by
computing the statistics of a constrained field
(section~\ref{sec:valid-alph-using}).
\subsection{Validation for  $\sigma_\star$}
\label{sec:valid-sigm}

\begin{figure}
  \centering
  \includegraphics[width=\columnwidth]{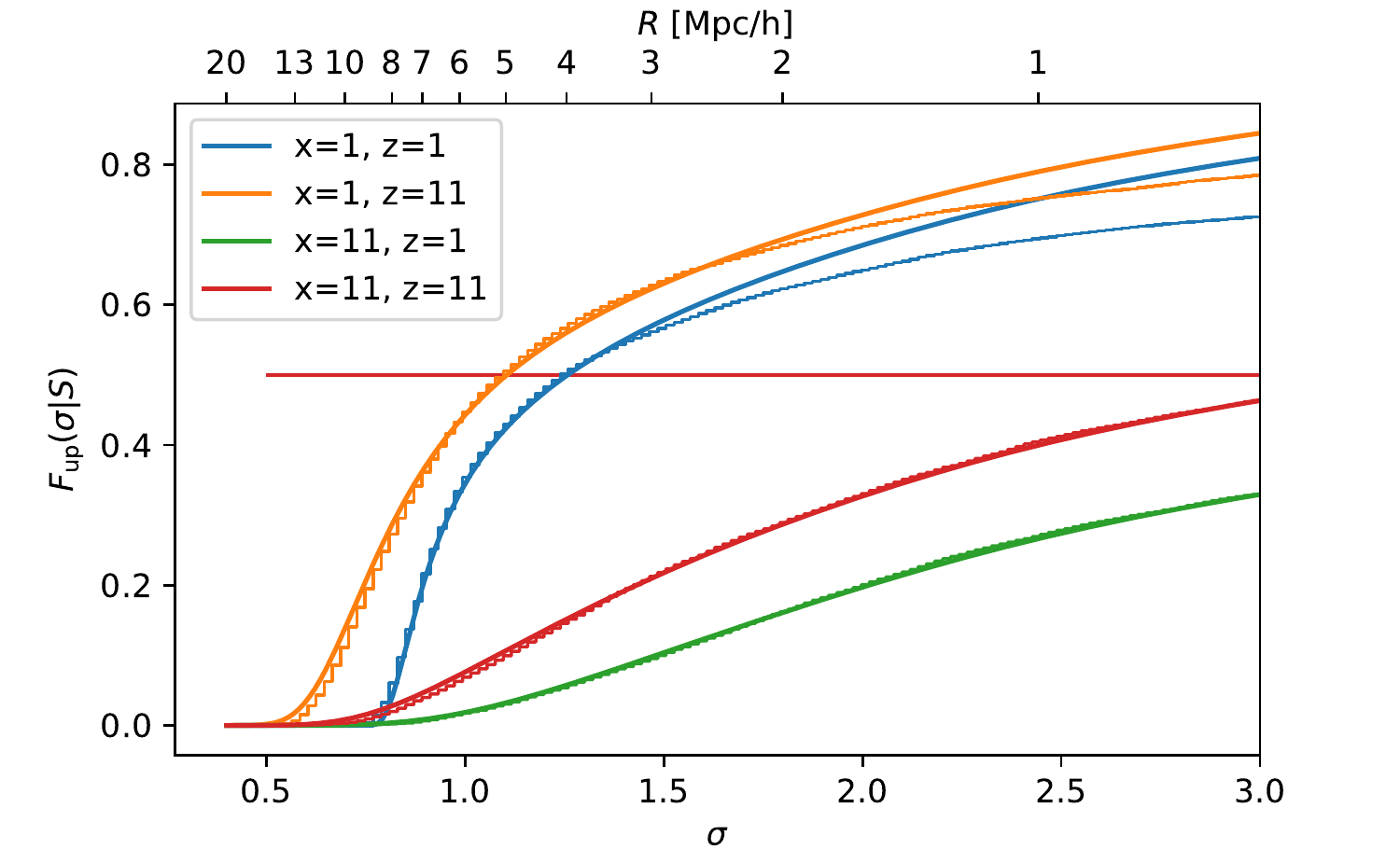}
  \caption{Theoretical CDF of $\sigma$ at upcrossing (bold lines) and
    numerical CDF (steps) at first-crossing at four locations around
    the saddle point (the distances are in $\mathrm{Mpc/h}$ in the $x$
    (void) and $z$ (filament) directions). The CDF have been
    normalized to share the same \SI{50}{\%} quantile (the horizontal
    line). See text for the details of the normalization.}
  \label{fig:sigma_CDF}
\end{figure}

The procedure is the following: i) \num{4000} cubes of size $(128)^3$
and width $L_\mathrm{box} = \SI{200}{Mpc/h}$ centred on a saddle
point were generated following a $\Lambda$CDM power spectrum; ii) each
cube has been smoothed using a Top-Hat filter at 25 different scales
ranging from \SI{0.5}{Mpc/h} to \SI{20}{Mpc/h}; iii) for each point of
each cube, the {\em first-crossing} point $\sigma_\first$ was
computed; iv) the \num{4000} realizations were stacked to get a
distribution of $\sigma_\first$ and to compute the median value. It is
worth noting that the value of $\Gamma(\sigma(R))$ in the GRF is not
the same as in theory. This is a well-known effect \cite[see
e.g.][]{sousbie08} that arise on small scales due to the finite
resolution of the grid and on large-scale because of the finite size
of the box. The $\Gamma$ measured in a GRF is correct at scales
verifying $\Delta L \lessapprox R \ll L_\mathrm{box}$, where
$\Delta L$ is the grid spacing. In our case, the largest smoothing
scale is $\SI{20}{Mpc/h} = L_\mathrm{box}/10$. However, the smallest
scale is comparable to the grid spacing. To attenuate the effect of
finite resolution, we have measured $\Gamma(\sigma(R))$ in the GRF and
used its value to compute the theoretical CDF. The results of the
measured CDF $F_\first$ and theoretical CDF $F_\up$ (with the measured
$\Gamma$) at four different positions are shown on
Fig.~\ref{fig:sigma_CDF}. The measured CDFs have been normalized so
that $F_\first^{-1}(0.5) = F_\up^{-1}(0.5)$: we impose that the CDF
match at the ``median'' (defined as the $\sigma$ such that
$F(\sigma) = 0.5$\footnote{This definition matches the classical one for
  distributions that have a normalized CDF, which is not true for
  $F_\up$.}). As shown on Fig.~\ref{fig:alpha_star} the abscissa of
the peak of the PDF in the direction of the void is around
$\sigma\approx 2.7$. As $\sigma(R_\mathrm{min})\approx 3$, it means
that in the direction of the void, the PDF is only sampled up to its
peak. The experimental CDF at such location is hence only probing less
than \SI{50}{\%} of the distribution and the median is not reached. In
this case, we are normalizing the experimental CDF to have the same
value at the largest $\sigma$ as the theoretical CDF. As shown on
Fig.~\ref{fig:sigma_CDF} the experimental and theoretical CDFs start
diverging at $F\gtrapprox 0.5$. At larger $\sigma$, the up-crossing
approximation used in the theory breaks as more and more trajectories
cross multiple time the barrier (they are counted once for the first
crossing and multiple times for up-crossing). The orange and blue
lines, in the direction of the filament show this clearly as they
diverge one from each other at large $\sigma$. As $\sigma_\star$ is a
measure of the location of the peak of the PDF (which is where the CDF
is the steepest), it is sufficient that the experimental and
theoretical CDF match up to their flat end to have the same
$\sigma_\star$ values.

\subsection{Validation for $\alpha_\star$ using constrained fields }
\label{sec:valid-alph-using}
\begin{figure}
  \centering
  \includegraphics[width=\columnwidth]{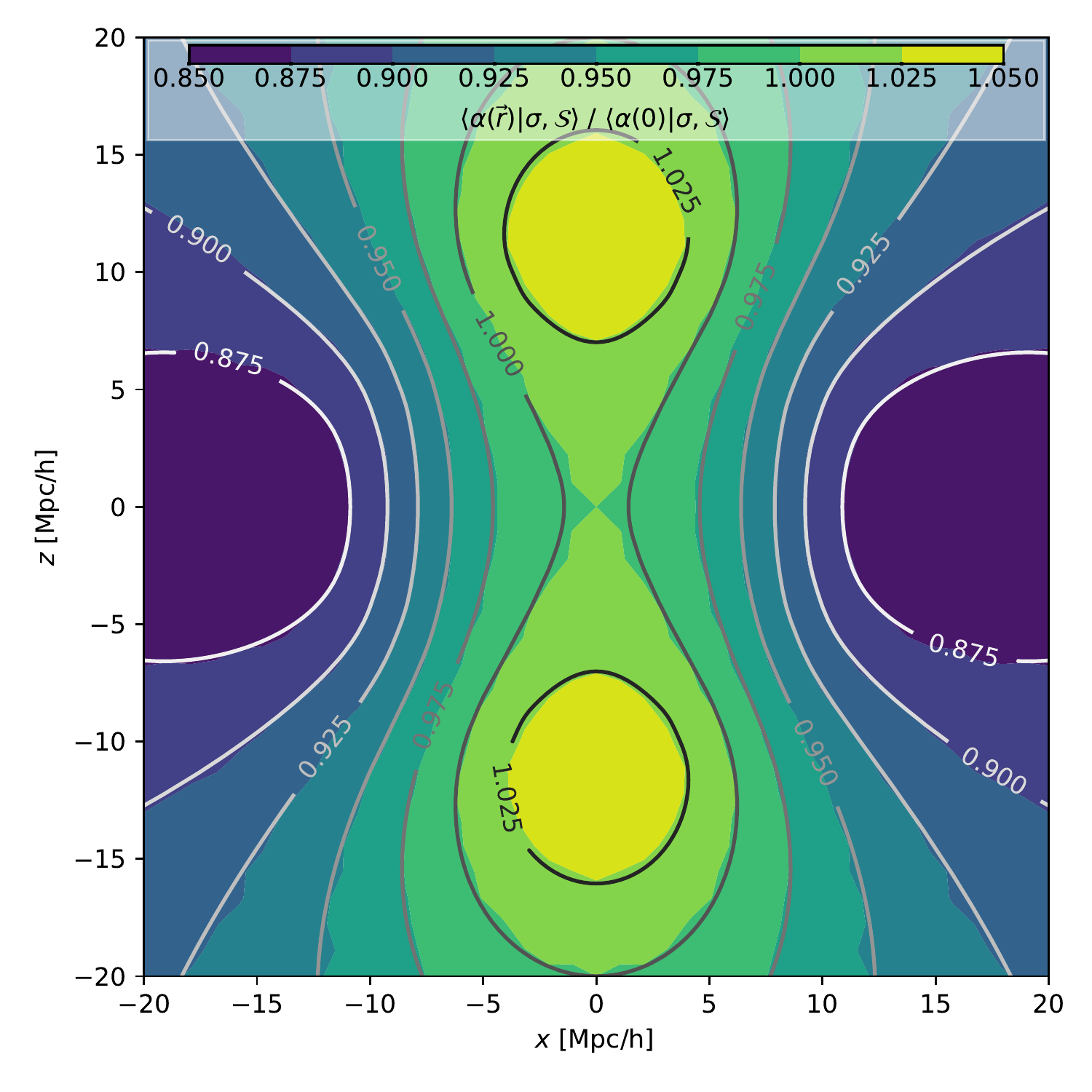}
  \caption{Mean value of $\alpha$ using a numerical method (purple to
    yellow) vs. its theoretical value (grey contours). Both are
    normalized by the theoretical value at the saddle point.}
  \label{fig:alpha_star}
\end{figure}

A second check was implemented on the accretion rate as follows: i)
for each location the covariance matrix of
$\nu, \delta', \nus, \bar{q}_{ij}, g_i$ was computed at finite
distance. These quantities all have a null mean; ii) the covariance
matrix and the mean of $\nu, \delta'$ conditioned to the value at the
saddle point was computed using the values of
Table~\ref{tab:qijbar_nu_values}; iii) the variance and mean of
$\nu, \delta'$ were computed given $\nu=\nu_c$ and the saddle point;
iv) a sample of \num{10^6} points were then drawn from the
distribution of $\delta'>0$ (up-crossing). v) The values of
$\alpha\propto 1/\delta'$ were computed to obtain a sample of
$\alpha$. Each draw was weighted by $1/\alpha$ (the Jacobian of the
transform from $\delta'$ to $\alpha$). Finally, the numerical value of
$\condmean{\alpha}{\sigma, {\cal S}}$ was estimated from the samples
and compared with the theoretical value. The results are shown on
Fig.~\ref{fig:alpha_star} and are found to be in very good agreement.

\begin{figure}
  \centering

  \includegraphics[width=\columnwidth]{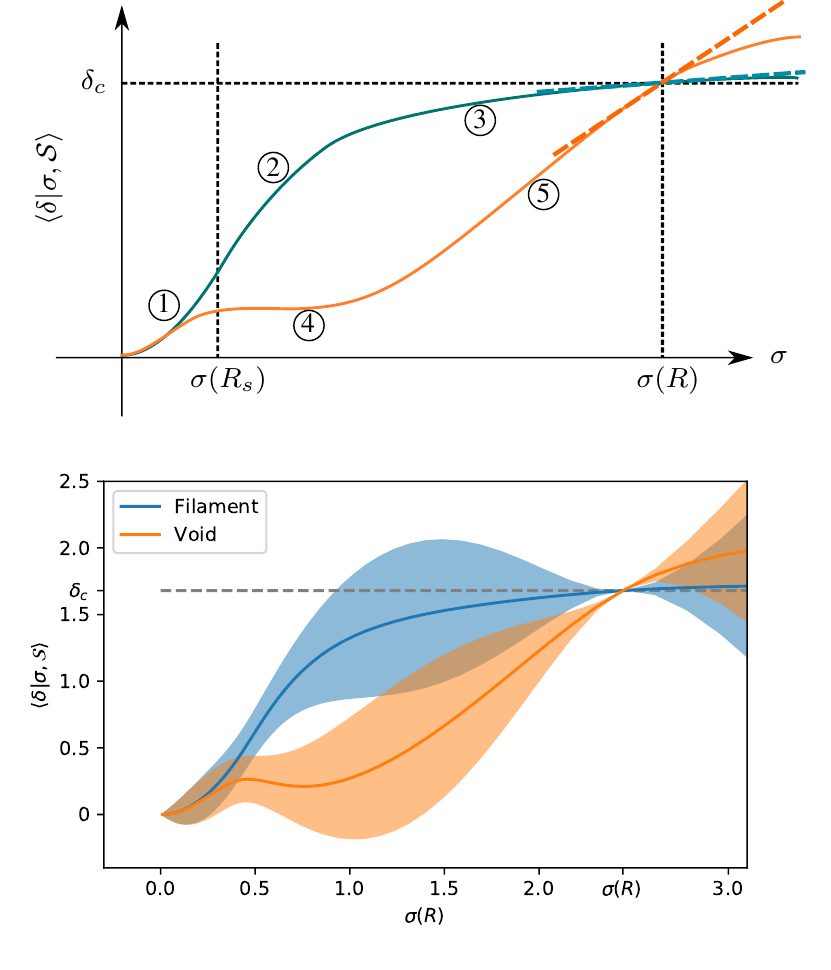}
     \caption{{\em Top}: scheme of the mean value of the density in
       the direction of a filament (red) and void (blue) close to a
       saddle point smoothed at $\sigma = \sigmas$ with the
       constrain that $\delta(\sigma(R))=\delta_c$. (1) The value of
       the density imposed at the saddle point forces both mean
       densities to increase. (2) In the direction of the filament, a
       large-scale overdensity, the mean density at a given point
       increases quickly but (3) the constrain
       $\delta(\sigma)=\delta_c$ prevents any further increase at
       $\sigma\lesssim\sigma(R)$, hence the slope $\delta'$ is small
       at upcrossing. (4) In the direction of the void, a large-scale
       under-density, the mean density at a given point cannot increase
       with $\sigma$. (5) At $\sigma\lesssim\sigma(R)$, the upcrossing
       constrain forces a sharp increase of the density to reach
       $\delta(R)=\delta_c$, hence the slope is high at
       upcrossing. {\em Bottom:} a validation using constrained GRF at
       a distance of \SI{10}{Mpc/h} in the direction of the filament
       (blue) and of the void (orange). See the text for the details.}
   \label{fig:saddleES}
 \end{figure}

We computed Fig.~\ref{fig:saddleES} by following steps i) to iii) at
\SI{10}{Mpc/h} in the direction of the filament (blue) and of the void
(orange) and plotting the mean and standard deviation of $\delta$
given the saddle and the threshold. Fig.~\ref{fig:constrained-slope}
was computed by following steps i) to iii) at the saddle point
($r=0$). An extra constrain on the value of $\delta'$ was then
added to compute the different curves.

\section{Other critical points }
\label{sec:other-critical}
For the sake of generality, let us discuss here the conditional excursion set
expectations in the vicinity of other critical points of the potential.
At the technical level, all the formulae we derived in Section~\ref{sec:conditional} depend on the eigenvalues of $q_{ij}$ with no a priori assumption on their sign. The expressions will thus remain formally the same, with all information about the environment being channelled through the values of $\nus$ and $\hat r_i \bar q_{ij} \hat r_j$.
For instance, the typical quantities $M_\star$, $\dot M_\star$ and $z_\star$ parametrizing the PDFs of interest will be defined in exactly the same way as in equations \eqref{eq:M*r}, \eqref{eq:mdot*r} and \eqref{eq:z*r}. However, their level curves will have different profiles in different environments.

As physical intuition suggests, and equation~\eqref{eq:smallr} explicitly shows, the dependence of the various halo statistics on the distance from the stationary point (whether the probability of a given halo property increases or decreases with separation) is encoded in the signs of the eigenvalues $q_i$ of $q_{ij}$. Besides filaments (having two positive eigenvalues), one may thus be interested in wall-type saddles (one positive eigenvalue), maxima (all negative) and minima (all positive), corresponding to voids and nodes respectively. In general, $q_1+q_2+q_3=\nus$ parametrizes the mean variation with distance (averaged over the angles), whereas the traceless shear $\bar q_{ij}$ is responsible for the angular variation at fixed distance.

In all cases, however, for a given direction $M_\star$, $\dot M_\star$ and $-z_\star$ will either all increase (if $r_iq_{ij}r_j<0$) or all decrease (if $r_iq_{ij}r_j>0$). Their increase will be fastest (or their decrease slowest) in the direction of $\bar q_3$, the least negative eigenvalue, and slowest in that of $\bar q_1$.
The rationale of this behaviour will always be that an increase of the conditional mean density will make it easier for excursion set trajectories to reach the threshold. Upcrossing will happen preferentially at smaller $\sigma$, corresponding to the formation of haloes of bigger mass. At fixed mass (fixed crossing scale $\sigma$), the crossing will happen preferentially with shallower slopes, corresponding to higher accretion rates and more recent formation (i.e. assembly of half mass).

\subsection{Walls}
\begin{figure}
  \centering
  \includegraphics[width=.9\columnwidth]{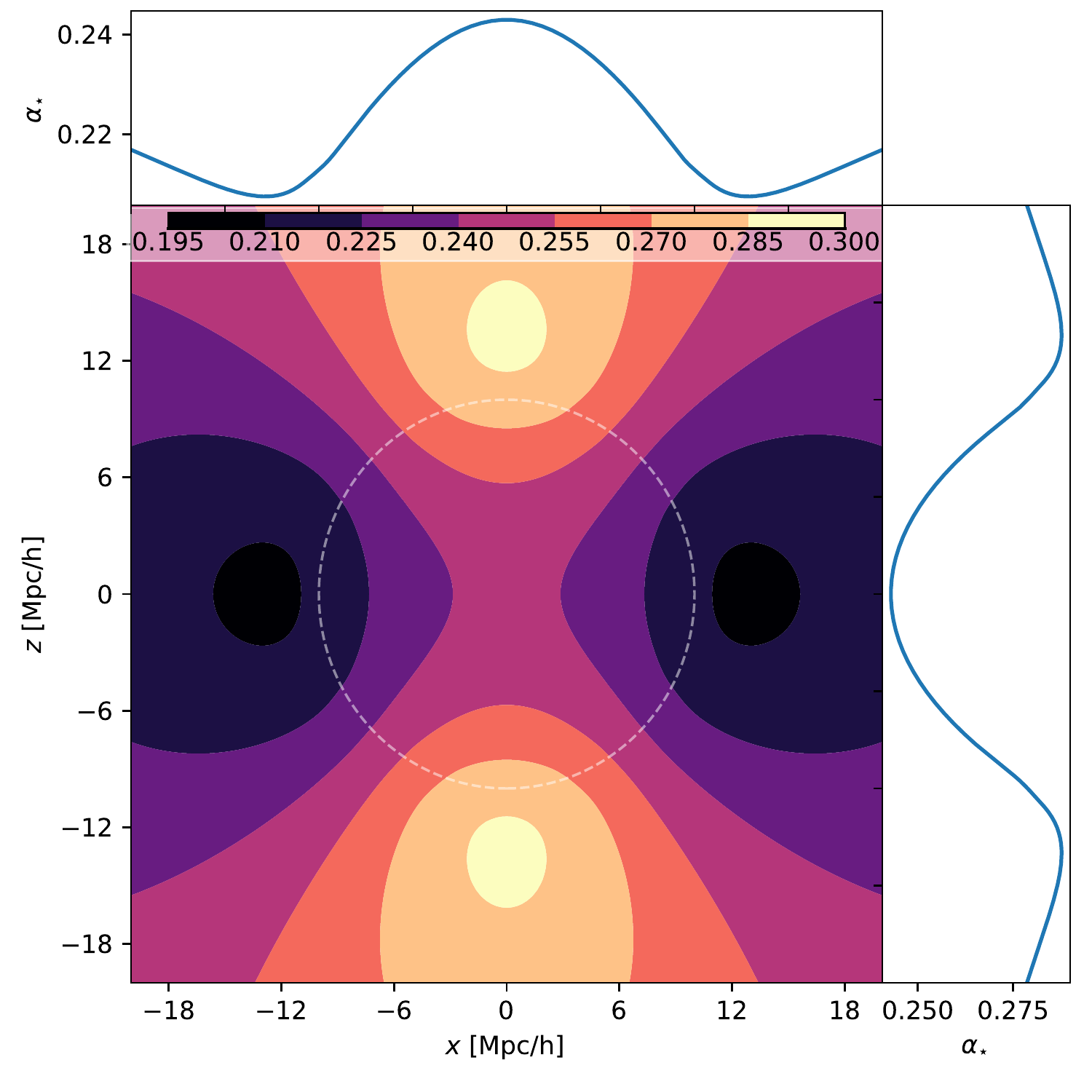}
  \caption{Isocontours in the $x-z$ plane of the typical accretion
    rate $\alpha_\star$ around a wall-type saddle point (at $(0,
    0)$). The saddle point is defined using the values of
    table~\ref{tab:qijbar_nu_values}. The profiles in the main
    direction of the wall ($z$ direction) and of the void ($x$
    direction) are plotted on the sides. The smoothing scale is
    $R=\SI{1}{Mpc/h}$.  The typical accretion
    rate is computed using a $\Lambda$CDM power
    spectrum. Similarly to what happens in filaments, haloes accrete
    more in the direction of the wall than in the direction of the
    void.}\label{fig:alpha_star_wall}
\end{figure}

A wall will form in correspondence of a saddle point of the potential
filtered on scale $\Rs$, for which $q_1< q_2<0<q_3$. This combination
of eigenvalue signs generates collapse in one spatial direction and
expansion in the other two. As argued, a saddle point of the potential
induces a saddle point of the opposite type in $M_\star$,
$\dot M_\star$ and $-z_\star$, which will increase along two space
directions following the increase of the mean density, and decrease
along one. Since for walls (like for filaments) the value of $\nus$
is likely to be smaller than $\sqrt{\tr(\bar q^2)}$, they will have
tend to have an angular modulation larger than the radial
angle-averaged variation. Walls are thus likely to be highly
anisotropic configurations also of the accretion rate and of the
formation time. This is illustrated for example in
Fig.~\ref{fig:alpha_star_wall} for the accretion rate. On average,
$\nus$ will be smaller for a wall-type saddle (which has two negative
eigenvalues) than for a filament-type one. Thus, haloes in walls tend
to be less massive, and at fixed mass they tend to have smaller
accretion rates and earlier assembly times.

\subsection{Voids}
A void will eventually form (although not necessarily by $z=0$) when  $\rr=0$ is a local maximum of the potential filtered on scale $\Rs$ (from which matter flows away), for which $q_1< q_2<q_3<0$. The centre of the void is a minimum of $M_\star$, $\dot M_\star$ and $-z_\star$. All these quantities will gradually increase with the separation.
As $|\nus|$ may be large (in particular for a large, early forming void), halo statistics in voids may not show a large anisotropy relative to their radial variation.
However, because voids have the most negative $\nus$, they are the environment with the least massive haloes, the smallest accretion rates and the earliest formation times (at fixed mass).

\subsection{Nodes }
Nodes form out of local minima of the gravitational potential, for which $0<q_1<q_2<q_3$ (corresponding to three directions of infall). The centre of the node is thus a maximum of $M_\star$, $\dot M_\star$ and $-z_\star$, all of which decrease with radial separation. Like voids, large early forming nodes (whose density $\nus$ must reach $\nuc$ when $\sigmas$ is very small) are relatively less anisotropic, since the relative amplitude of the angular variation induced by $\bar q_{ij}$ is likely to be small compared to the radial variation. Since $\nus$ is the largest for nodes, they host the most massive haloes, and at fixed mass those with the largest accretion rates and the latest formation times.

\section{PDF of saddles}
\label{sec:pdf-saddles}

\begin{table}
  \centering
  \renewcommand{\arraystretch}{1.1}
  \begin{tabular}{c|p{0.3cm}ccp{0.3cm}cc}
    & \multicolumn{3}{c}{Traceless tide} & Height & Scale & Saddle type\\
    \hline
    Quantity & $\bar{q}_1$ & $\bar{q}_2$ & $\bar{q}_3$ &  $\nus$  & $\Rs$ & \\
    \hline
    Value & $-0.7$ & $0.1$ & $0.6$ & $1.2$ & \SI{10}{Mpc/h} & filament-type \\
    Value & $-0.6$ & $-0.2$ & $0.8$ & $0$ & \SI{10}{Mpc/h} & wall-type
  \end{tabular}
  \caption{Eigenvalues $\bar q_i = q_i - \nus/3$ of the traceless tidal tensor
    $\bar{q}_{ij}$, height $\nus$ and smoothing scale used to define
    the saddle points. See Appendix~\ref{sec:pdf-saddles} for details.
}  \label{tab:qijbar_nu_values}
\end{table}

This section presents the distribution of the eigenvalues
of the anisotropic (i.e traceless) part of the tidal tensor at critical points of the potential
field. By definition, a critical point is such that $g_i=0$ and its
kind is given by the signature (the signs of the eigenvalues of the
hessian of the potential, $q_{ij}$): $+\!+\!+$ for a peak, $-\!+\!+$ for a
filament-type saddle point, $-\!-\!+$ for a wall-type saddle point and
$-\!-\!-$ for a void. Because the anisotropic tidal tensor reads
$\bar{q}_{ij} = q_{ij} - \delta_{ij} \nus /3$, the type of
the critical point is then given by the number of eigenvalues of
$\bar{q}_{ij}$ above $-\nus /3$.

The distribution of the eigenvalues of the (normalized) tidal tensor denoted $q_{1}<q_{2}<q_{3}$ is described by the Doroshkevich formula \citep{doroshkevich70,Pogosyanetal1998}
\begin{equation}
{p}(q_{i})=
\!
\frac{675 \sqrt 5
}{8\pi}
\!
\exp
\!\left[\!\frac {15} 2 I_{2}\!-\!3 I_{1}^{2}\!\right]\!
(q_{3}\!-\!q_{1})(q_{3}\!-\!q_{2})(q_{2}\!-\!q_{1}),
\end{equation}
where $\{I_{n}\}$ denotes the rotational invariants which define the characteristic polynomial of $q_{ij}$, namely its trace
$I_{1}=q_{1}+q_{2}+q_{3}$,
trace of the co-matrix
$I_{2}=q_{1}q_{2}+q_{2}q_{3}+q_{1}q_{3}$
and determinant $I_{3}=q_{1}q_{2}q_{3}$.
Subject to a filament-type saddle-point constraint, this PDF becomes
\begin{equation}
\label{eq:pqbarisad}
{p}(q_{i}|-\!+\!+)\!=\!\frac{540\sqrt{5\pi}}{29\sqrt 2+12 \sqrt 3}
q_{1}q_{2}q_{3}\vartheta(q_{2})\vartheta(-q_{1}){p}(q_{i})
,
\end{equation}
after imposing the condition of a saddle  $|\det q_{ij}|\delta_{D}(g_{i})\vartheta(q_{2})$ $\vartheta(-q_{1})$ for which as the acceleration is decoupled from the tidal tensor, only the condition on the sign of the eigenvalues and the determinant contribute.
From this PDF, it is straightforward to compute the distribution of saddles of heights $\nus=q_{1}+q_{2}+q_{3}$
\begin{equation}
{p}(\nus|-\!+\!+)={p^{+}}(\nus)\vartheta(\nus)+{p^{-}}(\nus)\vartheta(-\nus),
\end{equation}
with
\begin{align}
\nonumber
\!{p^{+}}\!(\nus)\!&=\!\!\frac{5 \sqrt{10 \pi } e^{\!-\!\frac{\nus^{2}}{2}}\!\!
   \left(3\nus\!-\!\nu^{3}_{s}\right) \!\text{Erfc}\!\left(\!\frac{\sqrt 5 \nus}{2\sqrt 2}\!
\right)\!\!+\!e^{-\frac{9 \nu^2_{s}}{8}}\!\!\left(32\!+\!155 \nu^2_{s}\right)\!}{\left(29
   \sqrt{2}\!+\!12 \sqrt{3}\right) \sqrt{\pi }},
\nonumber\\
\!{p^{-}}\!(\nus)\!&=\!\!\frac{ 5 \sqrt{10 \pi }  e^{\!-\!\frac{\nus^{2}}{2}}\!\!
   \left(3\nus\!-\!\nu^{3}_{s}\right) \!
   \text{Erfc}\!\left(\!\frac{-\sqrt 5 \nus}{\sqrt 2}\!
\right)\!\!+\!e^{-3 \nu^2_{s}}
\!\!\left(
32\!-\!10\nus^{2}\right)\!}{ (29 \sqrt{2}+12
   \sqrt{3}) \sqrt{\pi }}. \notag
\end{align}
In particular, the height of filament-type saddles has mean and standard deviation given by
\begin{align}
\mean{\nus|-\!+\!+} &={250}({3 \left(29 \sqrt{2}+12 \sqrt{3}\right) \sqrt{\pi }})^{-1}\approx 0.76\,,
\nonumber\\
{\rm Std}(\nus|-\!+\!+) &={\frac{\sqrt{\!696 \sqrt{6}\!+\!75\pi \left(10\!-\!3
   \sqrt{6}\right)\!-\!2114}}{15 \sqrt \pi}}\approx 0.55. \notag
   \end{align}
For other types of critical points, a similar calculation can be done. As expected, the heights of wall-type saddle points
 follow the same distribution as $-\nus$. Peak and void heights have mean $\pm \sqrt{{2114+696 \sqrt{6}}}/15\sqrt \pi\approx \pm 2.3$ and standard deviation $\sqrt{75\pi \left(10+3 \sqrt{6}\right)-({2114+696
   \sqrt{6}})}/15\sqrt \pi\approx 0.62$.

\begin{figure}
  \centering
     \includegraphics[width=0.9\columnwidth]{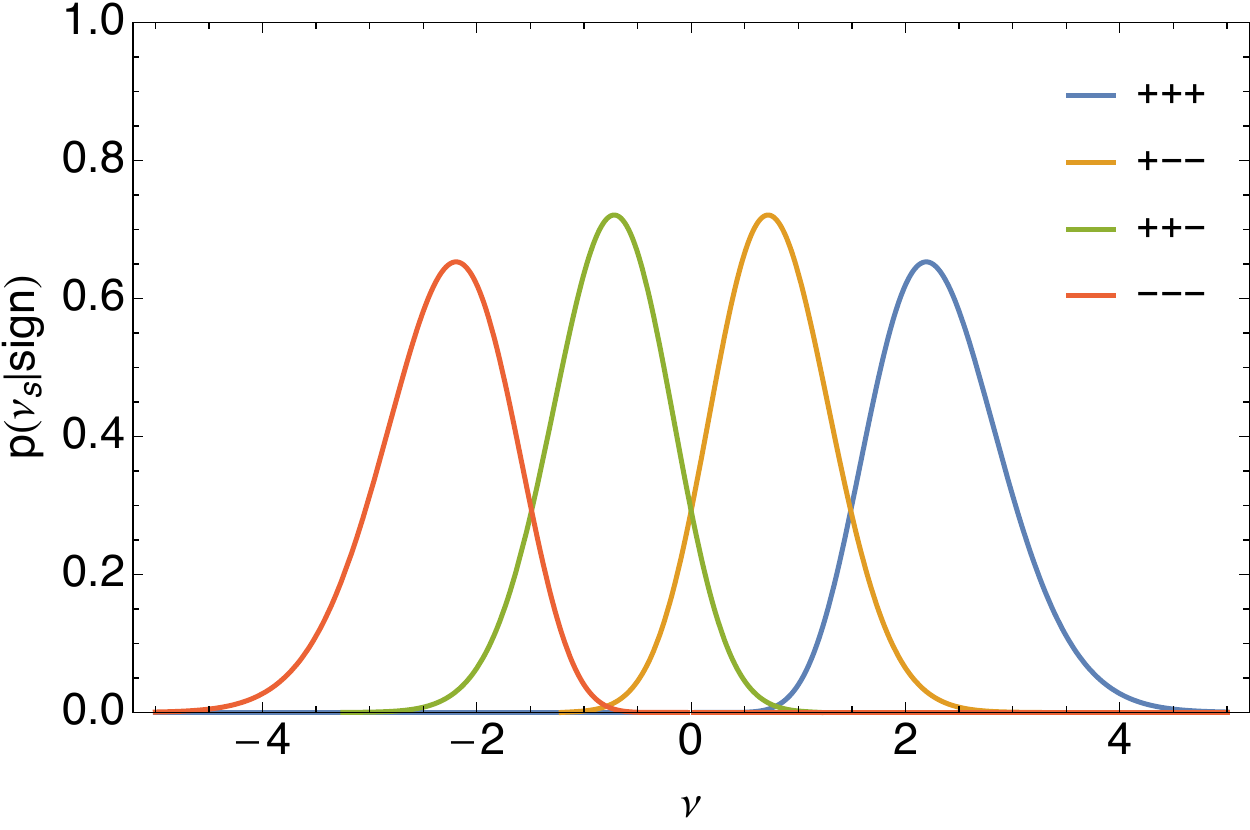}
    \includegraphics[width=0.9\columnwidth]{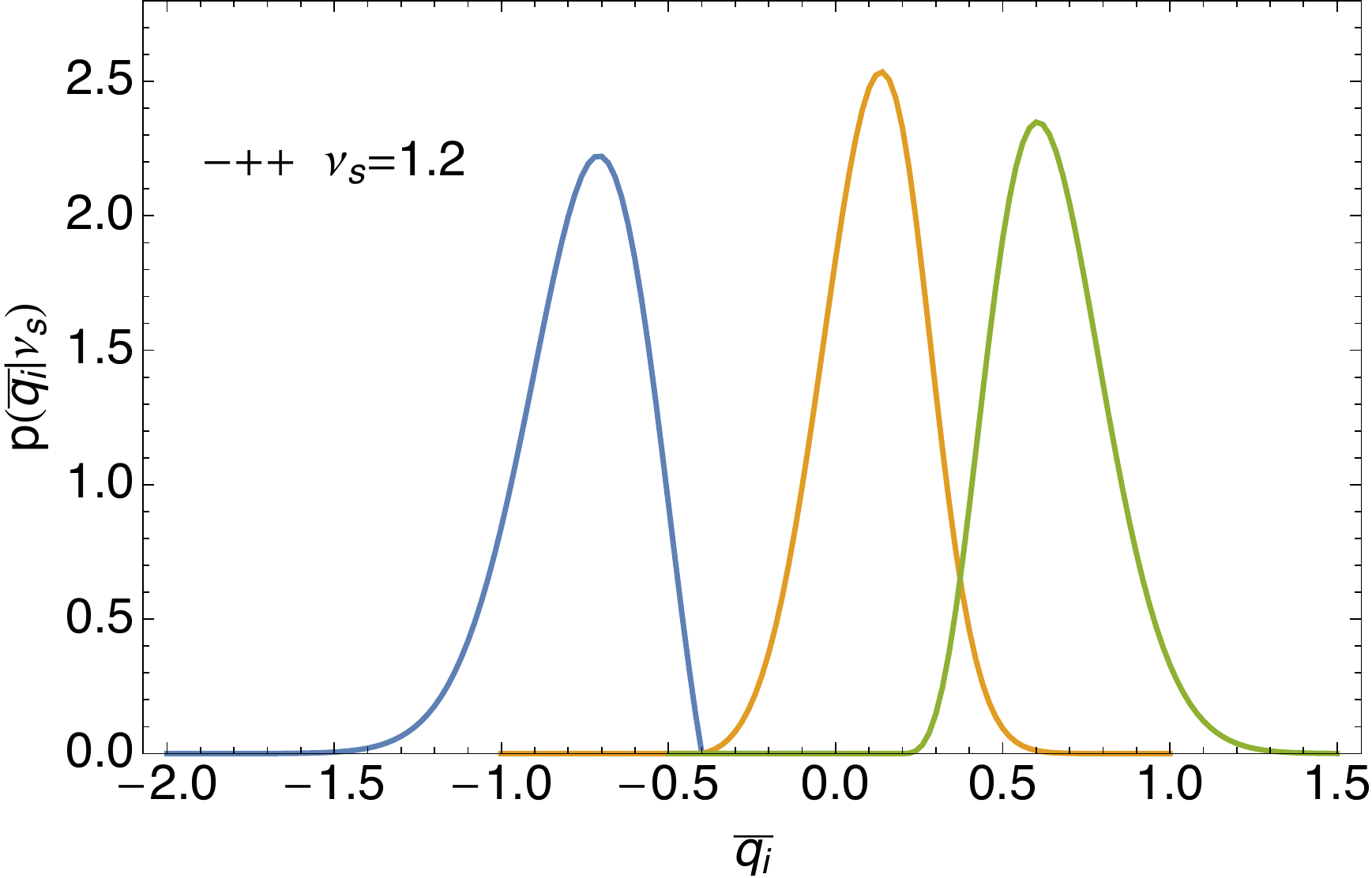}
  \includegraphics[width=0.9\columnwidth]{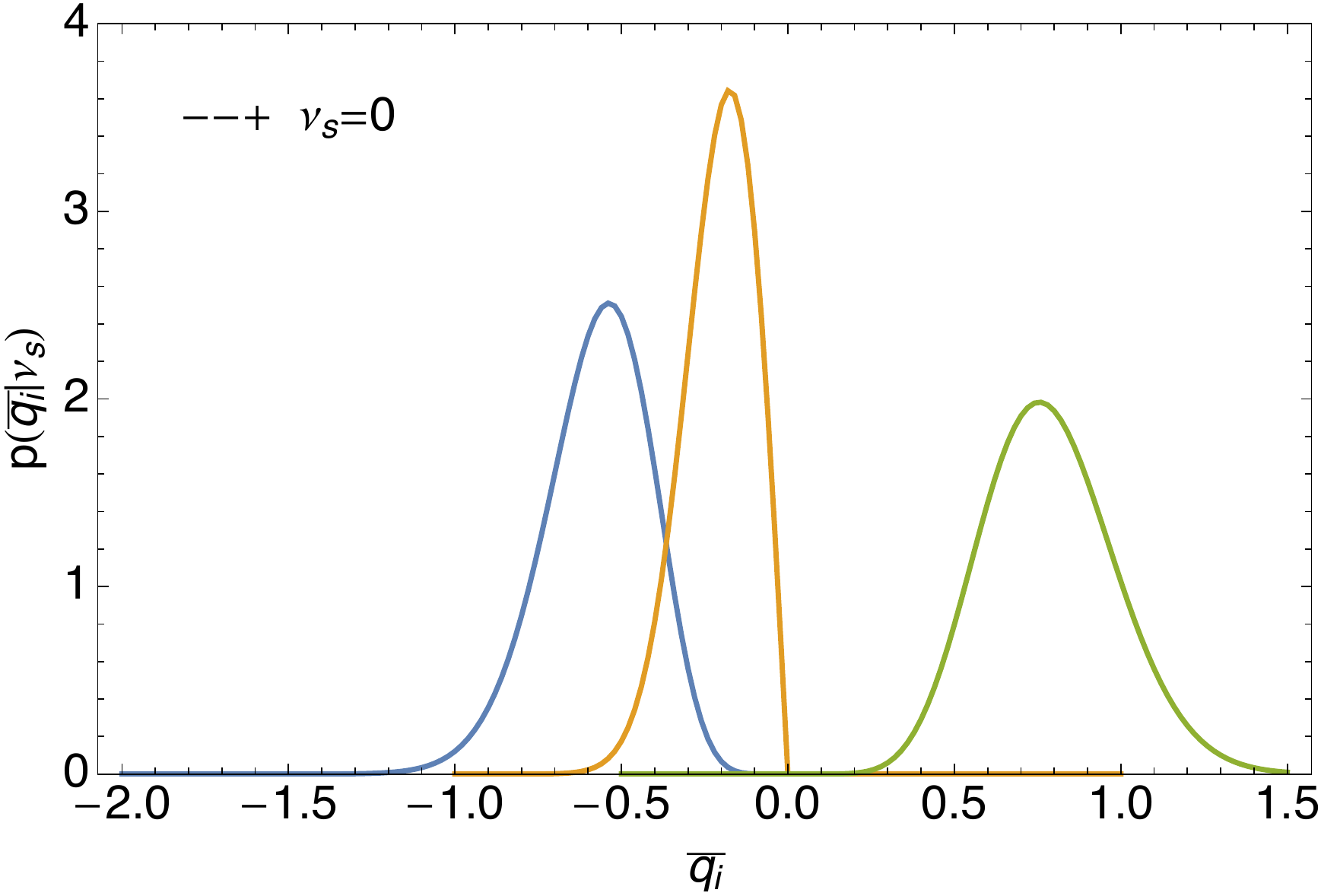}
  \caption{\emph{Top panel:} Distribution of heights of critical points of various signatures (peaks, filament-type saddles, wall-type saddles and voids) for GRF with any power spectrum. \emph{Middle panel:} PDF of the eigenvalues, $\bar q_1$ (blue), $\bar q_2$ (yellow) and $\bar q_3$ (green), of the anisotropic tidal tensor
 given a filament-type constraint at $\nus=1.2$. \emph{Bottom panel:} same as middle panel
for a wall-type constraint at $\nus=0$.}
  \label{fig:pdflambda}
\end{figure}

This work picks a typical value for the filament-type saddle at roughly one sigma from the mean  $\nus = 1.2$. For wall-type saddles,  $\nus = 0$ is chosen.
The distribution of eigenvalues of the anisotropic tidal tensor $\bar q_{i}$ for a filament-type saddle-point with a given positive\footnote{A similar expression can be obtained for negative heights.} height can then be easily obtained from equation~(\ref{eq:pqbarisad})
\begin{equation}
\nonumber
{p}(\bar q_{1}|\nus)\!=\!\frac{15(3\bar q_{1}\!+\!\nus)\!\left[a_{1}e^{-\frac{4\nus^{2}}{3}+\frac{5}{2}\bar q_{1}\nus-\frac{15\bar q^{2}_{1}}{2}}\!-\!a_{2}e^{-\frac{\nus^{2}}{2}\!-\!\frac{45\bar q^{2}_{1}}{8}}\right]\!}{16(29 \sqrt{2}+12
   \sqrt{3}) \sqrt{\pi }{\cal P^{+}}\!(\nus)}\,,
\end{equation}
where $\bar q_{1}<-\nus/3$ and $a_{1}$ and $a_{2}$ are two polynomials of $\bar q_{1}$ and $\nus$ given by
\begin{equation}\nonumber
a_{1}(\bar q_{1},\nus)=32\left[5|\nus-6 \bar q_{1}| (3 \bar q_{1}+\nus)+12\right]\,,
\end{equation}
and
\begin{equation}\nonumber
a_{2}\!=\!6075 \bar q^{4}_{1}-8100 \bar q^{3}_{1} \nus+900 \bar q^{2}_{1} (3 \nus^{2}-4)+480 \bar q_{1} \nus-160 \nus^{2}+384.
\end{equation}
Similarly, the PDF of the intermediate and major eigenvalues are respectively given by
\begin{equation}
\nonumber
{p}(\bar q_{2}|\nus)\!\!=\!\!\frac{15(3\bar q_{2}\!+\!\nus)a_{1}e^{-\frac{11}{12}\nus^{2}+\frac{5}{4}\bar q_{2}\nus-15\bar q^{2}_{2}-\frac 5 {12}(\nus+3\bar q_{2})|\nus-6\bar q_{2}|}}{16(29 \sqrt{2}+12
   \sqrt{3}) \sqrt{\pi }{\cal P^{+}}\!(\nus)}
\end{equation}
where $\bar q_{2}>-\nus/3$ and $a_{1}=a_{1}(\bar q_{2},\nus)$, and
\begin{equation}
\nonumber
{p}(\bar q_{3}|\nus)\!=\!\frac{15(3\bar q_{3}\!+\!\nus)\!\left[a_{1}e^{-\frac{\nus^{2}}{2}-\frac{45\bar q^{2}_{3}}{2}}\!+\!\bar a_{1}e^{-\frac{4\nus^{2}}{3}+\frac{5}{2}\bar q_{3}\nus-\frac{15\bar q^{2}_{3}}{2}}\right]\!}{16(29 \sqrt{2}+12
   \sqrt{3}) \sqrt{\pi }{\cal P^{+}}\!(\nus)}
\end{equation}
where $\bar q_{3}>\nus/6$, having defined $a_{1}=a_{1}(\bar q_{3},\nus)$
  and
$\bar a_{1}(\bar q_{3},\nus)$ $=\!-a_{1}(\!-\bar q_{3},\!-\nus)$.
Similar expressions can be obtained for wall-type saddles (together with peaks and voids).
The top
panel of Fig.~\ref{fig:pdflambda} shows the distribution of eigenvalues for a filament-type saddle point of height $\nus=1.2$ and the
bottom panel the distribution for a wall-type saddle point of height $\nus=0$. Typical values
of $\bar{q}_{ij}$ were selected to correspond roughly to the maximum of the above-mentioned distributions of $\bar{q}_1,\bar{q}_2,\bar{q}_3$ and are reported in table~\ref{tab:qijbar_nu_values}. Note that all the results obtained in this section are independent of the power spectrum. The only assumption is that the density is a gaussian random field.

\section{Covariance matrices}
\label{sec:covariance}

Let us present here the covariance matrix of all
variables introduced in the main text.
The density $\delta$ and slope $\delta'$  are
evaluated at position $\rr$ and smoothed on the halo scale
$R$, the half-mass density $\delta_{\half}$ is also evaluated at the halo position $\rr$ but smoothed on $R_{\half}= 2^{-{1/3}}R$ while the saddle rareness $\nus$, acceleration $g_i$ and detraced tidal tensor $\bar q_{ij}$ are evaluated
at the origin and smoothed on a scale $\Rs\gg R$.  The
correlation matrix of
$\mathbf{X}\equiv\left\lbrace \delta, \delta', \nu_{\half},\nus, g_i, \bar\shear_{ij}\right\rbrace $, a vector with 12 Gaussian components, is
\begin{equation}
  \mathbf{C} =
  \begin{pmatrix}
    \sigma^2        & \sigma          &       \omega      & \mathbf{C}_{14} & \mathbf{C}_{15} & \mathbf{C}_{16}\\
    \sigma          & \mean{\delta'^2} &  \omega'     & \mathbf{C}_{24} & \mathbf{C}_{25} & \mathbf{C}_{26}\\
   \omega            &  \omega'       &   \sigma_{\half}^2   &  \mathbf{C}_{34} & \mathbf{C}_{35} & \mathbf{C}_{36} \\
    \mathbf{C}_{14} & \mathbf{C}_{24} & \mathbf{C}_{34}   &          1           &  0              & 0 \\
    \mathbf{C}_{15}^T & \mathbf{C}_{25}^T & \mathbf{C}_{35}^T & 0               & \mathbf{C}_{55} & 0\\
    \mathbf{C}_{16}^T & \mathbf{C}_{26}^T & \mathbf{C}_{36}^T & 0               & 0               & \mathbf{C}_{66}
  \end{pmatrix}\,,
\label{eq:Cmat}
\end{equation}
with $ \omega = \mean{\delta\nu_\half}$, $\omega' = \mean{\delta'\nu_\half}$ and
\begin{align}
  \mathbf{C}_{14} &= \mean{\delta\nus} = \xi_{00},\quad
  \mathbf{C}_{15} = \mean{\delta g_i} =
  \frac{r_{i}}{R_{\star}}\xi_{11}, \\
  \mathbf{C}_{16} &= \mean{\delta\bar\shear_{ij}} =
  \left(\frac{\kron_{ij}}{3} - \hat{r}_i\hat{r}_j\right)\xi_{20}, \\
  \mathbf{C}_{24} &= \mean{\delta'\nus} = \xi'_{00}, \quad
  \mathbf{C}_{25} = \mean{\delta' g_{i}}
  = \frac{r_{i}}{R_{\star}}\xi'_{11}, \\
  \mathbf{C}_{26} &= \mean{\delta'\bar\shear_{ij}} =
  \left(\frac{\kron_{ij}}{3} - \hat{r}_i\hat{r}_j\right)\xi'_{20}, \\
  \mathbf{C}_{34} &= \mean{\nu_{\half}\nus}=
  \frac{\xi_{00}^{(\half)}}{\sigma_\half}, \,
  \mathbf{C}_{35} \!=\! \mean{\delta_{\half}g_{i}} \!=\!
  \frac{r_{i}}{R_{\star}}\frac{\xi_{11}^{(\half)}}{\sigma_\half}, \\
  \mathbf{C}_{36} &= \mean{\delta_{\half}\bar\shear_{ij}} =
  \left(\frac{\kron_{ij}}{3} - \hat{r}_i\hat{r}_j\right)\frac{\xi_{20}^{(\half)}}{\sigma_\half}, \\
  \mathbf{C}_{55} &= \mean{g_ig_{j}} = \frac{\kron_{ij}}{3}, \quad
  \mathbf{C}_{66} = \mean{\bar\shear_{ij}\bar\shear_{kl}} =
  \frac{2P_{ij,kl}}{15}\,.\label{eq:qijqkl}
 \end{align}
Hence, $\mathbf{C}_{14}$, $\mathbf{C}_{24}$ and $\mathbf{C}_{34}$ are scalars, $\mathbf{C}_{15}$, $\mathbf{C}_{25}$ and $\mathbf{C}_{35}$ are 3-vectors, $\mathbf{C}_{16}$, $\mathbf{C}_{26}$ and $\mathbf{C}_{36}$ are $3\times3$ traceless matrices (or 5-vectors in the space of symmetric traceless matrices), $\mathbf{C}_{55}$ is a $3\times3$ matrix,
$\mathbf{C}_{66}$ is a $5\times5$ matrix. The matrix   $\mathbf{C}_{66}$  involves
\begin{equation}
  P_{ij,kl}\equiv
  \frac{\delta_{ik}\delta_{jl}+\delta_{il}\delta_{jk}}{2}
  -\frac{\delta_{ij}\delta_{kl}}{3}\,,
\end{equation}
a projector that removes the trace and the anti-symmetric part from
a matrix. Since $P_{ij,ab}P_{ab,mn}=P_{ij,mn}$ and so
$P_{ij,mn}^{-1}= P_{ij,mn}$, it acts as the identity in the space of
symmetric traceless matrices. $P_{ij,kl}$ can be written in its matrix
form by numbering the pairs
$\lbrace (1,1), (2,2), (1,2), (1,3), (2,3) \rbrace$ from $1$ to $5$,
the dimensionality of the space, resulting in a $5\times 5$
matrix. The element $(3, 3)$ has been dropped because it is linearly
linked to $(1,1)$ and $(2, 2)$. The explicit value of
$\mathbf{C}_{66}$ is therefore
\begin{equation}
  \mathbf{C}_{66} = \frac{1}{45}
  \begin{pmatrix}
    4 & -2 & 0 & 0 & 0 \\
    -2 & 4 & 0 & 0 & 0 \\
    0 & 0 & 3 & 0 & 0 \\
    0 & 0 & 0 & 3 & 0 \\
    0 & 0 & 0 & 0 & 3 \\
  \end{pmatrix}.
\end{equation}

\begin{figure*}
  \centering
  \includegraphics[width=\textwidth]{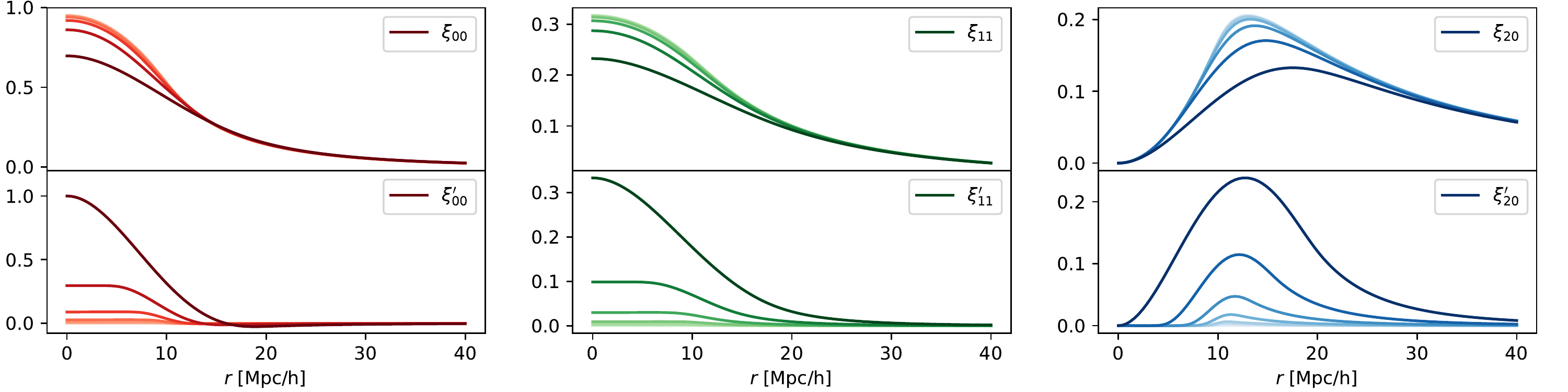}
  \caption{Plot as a function of $r$ of the correlation functions
    defined in equation~\eqref{eq:xi_xiprime}. From \emph{left to right on
    the top row} $\xi_{00}$, $\xi_{11}$ and $\xi_{20}$. The \emph{bottom row}
    shows the same quantities derived w.r.t. $\sigma$. The correlation
    functions are evaluated at $\Rs=\SI{10}{Mpc/h}$ for different
    values of $R$ logarithmically spaced between \SI{10^{-1}}{Mpc/h}
    (light colour) and \SI{10}{Mpc/h} (dark colours) with a $\Lambda$CDM
    power spectrum and plotted as a function of the distance $r$.}
  \label{fig:xifunctions}
\end{figure*}

The finite separation correlation
functions $\xi_{\alpha\beta}(r, R, \Rs)$ and $\xi'_{\alpha\beta}(r, R, \Rs)$ are defined as
\begin{align}
\label{eq:xi_xi}
   \xi_{\alpha\beta} &\equiv \int \d k \frac{k^2P(k)}{2\pi^2}
  W(kR)\frac{W(k\Rs)}{\sigmas}
  \frac{j_\alpha(kr)}{(kr)^\beta}, \\ 
  \xi'_{\alpha\beta} & \equiv
  \int \d k \frac{k^2P(k)}{2\pi^2} W'\!(kR)
  \frac{W(k\Rs)}{\sigmas}\frac{j_\alpha(kr)}{(kr)^\beta}\,,
  \label{eq:xi_xiprime}
\end{align}
where $W'(kR)=[\d W(kR)/\d R]/(\d\sigma/\d R)$.
Similarly, the correlation functions at the two different mass scales $M$ and $M/2$ are
\begin{equation}
  \xi_{\alpha\beta}^{(\half)} \equiv \xi_{\alpha\beta}(r, R_\half, \Rs)\,,
\end{equation}
where $R_\half\equiv R/2^{1/3}$.
At null separation ($r=0$), it yields
\begin{align}
  &\omega = \frac{\mean{\delta\delta_\half}}{\sigma_{\half}}
  = \int\d k \frac{k^2P(k)}{2\pi^2}
  W(kR)\frac{W(kR_{\half})}{\sigma_{\half}}\,,
\label{eq:omega}
  \\
  &\omega'\!= \frac{\mean{\delta'\!\delta_\half}}{\sigma_{\half}}
  = \int\d k \frac{k^2\!P(k)}{2\pi^2}
  W'(kR)\frac{W(kR_{\half})}{\sigma_{\half}}\,.
\label{eq:omegaprime}
\end{align}
Recall that for a Top-Hat filter one has
\begin{equation}
  W(kR) =\frac{3j_1(kR)}{kR} \quad \mathrm{and} \quad
  W'(kR) = \frac{3j_2(kR)}{R |\d\sigma/\d R|} \,,
\label{eq:WTH}
\end{equation}
and notice that $W'(kR)$ is suppressed by a factor of $k^2R^2$ w.r.t. $W(kR)/\sigma$ when $k\ll1/R$. In fact, in this limit $j_n(kR)\sim (kR)^n/(2n+1)!!$.
Hence, the action of $\dd/\dd\sigma$ is proportional to that of $R^2\nabla^2$, and $\sigma \xi'_{\alpha\beta}\propto R^2\nabla^2\xi_{\alpha\beta}\sim(R/\Rs)^2\xi_{\alpha\beta}$. It follows that for $R\ll \Rs$ one has $\sigma\xi_{\alpha\beta}'\ll\xi_{\alpha\beta}$. In presence of a strong hierarchy of scales, the terms containing $\xi_{\alpha\beta}'$ are negligible (see Fig.~\ref{fig:xifunctions}).

For a scale invariant power spectrum $P(k)=A(k/k_0)^{-n}$
$\xi_{\alpha\beta}$ and $\xi'_{\alpha\beta}$ have an analytical
expression that depends on the relation between $r, \Rs$ and $R$. For
example, when $\Rs > r+ R$:
\begin{gather}
  \begin{split}
    \frac{\xi_{\alpha\beta}(r, R, \Rs)}{\sigmas}\!=\! B
  \mathrm{F}_4\!\!\left(\frac{\!\alpha\!-\!\beta\!-n}{2}\!,\!\frac{3\!+\!\alpha\!-\!\beta\!-\!n}{2}\!;\!\frac{5}{2}\!,\!\alpha\!+\!\frac{3}{2}\!;\!\frac{R^2}{\Rs^2}\!,\!\frac{r^2}{\Rs^2}\right)\notag
 \end{split}
\end{gather}
and
\begin{gather}
  \begin{split}
     \xi'_{\alpha\beta}&(r, R, \Rs) = \frac{2 (\alpha \!-\!\beta \!-\!n\!+\!3) (n\!-\!\alpha \!+\!\beta) }{5 (n\!-\!3)} \left(\!\frac R {R_{\cal S}}\!\right)^{\!\!\!\frac{7\!-\!n}{2}}B\\
     &\times \mathrm{F}_4\!\!\left(\!\frac{2\!+\!\alpha\!-\!\beta\!-n}{2}\!,\!\frac{5\!+\!\alpha\!-\!\beta\!-n}{2}\!;\! \frac{7}{2}\!,\!\alpha\!+\!\frac{3}{2}\!;\!\frac{R^2}{\Rs^2}\!,\!\frac{r^2}{\Rs^2}\!\right)\notag\,,
  \end{split}
\end{gather}
where $\mathrm{F}_4$ is the Appell Hypergeometric function of the fourth kind
\citep[p.677]{gradshteyn2007}\footnote{\url{http://mathworld.wolfram.com/AppellHypergeometricFunction.html}},
while
\begin{equation}
  B = -\!\left(\!\frac{r}{\Rs}\!\right)^{\!\!\alpha\!-\!\beta}\!\!\!\!\!\frac{\pi(n\!+\!3) \csc\!\left(\frac{n \pi}{2}\right) \Gamma\!\left(\frac{3\!+\!\alpha\!-\!\beta\!-\!n}{2}\right)}
  {2^{\beta\!+\!2 n\!+\!2}3(n\!-\!1) \Gamma\!\left(\frac{3\!+\!2\alpha}{2}\right) \Gamma\!(-n\!-\!1) \Gamma\!\left(\frac{n\!-\!\alpha\!+\!\beta\!+\!2}{2}\right)}
\notag
\end{equation}
and
\begin{equation}
  \label{eq:sigmapowerlaw}
  \sigma^2(R) = \sigma_8^2\left(\frac{R}{R_8}\right)^{n-3},
    \ \ \frac{\dd\log\sigma^2}{\dd \log R} = n-3, 
\end{equation}
where $R_8=\SI{8}{Mpc/h}$ and $\sigma_8=\sigma(R_8)$ are normalization factors.
For the same power law power spectrum, setting $\alpha=1+n$ and $\beta = R_{\half}/R = 2^{-{1/3}}$, $\omega$ and $\omega'$  defined in \eqref{eq:omega} and \eqref{eq:omegaprime} have the analytical expressions
\begin{equation}
  \frac{\omega}{\sigma} =
  \frac{
 (1\!+\! {\beta})^{\alpha } \!\left( \beta^{2}\!-\!\alpha   {\beta}
    \!+\! 1\right) - ( 1\!-\! \beta)^{\alpha }\!
   \left( \beta^{2}\!+\!\alpha   \beta\!+\! 1\right)}{2^{\alpha }(2
  \!-\!   \alpha)  \beta^{\frac{\alpha+2 }{2}}}\,,
\end{equation}
and
\begin{multline}
  \omega' =
  \frac{ \left(3 \beta ^3\!+\!\beta n^2\!+\!3 \beta^2 n\!+\!n\right)\! (1\!-\!\beta )^n\!}{2^{n} \beta ^{\frac{n\!+\!3}{2} }(n\!-\!3) (n\!-\!1)} \\
  +\frac{\!\left(3 \beta ^3\!+\!\beta n^2\!-\!3 \beta
      ^2n\!-\!n\right)(1\!+\!\beta)^n }{2^{n} \beta ^{\frac{n\!+\!3}{2} }(n\!-\!3)
    (n\!-\!1)}\,.
\end{multline}

\section{Conditional statistics}
\label{sec:statistics}
The goal of this Section is to derive explicitly the conditional statistics needed in the paper. Assuming that the underlying density field obeys Gaussian statistics,  the PDF of the 12-dimensional vector $\mathbf{X}\equiv\left\lbrace \delta(\rr), \delta'(\rr), \nu_{\half}(\rr),\nus, g_i, \bar\shear_{ij}\right\rbrace $ already defined in Appendix~\ref{sec:covariance} involves inverting the $12\times12$ covariance matrix $\mathbf{C}\equiv \langle  \mathbf{X}\cdot \mathbf{X}^{\rm T}\rangle $, given by equation~\eqref{eq:Cmat}.
Since however the focus here is on conditioning heights and slopes, which are scalar quantities, their correlation with the saddle is the correlation with the 3 unit-variance Gaussian components
\begin{equation}
  \S(\hat\rr) \equiv 
  \{\nus,\sqrt{3}\hat r_ig_i r/R_\star,-\sqrt{5}(3\hat r_i\bar q_{ij} \hat r_j/2)\}\,.
\end{equation}
Hence the  6-dimensional vector $\mathbf{\tilde X}\equiv \lbrace \delta(\rr), \delta'(\rr), \nu_{\half}(\rr),\S \rbrace $ is sufficient, and has a 6$\times$6 covariance matrix given by
\begin{equation}
  \mathbf{\tilde C}(r) =
  \begin{pmatrix}
    \sigma^2   & \sigma    &   \omega    &  \xi(r) \\
    \sigma     & \mean{\delta'^2} & \omega'   &  \xi'(r) \\
    \omega     & \omega'   &  \sigma_{\half}^2  &  \xi_{\half}(r) \\
    \xi^T(r)   & \xi'^T(r)  & \xi_{\half}^T(r)   &  1\!\!1_{3\times3}
  \end{pmatrix},
\label{eq:tildeCmat}
\end{equation}
where
\begin{equation}
\begin{aligned}[c]
  \xi(r) &\equiv
  \{\xi_{00},\sqrt{3}\xi_{11}r/R_\star,\sqrt{5}\xi_{20}\}\,,
\\
  \xi'\!(r) &\equiv
  \{\xi_{00}',\sqrt{3}\xi_{11}'r/R_\star,\sqrt{5}\xi_{20}'\}\,,
\\
  \xi_{\half}(r) &\equiv
  \{\xi_{00}^{(\half)},\sqrt{3}r/R_\star\xi_{10}^{(\half)}, \sqrt{5}\xi_{20}^{(\half)}\}/\sigma_{\half} \,.
\end{aligned}
\label{eq:compact_notation}
\end{equation}
The PDF of $\mathbf{\tilde X}$ is the 6-variate Gaussian
\begin{equation}
  p_\G(\mathbf{\tilde X}) =
  \frac{1}{(2\pi)^3\sqrt{{\rm det} \mathbf{\tilde C}}}
  \exp\left(-\frac{1}{2} \mathbf{\tilde X}\cdot \mathbf{\tilde C}^{-1}\cdot \mathbf{\tilde X}\right) \,,
\end{equation}
so that in each case the task is to invert the appropriate section of the covariance matrix $\mathbf{\tilde C}\equiv \langle  \mathbf{\tilde X}\cdot \mathbf{\tilde X}^{\rm T}\rangle $, marginalizing over the variables that are not involved.

\subsection{The general conditional case}
\label{sec:condstats}

To speed up the computation of conditional statistics, rather than doing a brute force block inversion of $\mathbf{\tilde C}$, it is best to use the decorrelated variables
\begin{equation}
  \nu_v\equiv
  \frac{\delta-\condmean{\delta}{\{v\}}}{\sqrt{\Var{\delta}{\{v\}}}}\,,
  \quad\mathrm{and}\quad
  \nu_v'\equiv\frac{\dd\nu_v}{\d\sigma}\,,
\label{eq:genvar}
\end{equation}
where the possible $\{v\}$ considered in this work are $\nu_\half$, $\S$ or $\{\nu_\half,\S\}$.
By construction, $\nu_v$ and $\nu_v'$ are uncorrelated, because $\nu_v$ has unit variance. Furthermore, if each $v_I$ is independent of $\sigma$ (as it will be the case in the following), $\nu_v'$ does not correlate with the constraint either, since $\mean{\nu_v'v_I}=\mean{\nu_vv_I}'=0$. Then, being a linear combination of $\delta'$, $\nu$ and $\{v\}$ that does not correlate with $\nu$ nor $v_I$, $\nu_v'$ must be proportional to $\delta'-\condmean{\delta'}{\nu,\{v\}}$ (the only such linear combination by definition), and $\mean{\nu_v'^2}$ to $\Var{\delta'}{\nu,\{v\}}$. That is,
\begin{gather}
\begin{aligned}[b]
  \condmean{\delta'}{\nu,\{v\}} &= \delta'
  -\sqrt{\Var{\delta}{\{v\}}}\,\nu_v'\,, \\
  &= \condmean{\delta'}{\{v\}}
  +\frac{[\Var{\delta}{\{v\}}]'}{2\Var{\delta}{\{v\}}}
  \left(\delta-\condmean{\delta}{\{v\}}\right)\,,
\end{aligned}\label{eq:gencondmean}  \\
\begin{aligned}[b]
  \Var{\delta'}{\nu,\{v\}} &=
  \Var{\delta}{\{v\}}\mean{\nu_v'^2}\,,\\
  &= \Var{\delta'}{\{v\}}-
  \frac{[\Var{\delta}{\{v\}}]'^2}{4\Var{\delta}{\{v\}}}\,,
\end{aligned}\label{eq:gencondvar}
\end{gather}
providing the conditional statistics of $\delta'$ given $\nu$ and $\{v\}$ in terms of those of $\delta$ and $\delta'$ given $\{v\}$ alone. Since $[\Var{\delta}{\{v\}}]'=2\Cov{\delta,\delta'}{\{v\}}$, these formulae reduce to the standard results for constrained Gaussian variables, but taking derivatives makes their calculation easier.

To compute $\nu_v$ and $\nu_v'$ explicitly, one needs to insert (using Einstein's convention on repeated indices)
\begin{align}
 \condmean{\delta}{\{v\}} &= \psi_I C_{IJ}^{-1} v_J
\label{eq:genmeannu}\,,  \\
  \Var{\delta}{\{v\}} &= \sigma^2-\psi_I C_{IJ}^{-1} \psi_J
\label{eq:genvarnu} \,,
\end{align}
in equation~\eqref{eq:genvar},
where $C_{IJ}\equiv\mean{v_I v_J}$ is the covariance matrix of the constraint, and $\psi_I\equiv\mean{\delta v_I}$ is the mixed covariance. The conditional statistics obtained from equations \eqref{eq:gencondmean}-\eqref{eq:gencondvar} are then
\begin{align}
  \condmean{\delta'}{\nu,\!\{v\}} &= \psi_I' C_{IJ}^{-1} v_J +
  \frac{\sigma-\psi_I' C_{IJ}^{-1}\psi_J}{\sqrt{\sigma^2-\psi_I C_{IJ}^{-1}\psi_J}}\nu_v
\label{eq:genmeannuprime} \,, \\
\hskip -0.2cm   \Var{\delta'}{\nu,\!\{v\}} &\!=\! \langle\delta'^2\rangle - \psi_I' C_{IJ}^{-1} \psi_J'
  \!-\! \frac{(\sigma-\psi_I' C_{IJ}^{-1} \psi_J)^2}{\sigma^2-\psi_I C_{IJ}^{-1} \psi_J}
\label{eq:genvarnuprime},
\end{align}
(where $\nu_v$ is given by equation~\eqref{eq:genvar}) from which
one can evaluate \eqref{eq:genfup}-\eqref{eq:X}, after setting
$\delta=\dc$.  Since $\mean{\delta'|\nuc}=\nuc$ and
$\Var{\delta'}{\nuc}=1/\Gamma^2$, equation~\eqref{eq:fup2} is
recovered in the unconstrained case.  For later convenience, let us
also note that the conditional probability of $\nu$ and $\nu'$ given
the constraint $\{v\}$ is
\begin{equation}
  p_{\rm G}(\nu,\nu'|\{v\}) =  \sigma\,
  \frac{p_{\rm G}(\nu_v)\,p_{\rm G}(\delta'-\condmean{\delta'}{\nuc,\{v\}})}{\sqrt{1-\psi_I C_{IJ}^{-1} \psi_J/\sigma^2 }}\,,
\label{eq:gencondp}
\end{equation}
since by construction $\nu_v$ and $\delta'-\condmean{\delta'}{\nuc,\{v\}}\propto\nu_v'$ are independent.

\subsection{Conditioning to the saddle}
\label{sec:dgivenS}

Equation~\eqref{eq:genmeannu} and its derivative guarantee that conditioning on the values of $\S$ (that is, fixing the geometry of the saddle) returns
\begin{equation}
\begin{aligned}[c]
  \condmean{\delta}{\S} &= \xi\!\cdot\!\S
  \;,\quad
  \Var{\delta}{\S}= \sigma^2-\xi^2 \,,\\
  \condmean{\delta'}{\S}&= \xi'\!\cdot\!\S
  \;,\quad
  \Var{\delta'}{\S}= \mean{\delta'^2}-\xi'^2\,, \\
  \condmean{\nu_{\half}}{\S}&= \xi_{\half}\!\cdot\!\S
  \;,\quad
  \Var{\nu_{\half}}{\S}= 1-\xi_{\half}^2\,.
\end{aligned}
\label{eq:nu|S}
\end{equation}
To make the equations less cluttered, here and in the following,  scalar products of these vectors are denoted with a dot, rather than in Einstein's notation. Equation~\eqref{eq:nu|S} effectively amounts to replacing in all unconditional expressions
\begin{equation}
\begin{aligned}
  \delta &\to \delta-\xi\!\cdot\!\S \,,\\
  \delta' &\to \delta'-\xi'\!\cdot\!\S \,,\\
  \nu_{\half} &\to\nu_{\half}-\xi_{\half}\!\cdot\!\S \,,
\end{aligned}
\label{eq:effvarS}
\end{equation}
reducing the problem to three zero-mean variables that no longer correlate with $\S$ (but still do with each other!).
The covariance of $\delta$, $\delta'$ and $\nu_\half$
at fixed $\S$ reads
\begin{align}
  \Cov{\delta, \delta'}{\S}
  & = \sigma -\xi\!\cdot\!\xi'\,, \notag \\
  \Cov{\delta, \nu_{\half}}{\S}
  &= \omega -\xi\!\cdot\!\xi_{\half}\,, \label{eq:nux|S} \\
  \Cov{\delta', \nu_{\half}}{\S}
  &= \omega' -\xi'\!\cdot\!\xi_{\half}\,, \notag
\end{align}
with $\omega$ and its derivative $\omega'$ given by equation~\eqref{eq:omega} and \eqref{eq:omegaprime}.
The first equation in \eqref{eq:nux|S} is one half the derivative of $\Var{\delta}{\S}$ w.r.t. $\sigma$ from equation~\eqref{eq:nu|S}, consistently with taking the conditional expectation value of the relation $\delta\delta'=(1/2)\d\delta^2/\d\sigma$. The third is the derivative of the second, since $\xi_\half$ depends on $\sigma_\half$ and not on $\sigma$ (the relation between the two scales arising since $\sigma_{\half}=\sigma(M/2)$ should be imposed after taking the derivative).

\subsection{Slope given height at distance $\rr$ from the saddle}
\label{sec:upgivenS}

The saddle point being fixed, it can now be assumed that the excursion
set point is at the critical overdensity $\nu = \nuc$.
The conditional mean and variance of the slope are then
\begin{align}
  \condmean{\delta'}{\nuc,\S}
  &= \condmean{\delta'}{\S}
  + \frac{\Cov{\delta',\delta}{\S}}{\Var{\delta}{\S}}
  \left(\dc- \condmean{\delta}{\S} \right)
  \notag \\
  &=\xi'\!\!\cdot\!\S +
  \frac{\sigma -\xi\!\cdot\!\xi'}{\sigma^2-\xi^2}
  \,(\dc-\xi\!\cdot\!\S) \,,
\label{eq:condmeanx}
\end{align}
after using equations~\eqref{eq:nu|S} and \eqref{eq:nux|S}, and
\begin{align}
  \Var{\delta'}{\nuc, \S}
  &= \Var{\delta'}{\S}
    - \frac{\Cov{\delta',\nu}{\S}^2}{\Var{\nu}{\S}}\,,
    \notag \\
  &= \mean{\delta'^2} - \xi'^2
    - \frac{(\sigma - \xi\!\cdot\xi')^2}{\sigma^2-\xi^2}\,,
\label{eq:condvarx}
\end{align}
respectively. This result is equivalent to decorrelating the effective
variables $\delta-\xi\!\cdot\!\S$ and $\delta'-\xi'\!\cdot\!\S$ introduced in equation~\eqref{eq:effvarS}, whose covariance is in fact $\sigma-\xi'\!\cdot\!\xi$.

Equation~\eqref{eq:condmeanx} contains an angle dependent offset ${\hat r_i\bar\shear_{ij}\hat r_j}\xi_{20}$ and a density dependent one $\xi_{00}\nus$, entering through $\S$. On the contrary,  the conditional variance does not depend on the angle nor the height of the saddle.
At large distance from the saddle, when $\xi=\xi'=0$, equations \eqref{eq:condmeanx} and \eqref{eq:condvarx} tend as expected to the unconditional mean $\nuc$ and variance $1/\Gamma^2=\mean{\delta'^2}-1$.

From equations \eqref{eq:condmeanx} and \eqref{eq:condvarx} one can compute the effective upcrossing parameters presented in the main text
\begin{align}
  &\mu_{\S}(\rr) = \xi'\cdot \S +
  \frac{\sigma-\xi'\cdot\xi}{\sigma^2-\xi^2}
    (\dc-\xi\cdot\S)\,,
\label{eq:machin2} \\
  &X_{\S}(\rr) = \mu_S(\rr)/\sqrt{\Var{\delta'}{\nuc,\S}}\,.
\label{eq:bidule2}
\end{align}


\subsection{Upcrossing at $\sigma$ with given formation time but no saddle}
\label{sec:condhalfmass}

Recalling that $\omega=\mean{\delta\delta_\half}\!/\sigma_{\half}$ and $\omega'=\mean{\delta'\delta_\half}\!/\sigma_{\half}$, as defined by equations \eqref{eq:omega} and \eqref{eq:omegaprime}, the conditional statistics of $\delta$ and $\delta'$ given that $\nu_\half=\nuf$ are
\begin{gather}
  \condmean{\delta}{\nuf} = \omega\nuf \;,\quad
  \Var{\delta}{\nuf} = \sigma^2-\omega^2\,, \notag\\
  \condmean{\delta'}{\nuf} = \omega'\nuf \;,\quad
  \Var{\delta'}{\nuf} = \mean{\delta'^2}-\omega'^2\,, \\
  \Cov{\delta,\delta'}{\nuf} = \sigma-\omega\omega'\,.\notag
\end{gather}
Hence, the conditional mean and variance of $\delta'$ given $\nuc=\dc/\sigma$ and $\nuf$ are
\begin{gather}
  \condmean{\delta'}{\nuc,\nuf} = \omega'\nuf +
  \frac{\sigma-\omega'\omega}{\sigma^2-\omega^2} \,(\dc-\omega\nuf)\,,
\label{eq:truc} \\
  \Var{\delta'}{\nuc,\nuf} = \langle\delta'^2\rangle - \omega'^2
  - \frac{(\sigma-\omega'\omega)^2}{\sigma^2-\omega^2} \,.
  \label{eq:varnucnuf}
\end{gather}
which is equivalent to decorrelating the zero-mean effective variables $\delta-\omega\nuf$ and $\delta'-\omega'\nuf$, whose covariance is $\sigma-\omega'\omega$. From equations \eqref{eq:truc} and \eqref{eq:varnucnuf}, one can compute the parameters of the effective upcrossing problem
\begin{align}
  & \mu_{\form}(\Df) = \condmean{\delta'}{\nuc,\nuf} \,,
\label{eq:machin}  \\
  & X_{\form}(\Df) = \mu_{\form}(\Df)/\sqrt{\Var{\delta'}{\nuc,\nuf}}
\label{eq:bidule} \,,
\end{align}
introduced in Section~\ref{sec:multivar}.

\subsection{Upcrossing at $\sigma$ given formation time and the saddle}

Similarly, thanks to equation~\eqref{eq:nu|S} and \eqref{eq:nux|S}, the mean and covariance of $p_{\rm G}(\nu|\nuf,\S)$ are
\begin{gather}
\begin{split}
  \condmean{\delta}{\nuf,\S}
  &= \condmean{\delta}{\S} + \frac{\Cov{\delta,\nu_{\half}}{\S}}{\Var{\nu_{\half}}{\S}}\left(\nuf - \condmean{\nu_{\half}}{{\cal S}} \right) \,,\\
  &= \xi\!\cdot\!\S  + \Omega\,\nu_{\form,\S}\,,
\label{eq:toto}
\end{split}
\\
\begin{split}
   \Var{\delta}{\nuc, \S}
  &= \Var{\delta}{\S} - \frac{\Cov{\delta, \nu_{\half}}{\S}^2}{\Var{{\nu_{\half}}}{\S}}\,, \\
   &= \sigma^2-\xi^2- \Omega^2\,,
\label{eq:titi}
\end{split}
\end{gather}
where (recalling that $\xi$ has the dimensions of $\delta$ but $\xi_{\half}$ has those of $\nu$, see equation~\eqref{eq:compact_notation})
\begin{equation}
  \nu_{\form,\S} \equiv \frac{(\nuf - \xi_{\half}\!\cdot\!\S)}{\sqrt{1-\xi_{\half}^2}}\;,\quad
  \Omega\equiv \frac{\omega-\xi\!\cdot\!\xi_{\half}}{\sqrt{1-\xi_{\half}^2}}\,.
\label{eq:Omega}
\end{equation}

As discussed in Appendix~\ref{sec:condstats}, the statistics of $p_{\rm G}(\delta'|\nuc,\nuf,\S)$ can be derived from those of $p_{\rm G}(\delta|\nuf,\S)$ as follows:
\begin{align}
\hskip -0.1cm  \condmean{\delta'}{\nuc,\nuf,\S} \!=\! \condmean{\delta}{\nuf,\S}'
  \!+\! \frac{\Var{\delta}{\nuf,\S}'}{2\Var{\delta}{\nuf,\S}}(\dc\!-\!\condmean{\delta}{\nuf,\S})
\end{align}
thanks to the relations $\condmean{\delta}{\nuf,\S}'=\condmean{\delta'}{\nuf,\S}$ and $\Var{\delta}{\nuf,\S}'=2\Cov{\delta\delta'}{\nuf,\S}$, and \begin{equation}
  \Var{\delta'}{\nuc,\nuf,\S} = \Var{\delta'}{\nuf,\S}
  - \frac{[\Var{\delta}{\nuf,\S}']^2}{4\Var{\delta}{\nuf,\S}}\,.
\end{equation}
Hence, taking derivatives of equations \eqref{eq:toto} and \eqref{eq:titi} gives
\begin{multline}
  \condmean{\delta'}{\nuc,\nuf,\S} = \xi'\!\cdot\!\S
  + \Omega'\nu_{\form,\S} \\
  +\frac{\sigma-\xi'\!\cdot\!\xi-\Omega'\Omega}{\sigma^2-\xi^2-\Omega^2}(\dc-\xi\!\cdot\!\S  - \Omega\,\nu_{\form,\S})\,,
\label{eq:meandp_nucnufS}
\end{multline}
and
\begin{multline}
 \hskip -0.4cm \Var{\delta'}{\nuc,\nuf,\S} \!=\! \mean{\delta'^2} \!-\! \xi'^2 \!-\Omega'^2
 -\frac{(\sigma\!-\!\xi'\!\cdot\!\xi\!-\!\Omega'\Omega)^2}{\sigma^2\!-\!\xi^2\!-\!\Omega^2},
\label{eq:vardp_nucnufS}
\end{multline}
where
\begin{equation}
  \Omega' = \frac{\omega'-\xi'\!\cdot\!\xi_{\half}}{\sqrt{1-\xi_{\half}^2}}\,,
\label{eq:Omegap}
\end{equation}
which can finally be used to compute the effective slope parameters
\begin{align}
  \mu_{\form,\S}(\Df,\rr)
  &= \condmean{\delta'}{\nuc,\nuf,\S}\,,
\label{eq:machin3} \\
  X_{\form,\S}(\Df,\rr)
  &= \mu_{\form,\S}(\Df,\rr)/\sqrt{\Var{\delta'}{\nuc,\nuf,\S}}\,.
\label{eq:bidule3}
\end{align}
\balance
\section{Generic and moving barrier}
\label{sec:moving-barr-gener}
The results presented hereby hold for a constant barrier, however, one can easily recover the results for a non-constant one -- where the upcrossing conditions becomes $\dc>\dc'$ -- by replacing $\mu_v$ by $\mu_v - \dc'$ in the general formula of Equations~\eqref{eq:genfup} and~\eqref{eq:X}, yielding
\begin{equation}
  \mu_v \equiv \condmean{\delta'}{\nuc,\{v\}}-\dc'\,,
\end{equation}
and by taking into account contributions from $\dc'$ in $\nuc'$
\begin{equation}
  \nuc' = \frac{\dc'}{\sigma} - \frac{\dc}{\sigma^2}\,,
\end{equation}
and in the definition of accretion rate
\begin{equation}
  \alpha = \frac{\dc}{\sigma(\delta'-\dc')}
\end{equation}
in equation \eqref{eq:accrate}.
In practical terms, dealing with a moving barrier simply amounts to replacing
\begin{align}
  \mu &\to \mean{\delta'|\nuc} -\dc'\,, \\
  \mu_\form &\to \mean{\delta'|\nuc,\nuf} -\dc'\,, \\
  \mu_\S &\to \mean{\delta'|\nuc,\S} -\dc' \,,\\
  \mu_{\form,\S} &\to \mean{\delta'|\nuc,\nuf,\S} -\dc'\,,
\end{align}
in equations \eqref{eq:X0}, \eqref{eq:Xf}, \eqref{eq:XS} and \eqref{eq:XfS}, which automatically affects also the corresponding $X$, $X_\form$, $X_\S$ and $X_{\form,\S}$, as well as $Y_{\alpha}$ and $Y_{\alpha,\S}$ in equation \eqref{eq:Yalpha} and \eqref{eq:defYalpha}.

For instance, for a barrier of the type $\dc+\beta\sigma\bar q_{ij,R}\bar q_{ij,R}$  \citep{Castorina:2016vg},
where $\bar q_{ij,R}$ is the traceless tidal tensor smoothed on scale $R$, and $\beta$ is some constant,
one would use
\begin{equation}
  \dc' \to \beta
  (\bar q_{ij,R}\bar q_{ij,R}+2\sigma\bar q_{ij,R}'\bar q_{ij,R}) \,.
\end{equation}
More generally, barriers should involve $\{I_{n}\}$  the rotationally invariants of $\bar q_{ij,R}$  defined in Section~\ref{sec:pdf-saddles}.

\section{Implied galactic colours}
\label{sec:speculations}
\begin{figure}
  \includegraphics[width=0.5\columnwidth]{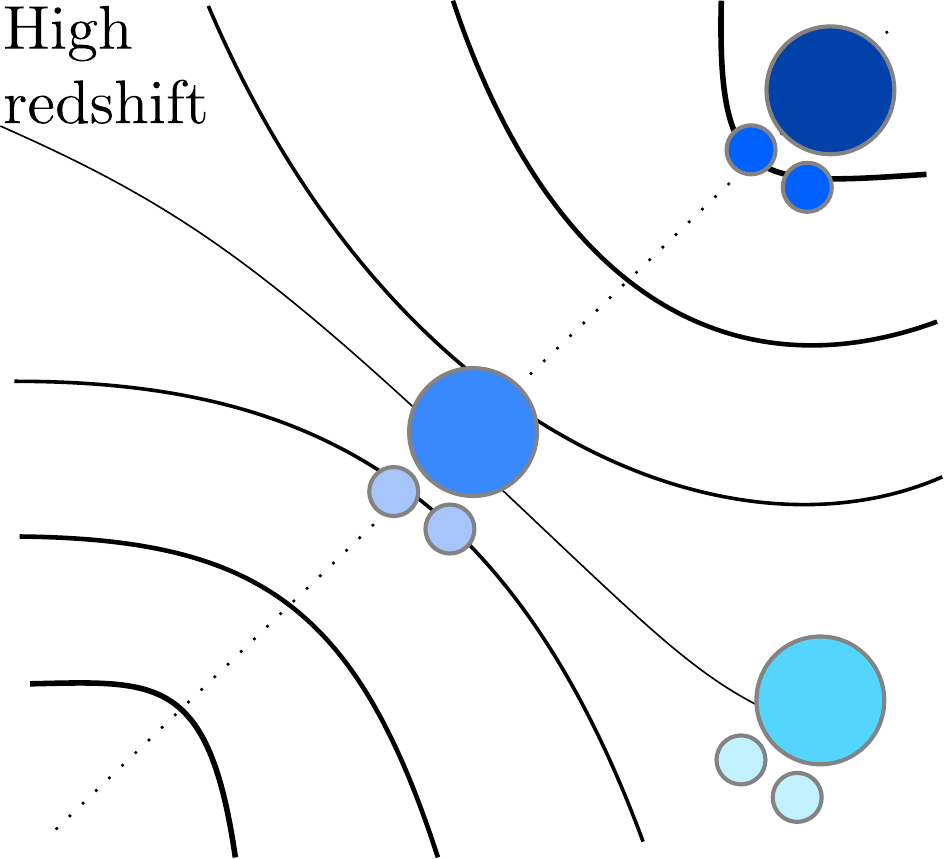}
  \hskip 0.1cm
  \includegraphics[width=0.5\columnwidth]{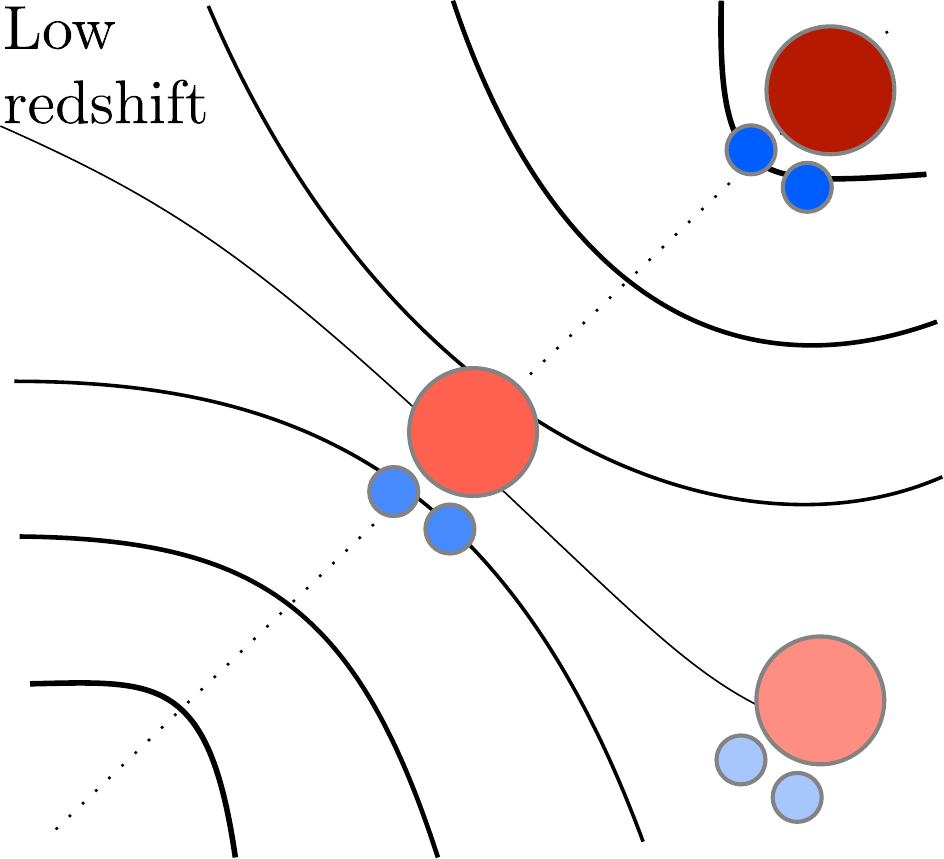}
  \caption{Scheme of the intensity of expected colour/SFR at different
    location near a filament-type saddle for different final halo
    mass. The displayed colour encodes galactic colour (or
    equivalently sSFR from high-blue to low-red).  Massive galaxies in
    the filament (resp. nodes) are expected to accrete more cold
    baryonic matter at high redshift and be bluer than less massive
    ones and than their counterparts in voids (resp. filaments). At
    lower redshifts, AGN feedback is expected to quench cold gas
    accretion hence redden the massive ones -- they are more likely to
    be be central ones. The impact on smaller ones -- they are more
    likely to be satellites --- may also depend on the efficiency of
    starvation, ram pressure stripping etc.}
      \label{fig:color-scheme}
\end{figure}
Let us in closing attempt to convert the position dependent accretion rates computed in the main text in terms of colour and specific star formation rate
modulo some reasonable assumption on the resp. role of AGN et star formation rate at low and high redshift.
Simply put, colour is directly proportional to recent star formation, which in turn is driven by the availability of pristine gas.
In filaments, one could expect that gas infall is proportional to dark matter infall.
One further complication comes from the impact of feedback on heating cold gas.
Indeed, hydrodynamical simulations which include sub-grid physics modeling the role of supermassive black holes suggests that
at intermediate and low redshift,  merger triggers AGN feedback, which in turn heat up the CGM and prevent subsequent cold flows from feeding central galaxies.  Conversely, at higher redshift,  these cold flows reach the centres of dark halos unimpaired and matter infall translates into bluer
galaxies.
  Fig.~\ref{fig:color-scheme} sketches these ideas, while distinguishing low and high mass halos.
As argued in the main text, this scenario remains speculative, if
only because the impact of AGN feedback is still a fairly debated
topic. For instance ram pressure stripping on satellites plunging into clusters is known to induce reddening, but its efficiency  within filaments is  unclear.
Fig.~\ref{fig:conclusive-scheme}	encodes the robust result of the present investigation.
\end{document}